Corresponding Author: Dr. Kilho Eom, Ph.D.

Corresponding Author's Institution: Korea University

First Author: Kilho Eom, Ph.D.

Order of Authors: Kilho Eom, Ph.D.; Harold S Park; Dae Sung Yoon; Taeyun Kwon

# Nanomechanical Resonators and Their Applications in Biological/Chemical Detection: Nanomechanics Principles

Kilho Eom[1,*], Harold S. Park[2,†], Dae Sung Yoon[3,‡], and Taeyun Kwon[3,§]
[1]Department of Mechanical Engineering, Korea University, Seoul 136-701, Republic of Korea
[2]Department of Mechanical Engineering, Boston University, Boston, MA 02215, USA
[3]Department of Biomedical Engineering, Yonsei University, Kangwon-do 220-740, Republic of Korea

Correspondence should be addressed to K.E. (kilhoeom@korea.ac.kr), H.S.P. (parkhs@bu.edu), D.S.Y. (dsyoon@yonsei.ac.kr), and T.K. (tkwon@yonsei.ac.kr).

[*]Address: Department of Mechanical Engineering, Korea University, Seoul 136-701, Republic of Korea
E-mail: kilhoeom@korea.ac.kr or kilhoeom@gmail.com
Phone: +82-2-3290-3870
Fax: +82-2-6008-3855

[†]Address: Department of Mechanical Engineering, Boston University, Boston, MA 02215, USA
E-mail: parkhs@bu.edu
Phone: +1-617-353-4208
Fax: +1-617-353-5866

[‡]Address: Department of Biomedical Engineering, Yonsei University, Kangwon-do 220-740, Republic of Korea
E-mail: dsyoon@yonsei.ac.kr
Phone: +82-33-760-2479
Fax: +82-33-760-2579

[§]Address: Department of Biomedical Engineering, Yonsei University, Kangwon-do 220-740, Republic of Korea
E-mail: tkwon@yonsei.ac.kr
Phone: +82-33-760-2648
Fax: +82-33-760-2579

**Abstract**
Recent advances in nanotechnology have led to the development of nano-electro-mechanical systems (NEMS) such as nanomechanical resonators, which have recently received significant attention from the scientific community. This has not only been for their capability for the label-free detection of bio/chemical-molecules at single-molecule (or atomic) resolution for future applications such as the early diagnostics of diseases such as cancer, but also for their unprecedented ability to detect physical quantities such as molecular weight, elastic stiffness, surface stress, and surface elastic stiffness for adsorbed molecules on the surface. Most experimental works on resonator-based molecular detection have been based on the principle that molecular adsorption onto a resonator surface increases the effective mass, and consequently decreases the resonant frequencies of the nanomechanical resonator. However, this principle is insufficient to provide fundamental insights into resonator-based molecular detection at the nanoscale; this is due to recently proposed novel nanoscale detection principles including various effects such as surface effects, nonlinear oscillations, coupled resonance, and stiffness effects. Furthermore, these effects have only recently been incorporated into existing physical models for resonators, and therefore the universal physical principles governing nanoresonator-based detection have not been completely described. Therefore, our objective in this review is to overview the current attempts to understand the underlying mechanisms in nanoresonator-based detection using physical models coupled to computational simulations and/or experiments. Specifically, we will focus on issues of special relevance to the dynamic behavior of nanoresonators and their applications in biological/chemical detection: the resonance behavior of micro/nano-resonators; resonator-based chemical/biological detection; physical models of various nanoresonators such as nanowires, carbon nanotubes, and graphene. We pay particular attention to experimental and computational approaches that have been useful in elucidating the mechanisms underlying the dynamic behavior of resonators across multiple and disparate spatial/length scales, and the resulting insight into resonator-based detection that has been obtained. We additionally provide extensive discussion regarding potentially fruitful future research directions coupling experiments and simulations in order to develop a fundamental understanding of the basic physical principles that govern NEMS and NEMS-based sensing and detection applications.

**Table of Contents**

1. Introduction

The past decade has witnessed the emergence of nanotechnology that enables the development of nanoscale functional devices designed for specific aims such as nanoscale actuation, sensing, and detection [1-3]. For instance, micro/nano-electro-mechanical system (MEMS/NEMS) devices have allowed the sensitive detection of physical quantities such as spin [4, 5], molecular mass [6-10], quantum state [11] (see also Refs. [12-14] which discuss the prospect of NEMS for studying quantum mechanics), thermal fluctuation [15-18], coupled resonance [19-21], and biochemical reactions [22-26]. Among MEMS/NEMS devices, nanomechanical resonators have been recently highlighted for their unprecedented dynamic characteristics as they can easily reach ultra-high-frequency (UHF) and/or very-high-frequency (VHF) dynamic behavior up to the Giga Hertz (GHz = $10^9$ Hz) regime [3, 27-29]. Reaching this frequency range is critical as it implies that nanoresonators can be directly utilized as an electronic device for radio communications. This high-frequency dynamic behavior is achieved by scaling down the size of the resonator because the resonant frequency is proportional to $L^{-2}$, where $L$ is the length of a device. Therefore, if the resonator length is decreased by an order of magnitude, then its resonant frequency is increased by two orders of magnitude. Furthermore, the ability of the resonator to sense or detect physical quantities (i.e. mass, force or pressure) is closely related to its resonant frequency. For example, for sensing mass that is added onto a resonator, the detection sensitivity is given by the relation $\Delta f_n/\Delta m = (1/2m)f_n$ [30-32], where $f_n$ and $m$ represent the resonant frequency and the effective mass of a device, respectively, while $\Delta f_n$ and $\Delta m$ indicate the resonant frequency shift and the added mass. Clearly, this relationship suggests that as the frequency of the resonator increases, so does its ability to sense or detect ever smaller masses, which implies that UHF/VHF resonators are suitable for ultra sensitive detection, where the eventual limit of a single atom or molecule is experimentally within reach.

An example of the incredible potential of NEMS resonators can be found in recent works by Roukes and coworkers [6, 7, 10], who first showed the possibility of nanoscale mass spectrometers that enable the measurement of the molecular weight of specific molecules. This implies not only that nanomechanical resonators could be a viable alternative to conventional mass spectrometry techniques such as matrix-assisted desorption/ionization time-of-flight (MALDI-TOF), but also that mass spectrometry could be realized in a lab-on-a-chip [33]. It should be emphasized that NEMS-based sensing is not restricted to small, molecular masses; other important physical quantities such as quantum state [11], spin [4, 5], and force [34-36] can also be detected using NEMS, which suggests that nanomechanical resonators may allow the realization of lab-on-a-chip sensing toolkits for detecting other relevant physical quantities [1, 3].

In recent years, micro/nano-mechanical resonators have also received significant attention for their capability of label-free detection of specific biological molecules [10, 22, 37, 38] and/or cells [25, 39, 40], even at low concentrations, that are relevant to specific diseases such as cancer [41, 42]. However, current biosensing tools such as enzyme-linked immunosorbent assay (ELISA) exhibits a key restriction in that they are unable to accurately detect marker proteins (relevant to specific cancers) in the concentration of ~1 ng/ml, which is known as the "diagnostic gray zone" [43], in blood serum. On the other hand, micro/nano-mechanical resonators are able to easily overcome the "diagnostic gray zone" limitation because of their unprecedented detection sensitivity even at single-molecule resolution [9, 10, 44], which shows that nanomechanical resonators can serve as lab-on-a-chip biosensors enabling the early diagnostics of important diseases such as cancer. Moreover, nano/micro-mechanical resonators show the promising ability to provide the detailed mechanisms of biochemical reactions [23, 24, 26, 45, 46] and/or cell functions [47].

As stated earlier, the excellent performance of nanomechanical resonators for sensing applications is highly correlated with their dynamic characteristics [30-32]. It is therefore essential to characterize and understand the dynamic behavior of nanomechanical resonators for the novel design of resonator-based sensing toolkits. Furthermore, it has been suggested that nanoresonators are easily able to exhibit unique dynamic features such as nonlinear oscillations [48, 49] and/or coupled resonance [19, 20]. For instance, nanostructures can easily be tuned to oscillate nonlinearly by modulating the actuation force so as to drive the geometric nonlinear deformation of doubly-clamped nanostructures [28, 50, 51]; we note that this would also enable fundamental investigations into various theories underlying the field of nonlinear vibrations. In addition, it has been reported that coupled nanomechanical resonators not only provide unique dynamic features such as coupled oscillation [19, 20], but also enable ultra sensitive mass detection [21]. These show that unique dynamic features such as nonlinear oscillations and/or coupled resonance could be a new avenue to improve the detection sensitivity of nanoresonators.

Moreover, because nanoresonators are characterized by a large surface-to-volume ratio, surface effects play a critical, but currently not well-understood role on their dynamic characteristics. Specifically, as the resonator size is scaled down, its surface area is increased which leads to an increase of its surface energy [52-54], which is defined as the energetic cost to create a new surface. In general, the surface energy $U_S$ depends on the deformation of the surface, and consequently the surface stress can be defined as $\tau = \partial U_S/\partial \varepsilon_s = \tau_0 + S\varepsilon_s + O(\varepsilon_s^2)$ [55, 56], where $\tau_0$, $\varepsilon_s$, and $S$ represents the constant surface stress, surface strain, and surface elastic stiffness, respectively. The surface stress is also inherent to nanostructures due to the fact that surface atoms have fewer bonding neighbors than do bulk atoms; because they are therefore not at equilibrium, they are subject to a surface stress [57], which causes deformation of the surface in the absence of external forces [58-60]. Both of these explanations strongly suggest that surface effects such as surface stress play a key role on the mechanical properties and thus the resonant frequencies of a nanostructure. Furthermore, because the detection sensitivity is correlated with the resonant frequency of a nanomechanical resonator, the surface stress will inherently have a significant effect on the detection sensitivity for nanoresonators. In addition, for sensing applications, the adsorption of molecules onto the surface of a nanoresonator induces changes of the surface state, and consequently the surface energy (or equivalently the surface stress) by changing the bonding configuration for the surface atoms. Therefore, the effect of surface stress changes that result due to molecular adsorption has to be carefully considered for gaining insight into the detection principle for NEMS. We note that surface effects on the resonance behaviors as well as the detection mechanisms of nanoresonators are a very active and on-going area of research.

The purpose of our review article is to present not only the current state-of-art in the development of nanomechanical resonators and their applications in chemical/biological sensing, but also the physical principles that have enabled fundamental insights into the underlying mechanisms for the dynamic characteristics of nanomechanical resonators as well as the detection principles. Specifically, we describe not only the current state-of-art experimentally, but also the on-going development in theoretical and computational techniques that are being used to develop a fundamental understanding of NEMS-based resonators across multiple spatial/length scales ranging from atomistics to continua. We first overview the experimental approaches that are used to characterize the dynamic behavior of micro/nano-resonators as well as their applications to sensing chemical and/or biological species. Subsequently, we review the continuum elastic models that are able to explain the fundamental physics of nanomechanical resonators such as nonlinear oscillations, coupled resonance and detection principles. However, these classical continuum elastic models are unable to capture the underlying physics of nanoscale surface effects, which leads us to review recent molecular/multiscale modeling of nanoresonators such as carbon nanotube, nanowire, and graphene in order to understand their nanoscale driven dynamic behavior and detection principles for resonator-based sensing.

We anticipate that the theoretical and computational models reviewed here will allow one to gain insight into not only the anomalous dynamics experimentally observed (e.g. surface effect-coupled resonance behavior) but also the fundamental, novel detection principles for sensing applications. For instance, theoretical and computational models that take surface effects into account are able to explain the resonance behavior of nanocantilevers, which cannot be understood by conventional continuum models that do not account for surface effects. Furthermore, the experimentally observed resonance response of a nanocantilever to biomolecular adsorption has not been well described by conventional detection principles, because the resonant frequency shift for nanoresonators due to biomolecular adsorptions depends on not only the mass of adsorbates but also other unexpected effects, which were not considered in conventional detection principles, such as elastic stiffness of adsorbates [61, 62], the change of surface elastic stiffness [63-65], and surface stress [63-66] due to adsorbates. These indicate that theoretical and computational models are complimentary to the experimentally observed unique behavior of nanomechanical resonators and their sensing applications. In addition, theoretical and/or computational models are able to provide milestones for further guidance and design of novel nanoresonators and their sensing applications.

Our review article is organized as follows: Section 2 reviews the current progress in the development of micro/nanoresonators and their related harmonic dynamic behavior such as resonant frequencies and Q-factors within the physical background of continuum elasticity. In Section 3, we review micro/nanoresonator-based sensing applications such as chemical and/or biological detection. Furthermore, we provide perspectives and challenges of the current state-of-art in resonator-based chemical/biological sensing applications. Section 4 overviews the continuum elastic models that enable the understanding of not only the unique dynamic features such as nonlinear oscillations and coupled resonance but also the novel detection principles such as mass sensing based on coupled resonance and/or nonlinear oscillations. Section 4 also discusses the controversial issues surrounding continuum elastic models for gaining insight into surface stress effects on the resonant frequencies of nanoresonators. Section 5 reviews the current computational approaches based on molecular/multiscale models that have been utilized in order to understand the surface-induced dynamic behavior of nanoresonators that cannot be captured by the standard continuum models discussed in Section 2, and their related sensing applications. We present a future outlook in Section 6, while Section 7 concludes our review with closing remarks.

## 2. Dynamics of Micro/Nanomechanical Resonators

As previously discussed, the operational principle underlying micro and nanoresonators is that they can be used to detect minute forces, masses, biological/chemical species through the resulting changes that are induced in their resonant frequencies. Because this shows that characterization of the resonant frequency and/or its shift due to mass is essential to elucidate the sensing performance for a resonator, we now overview the theories that have been utilized to not only analyze the results obtained experimentally, but also to predict new behavior and properties of micro/nanoresonators.

In general, it has been accepted that the resonance behavior of a MEMS/NEMS device is well described by continuum elasticity theory. For instance, the dynamic behavior of microcantilevers has been shown experimentally to be accurately represented by the harmonic, flexural motion that is assumed in classical linear elastic Bernoulli-Euler beam theory. In this section, the resonance behavior of MEMS/NEMS devices is reviewed. First, we will review the current experimental work on the development of MEMS/NEMS resonators as well as the quantitative description on their frequency behaviors. Second, the mechanism of energy dissipation as quantified using the quality (Q)-factors will be reviewed. Here, we categorize the Q-factors as a function of both intrinsic and extrinsic properties of MEMS/NEMS such as surface effects, damping effects, clamping losses, and thermoelastic damping. Finally, we review recent works using mechanical manipulation via stress and strain to tailor and tune the resonant frequencies and Q-factors of MEMS/NEMS.

Here, our review is restricted to experimentally observed resonance motion of micro/nanostructures, which is well described by harmonic resonance motions. That is, the nonlinear effect due to boundary conditions such as double-clamping is neglected, since the amplitude of the vibrational motion of micro/nanostructures reviewed here is small enough such that nonlinear effects do not play any role in the vibration. The nonlinear effects on the vibrational motion will be discussed later in Section 4.1.1. Furthermore, we do not take into account the effect of surface stress on the resonance motion, because the length scale of micro/nanostructures reviewed here is larger than the critical size at which surface stress effect appears. Specifically, various experimental studies on both metallic and semiconducting nanostructures have shown that when the cross-sectional dimension (i.e. diameter) of a nanobeam or nanoresonator becomes smaller than about 50-100 nm, surface effects begin to have a significant effect on the bending rigidity, and consequently the resonant frequencies and the effective elastic modulus [67]. The surface stress effects will be described later in Section 4.3.

## 2.1. Characterization of Resonance Behavior

### 2.1.1. Characterization of Resonance: Theory

We first describe the well-known Bernoulli-Euler beam theory [68-71], which captures the vibrational motion of MEMS/NEMS devices that have characteristic sizes ranging from micro- to nano-meters. In general, these devices (e.g. nanowire, nanotube, microcantilever) have a geometry such that the transverse dimensions (i.e. thickness and width) are much smaller than the longitudinal dimension (i.e. length). This indicates that the device can be modeled as a one-dimensional elastic beam which assumes that the vibrational motion is governed by flexural motion. Consequently, the vibrational motion of MEMS/NEMS can be described by the following equation of motion.

$$EI\frac{\partial^4 w(x,t)}{\partial x^4} + c\frac{\partial w(x,t)}{\partial t} + \rho A\frac{\partial^2 w(x,t)}{\partial t^2} = 0 \tag{1}$$

where $w(x, t)$ is the flexural deflection as a function of coordinate $x$ and time $t$, $c$ is a damping coefficient due to viscous damping effect, and $E$, $I$, $\rho$, and $A$ represent the elastic modulus, the cross-sectional moment of inertia, the mass density, and the cross-sectional area of the device, respectively. Again, we note that it is assumed that the device experiences vibrational motion with small deflection amplitude, i.e. it follows the harmonic approximation. In other words, in the theoretical model given by Eq. (1), the effect of geometric nonlinearity has been ignored, though we will consider such effect later in Section 4.1.1, while the mathematical approach to accounting for geometric nonlinearity is demonstrated below [e.g. see Eq. (7)]. Furthermore, for cases where the device vibrates in vacuum or air, damping effects on the resonant frequency can be neglected. By writing the flexural deflection in the form $w(x, t) = u(x) \times \exp[j\omega t]$, where $\omega$ and $u(x)$ indicate the resonant frequency and its corresponding deflection eigenmode, respectively, and $j$ is a unit of complex number, i.e. $j = (-1)^{1/2}$, the equation of motion can be transformed into the following eigenvalue problem:

$$\wp u = \lambda u \tag{2.a}$$

where

$$\wp = EI\frac{\partial^4}{\partial x^4} \quad \text{and} \quad \lambda = \omega^2 \rho A \tag{2.b}$$

When a specific boundary condition is prescribed, the resonant frequency and its corresponding deflection eigenmode can be straightforwardly obtained from the eigenvalue problem given in Eq. (2). Fig. 1 shows the vibrational behavior of microcantilevers, which are typically utilized in atomic force microscopy (AFM), with different geometries and/or force constants (spring constants). It is shown that the fundamental deflection eigenmode is well described by the theoretical model in Eq. (1), while the high-frequency deflection eigenmode is less accurately captured due to the coupling between flexural motion and torsional motion [72]. To capture mode coupling in the high-frequency behavior in an AFM cantilever, it is important to include terms related to torsional motion in the equation of motion. It is generally difficult to analytically solve the equation of motion for such a case, while it can be computationally tractable using finite element methods [73]. The topic of mode-coupling in the vibrational dynamics of micro/nanocantilevers has been recently received special attention because the mode-coupling has been found to improve the AFM imaging quality [74-76].

### 2.1.2. Resonance of Piezoelectric Microcantilever

We now consider the resonance behavior of micro- and/or nanocantilevers that are actuated by piezoelectricity. Unlike the former case, a piezoelectric cantilever consists of two layers, a piezoelectric layer and a substrate layer (for details, see Table 1). Herein, the piezoelectric layer plays the role on actuating the cantilever, while the substrate layer is used for nano- or micro-patterning. Recently, Kwon et al. [77, 78] developed a novel piezoelectric thick film, i.e. a $0.1Pb(Zn_{0.5}W_{0.5})O_3$-$0.9Pb(Zr_{0.5}Ti_{0.5})O_3$ (PZW-PZT) thick film, that allows the actuation of micro devices with large actuation forces. Using this PZW-PZT thick film, they developed a piezoelectric unimorph microcantilever that can vibrate with large actuation forces even in a viscous medium [79]. The resonance behavior of the piezoelectric unimorph microcantilever is well described by classical linear elastic Euler-Bernoulli beam theory. Specifically, the flexural rigidity for the microcantilever composed of two layers is given as [79, 80]

$$D = \frac{E_p^2 h_p^4 + E_s^2 h_s^4 + 2 E_p E_s h_p h_s \left(2 h_p^2 + 2 h_s^2 + 3 h_p h_s\right)}{12 \left(E_p h_p + E_s h_s\right)} \quad (3)$$

Here, $D$ is the flexural rigidity per unit length for a piezoelectric unimorph microcantilever, $E_p$ and $E_s$ represent the elastic moduli of the piezoelectric thick film and silicon substrate, respectively, while $h_p$ and $h_s$ indicate the thicknesses of piezoelectric thick film and silicon substrate, respectively. The effective mass per unit length is defined as $m^* = \rho_p h_p + \rho_s h_s$, where $\rho_p$ and $\rho_s$ are the mass densities of the piezoelectric thick film and silicon substrate, respectively. Accordingly, the resonant frequency of such a microcantilever operating in vacuum or air can be written as

$$f_n = \frac{\lambda_n^2}{2\pi L^2} \sqrt{\frac{D}{m^*}} \quad (4)$$

where $f_n$ is the $n$-th resonant frequency, and $\lambda_n$ is the $n$-th eigenvalue for the $n$-th flexural mode for the cantilevered boundary condition. By introducing the effective elastic modulus $E^*$, the resonant frequency becomes

$$f_n = \frac{\lambda_n^2}{2\pi\sqrt{12}} \left(\frac{h}{L^2}\right) \sqrt{\frac{E^*}{\rho^*}} \quad (5)$$

where $h$ is the thickness of the microcantilever given as $h \equiv h_p + h_s$, and $E^*$ and $\rho^*$ are the effective elastic modulus and the effective mass density defined as

$$E^* = \frac{E_p^2 r_p^4 + E_s^2 \left(1 - r_p\right)^4 + 2 E_p E_s \left(1 - r_p\right)\left\{2 r_p^2 + 2\left(1 - r_p\right)^2 + 3 r_p \left(1 - r_p\right)\right\}}{E_p r_p + E_s \left(1 - r_p\right)} \quad (6.a)$$

$$\rho^* = \rho_p r_p + \rho_s \left(1 - r_p\right) \quad (6.b)$$

where $r_p$ is defined as $r_p = h_p/h$. It is shown that the resonance behavior of the piezoelectric unimorph microcantilever is well characterized by the Euler-Bernoulli beam theory given in Eq. (4) or Eq. (5). Fig. 2 presents the resonance curve of the piezoelectric unimorph microcantilever, where the PZW-PZT thick film was used as an actuating layer. In addition, Table 2 summarizes the resonant frequencies which are measured from experiments and calculated from Euler-Bernoulli beam theory. It suggests that classical elasticity theories such as Euler-Bernoulli beam theory are able to characterize the resonance behavior of piezoelectric microcantilevers.

2.1.3. Resonance Behavior of Doubly-Clamped Nanobeams

We consider the resonance behavior of doubly-clamped nanostructures (e.g. nanowires, nanotubes, etc.) which are fabricated using either a bottom-up approach or top-down approach. Recently, a few research groups [28, 29, 81-86] have reported the ultrahigh resonance behavior of doubly-clamped nanostructures such as nanowires [28, 82], nanotubes [29, 83, 84], graphene [85, 86], and/or fabricated nanobeams [81]. The vibration dynamics of a doubly-clamped beam is described as

$$EI \frac{\partial^4 w(x,t)}{\partial x^4} - \left[\frac{EA}{2L} \int_0^L \left\{\frac{\partial w(x,t)}{\partial x}\right\}^2 dx\right] \frac{\partial^2 w(x,t)}{\partial x^2} + \mu \frac{\partial^2 w(x,t)}{\partial t^2} = f(x,t) \quad (7)$$

where $w(x, t)$ is the deflection of the nanowire as a function of coordinate $x$ and time $t$, $E$, $I$, and $\mu$ represent the elastic modulus, the cross-sectional moment of inertia, and the effective mass per unit length of a nanobeam, respectively, while $f(x, t)$ is the actuating force per unit length induced by an external field such as an electric field [83, 84, 87], magnetic field [82], and/or optical drive [88]. Herein, the effective mass per unit length for a nanobeam (without any mass adsorption) is given by $\mu = \rho A$, where $\rho$ and $A$ are the mass density and the cross-sectional area of the nanobeam. For small amplitude vibrational motion, the resonance behavior obeys the harmonic approximation, for which the resonant frequency of a nanowire is given by

$$f_n = \frac{\mu_n^2}{2\pi L^2}\sqrt{\frac{EI}{\rho A}} \tag{8}$$

where $\mu_n$ is the *n*-th eigenvalue for the *n*-th flexural mode with doubly-clamped boundary condition, e.g. $\mu_1$ = 4.73 (for the fundamental flexural mode), $\mu_2$ = 7.85 (for the second-harmonic flexural mode), etc. The harmonic vibrational behavior of doubly-clamped nanostructures is well described in Ref. [81] (see also Fig. 3), which shows the resonance behavior of a piezoelectric GaAs beam. In particular, an electric field is applied perpendicular to the longitudinal direction of a GaAs beam in order to induce its resonace motion. Here, it should be noted that a transversely applied electric field (i.e. d.c. bias with an a.c. driving amplitude fixed at 70 mV) produces the longitudinal strain in the GaAs beam due to the piezoelectric effct. In other words, due to the boundary condition, an electric field generates a residual stress, which effectively serves as an axial force acting on a doubly-clamped beam. Specifically, the residual stress induced by an electric field is given as $\sigma_r = Ed_{3j}V$, where $d_{3j}$ is an anisotropic piezoelectric constant and $V$ is the d.c. bias voltage applied transversely to the beam. Accordingly, with axial load $S$ driven by boundary condition, i.e. $S = \sigma_r A$, the resonant frequency of a doubly-clamped beam is given by

$$f_n^* = \frac{n^2\pi}{L^2}\sqrt{\frac{EI}{\rho A}\left(1 + \frac{d_{3j}VL^2}{n^2\pi^2 I}\right)} \tag{9}$$

Then, assuming that the frequency shift due to residual stress is small such that $d_{3j}VL^2/n^2\pi^2 I \ll 1$, the fundamental resonant frequency shift due to the transversely applied electric field, $V$, can be written as

$$\Delta f_1 \equiv f_1^* - f_1 \approx -\left(d_{3j}V/2\pi t^2\right)\sqrt{3E/\rho} \tag{10}$$

The relationship between the transversely applied electric field and the resonant frequency of a doubly-clamped nanobeam, which experiences the harmonic vibration, is well described by continuum mechanics model given in Eq. (10) [see Fig. 3c]. Specifically, as shown in Fig. 3c, for a beam fabricated along the [110] crystallographic direction, the increase in the resonant frequency due to d.c. bias is observed due to negative piezoelectric constant that is estimated as $d_{31}$ = –1.33 pm/V [81]. On the other hand, a beam fabricated along the [-110] crystallographic direction exhibits a positive piezoelectric constant, which is responsible for the decrease in the resonant frequency due to applied d.c. bias (Fig. 3c).

*2.1.4. Resonance Behavior in an Aqueous Environment*

Until now, we have only considered the vibration of MEMS/NEMS that are operated in vacuum, where the hydrodynamic effects that capture the interaction between MEMS/NEMS devices and the environment are neglected. However, a quantitative understanding of the role of the hydrodynamic effect in the vibration behavior of MEMS/NEMS is essential since the dynamic performance of many MEMS/NEMS devices is strongly dependent on the environment (e.g. gas, air, or liquid) in which the MEMS/NEMS device operates. In particular, the resonance behavior of MEMS devices that are required to operate in aqueous environment will be significantly impacted by the hydrodynamic effect. For instance, when a microcantilever acting as an AFM tip vibrates in an aqueous environment in order to image a biological sample in an aqueous environment [89, 90], the hydrodynamic effect significantly deteriorates the resonance behavior of microcantilever, and consequently restricts the resolution of the AFM image. Furthermore, cantilever sensors have recently been employed for *in situ* detection of biological molecules and/or biomolecular interactions in an aqueous environment [22, 23, 26, 45, 91], where the hydrodynamic effect significantly reduces the resonant frequencies, which consequently reduces the detection sensitivity. The hydrodynamic effect also has a significant impact on the resonance behavior of NEMS in air. For example, when a nanoscale resonator (e.g. carbon nanotube resonator) vibrates in a gaseous environment, gas damping is the largest source of energy dissipation in both air [92] and gas even for pressures as low as $10^{-3}$ torr [93, 94].

We consider environmental effects on the resonance behavior of MEMS/NEMS. As mentioned earlier, this effect (i.e. interction between environment and a resonator) is referred to as the hydrodynamic effect. Here, we revisit Sader's model [95], which provides an accurate theoretical description of hydrodynamic effects in the resonance behavior of MEMS/NEMS immersed in a viscous fluid. It is noted that Sader's model [95] has been rigorously verified by subsequent simulation work by Paul and Cross [92], finite element simulations by Raman and coworkers [96], and experimental measurements by Chon et al. [97]. The equation of motion for the vibration of MEMS/NEMS subject to the hydrodynamic effect is given by [95]

$$EI\frac{\partial^4 w(x,t)}{\partial x^4} + \rho A \frac{\partial^2 w(x,t)}{\partial x^2} = f_H(x,t) + f_{drive}(x,t) \tag{11}$$

where $E$, $I$, $\rho$, and $A$ indicate the Young's modulus, cross-sectional moment of inertia, density, and cross-sectional area of a resonator, respectively, $f_H(x, t)$ is the hydrodynamic force acting on the MEMS/NEMS, which arises from the surrounding environment, and $f_{drive}(x, t)$ is the driving force acting on a resonator. Here, it is noted that for theoretical convenience, we ignore vibrations that could arise due to geometrically nonlinear effects, and consider only small amplitude vibrations. Using a Fourier transform, the equation of motion depicted in Eq. (11) can be transformed to

$$\frac{EI}{L^4}\frac{d^4 \bar{W}(\tilde{x};\omega)}{d\tilde{x}^4} - \omega^2 \rho A \bar{W}(\tilde{x};\omega) = \bar{F}_H(\tilde{x};\omega) \tag{12.a}$$

where $\tilde{x}$ is the rescaled coordinate defined as $\tilde{x} = x/L$ with $L$ being the length of a resonator, and the Fourier transform is defined as

$$\bar{V}(x;\omega) = \int_{-\infty}^{\infty} v(x,t)e^{-j\omega t}\, dt \tag{12.b}$$

for any function $v(x, t)$ with $j$ being the unit of a complex number, i.e. $j = \sqrt{-1}$. The hydrodynamic force acting on the MEMS/NEMS can be computed from the Navier-Stokes equation (i.e. equation of motion for the fluids that compose the environment), whose form in Fourier space is given by $\nabla \cdot \bar{\mathbf{u}} = 0$ and $-\nabla \bar{p} + \eta \nabla^2 \bar{\mathbf{u}} = -j\omega \rho_f \bar{\mathbf{u}}$, where $\bar{\mathbf{u}}$ and $\bar{p}$ represent the velocity field of a fluid and the pressure expressed in Fourier space, respectively, and $\eta$ and $\rho_f$ are the viscosity and density of a fluid, respectively. In general, the nonlinear convective inertial term can be neglected for small amplitude vibrations, so that the hydrodynamic force becomes linearly proportional to the displacement field. The hydrodynamic force in Fourier space, $\bar{F}_H(x;\omega)$, is represented in the form [92, 95, 96]

$$\bar{F}_H(x;\omega) = \frac{\pi}{4}\rho_f \omega^2 b^2 \Gamma(\omega) \bar{W}(x;\omega) \tag{13}$$

where $b$ is the width of a resonator, and $\Gamma(\omega)$ is the dimensionless hydrodynamic function. In general, as shown in Eq. (13), the hydrodynamic force (and also dimensionless hydrodynamic function) depends on the cross-sectional dimension (e.g. width), and properties of a resonator (e.g. frequency $\omega$ that is implicitly dependent on the mechanical properties of a resonator), as well as the properties of the surrounding fluid (i.e. density and viscosity). For a circular cross-sectional shape, the dimensionless hydrodynamic function is given as [98]

$$\Gamma_{circ}(\omega) = 1 + \frac{4jK_1\left(-j\sqrt{j\mathrm{Re}}\right)}{\sqrt{j\mathrm{Re}}K_0\left(-j\sqrt{j\mathrm{Re}}\right)} \tag{14}$$

where Re is the Reynolds number defined as Re = $\rho_f \omega b^2/4\eta$, and $K_1$ and $K_0$ are modified Bessel functions of the third kind. For a rectangular cross-sectional shape, the dimensionless hydrodynamic function can be expressed as $\Gamma_{rect}(\omega) = \Omega(\omega)\Gamma_{circ}(\omega)$, where $\Omega(\omega)$ is the correction factor, and the procedure to evaluate the correction factor is well presented in Ref. [95]. It is noted that the hydrodynamic force $\bar{F}(x;\omega)$ can be decomposed into two components – the inertial force $\bar{F}_{inertial}(x;\omega)$ and the damping force $\bar{F}_{damp}$, i.e. $\bar{F}(x;\omega) = \bar{F}_{inertial}(x;\omega) + j\bar{F}_{damp}(x;\omega)$. Therefore, by using Eqs. (12) – (14), the equation of motion in Fourier space is given by

$$\frac{d^4 \overline{W}(\tilde{x};\omega)}{d\tilde{x}^4} - [G(\omega)]^4 \overline{W}(\tilde{x};\omega) = 0 \qquad (15.a)$$

where

$$G(\omega) = \lambda_1 \sqrt{\frac{\omega}{\omega_1^0}} \left[1 + \frac{\pi \rho_f b^2}{4\rho A} \Gamma(\omega)\right]^{1/4} \qquad (15.b)$$

Here, $\lambda_1$ is the boundary condition-dependent eigenvalue for the fundamental mode (e.g. $\lambda_1 = 1.87$ for a cantilever), and $\omega_1^0$ is the resonant frequency measured in vacuum condition, i.e. $\omega_1^0 = (\lambda_1/L)^2 \sqrt{EI/\rho A}$.

The dynamic behavior of a resonator operated in a specific environment (e.g. air, gas, and liquid) can thus be theoretically predicted by solving Eq. (15). However, in the works by Kirstein et al. [99], Kwon et al. [22], and Dareing et al. [100], an *added mass* and *fluid damping* terms were heuristically introduced based on the hydrodynamic force given by Eq. (13) in order to predict the resonance behavior of a cantilever with circular cross-sectional shape. In particular, based on the Navier-Stokes equation for a cylindrical cantilever, the *added mass* due to the hydrodynamic force is given as $m_a = \rho_f(\pi b^2/4)\Gamma_R$, while the *fluid damping coefficient* is expressed as $C_f = \omega \rho_f(\pi b^2/4)\Gamma_I$ [99], where $\Gamma_R$ and $\Gamma_I$ indicate the real and imaginary parts of dimensionless hydrodynamic function $\Gamma$, respectively, i.e. $\Gamma = \Gamma_R + j\Gamma_I$. For cantilevers with a rectangular cross-section, Kirstein et al. have heuristically approximated the *added mass* and *fluid damping coefficient* due to hydrodynamic force based on the assumption that hydrodynamic force is proportional to $W$ while the solutions are identical to the case of the cylindrical cantilever if $W = T$, where $W$ and $T$ indicate the width and thickness of a cantilever, respectively. Specifically, for a cantilever with a rectangular cross-section, the *added mass* and *fluid damping coefficient* are approximated as $m_a = (W/T)\rho_f(\pi W^2/4)\Gamma_R$ and $C_f = (W/T)\omega \rho_f(\pi W^2/4)\Gamma_I$, where the dimensionless hydrodynamic function $\Gamma$ is estimated based on the Reynolds number defined as $\text{Re} = \rho_f \omega W^2/\eta$ for rectangular cross-sections. It is noted that Kirstein et al. [99] have reformulated the equation of motion given by Eq. (11) by using the *added mass* and *fluid damping* due to hydrodynamic force such as

$$EI \frac{\partial^4 w(x,t)}{\partial x^4} + C_f \frac{\partial w(x,t)}{\partial t} + (\rho A + m_a) \frac{\partial^2 w(x,t)}{\partial t^2} = f_{drive}(x,t) \qquad (16)$$

It should be noted that, in the work by Kirstein et al. [99], the coefficients $m_a$ and $C_f$ are regarded as a constant, though they are not truly constants but are generally dependent on time $t$. This indicates that the Eq. (16) with constant *added* mass and *fluid damping* provides an approximate solution (i.e. approximate resonant frequency and/or $Q$-factor) rather than exact quantities of resonant frequency and/or $Q$-factor. Here, the $Q$-factor is defined as the ratio of the resonance frequency to the linewidth of the resonance response corresponding to the full-width at half maximum, i.e. $Q = f_n/\Delta f_n$, where $f_n$ is the resonant frequency and $\Delta f_n$ is the linewidth of resonance response corresponding to full-width at half maximum.

Now, as in Sader's model [95], we consider the case of thermal actuation (i.e. the bending motion of the resonator is excited by Brownian motion of the fluid) while assuming that the dissipative effects are small (i.e. the imaginary part of $G(\omega)$ is much smaller than the real part of $G(\omega)$). For this case, the variations in $G(\omega)$ near the resonance peak are dominantly governed as $O(\omega^2)$. As a consequence, in the neighborhood of the resonance peak, the hydrodynamic function $\Gamma(\omega)$ can be regarded as a constant that can be evaluated at the resonant frequency in the absence of dissipative effect. In the neighborhood of the resonance peak at the $n$-th normal mode, the function $G(\omega)$ can be approximated as [95]

$$G(\omega) \approx G_n(\omega) = \lambda_n \sqrt{\frac{\omega}{\omega_n^0}} \left[1 + \frac{\pi \rho_f b^2}{4\rho A} \left\{\Gamma_R(\omega_n^R) + j\Gamma_I(\omega_n^R)\right\}\right]^{1/4} \qquad (17)$$

where $\omega_n^0$ is the resonant frequency at mode *n* measured in vacuum, and $\omega_n^R$ is the resonant frequency at mode *n* in the absence of dissipative effect, i.e. $\Gamma_R \gg \Gamma_I$. From Eq. (17), the resonant frequency at mode *n* (in the absence of dissipative effect) is given by [95]

$$\frac{\omega_n^R}{\omega_n^0} = \left[1 + \frac{\pi \rho_f b^2}{4\rho A} \Gamma_R\left(\omega_n^R\right)\right]^{-1/2} \qquad (18)$$

It is noted that when $\mathrm{Re} \equiv \rho_f \omega_1^0 b^2 / \eta \to \infty$, the real component of hydrodynamic function approaches 1 (i.e. $\Gamma \to 1$), and consequently Eq. (18) becomes identical to the result of Chu [101]. As shown in Eq. (18), the resonant frequency of a device immersed in a viscous fluid is critically dependent on the dimensions and composition of the device and fluid. Using Eqs. (17) and (18), one can find that $G_n(\omega) = \lambda_n \sqrt{\omega/\omega_n^R} \left(1 + j/Q_n\right)^{1/4}$, where $Q_n$ is the quality factor represented in the form [95]

$$Q_n = \frac{\left(4\rho A / \pi \rho_f b^2\right) + \Gamma_R\left(\omega_n^R\right)}{\Gamma_I\left(\omega_n^R\right)} \qquad (19)$$

Eq. (19) clearly demonstrates the dependence of the linewidth of the resonance response (or equivalently, *Q*-factor) on the hydrodynamic effect, that is, both inertial and damping components of hydrodynamic force. In particular, it is clear from Eq. (19) that the *Q*-factor for a resonator immersed in a viscous fluid is determined by the ratio of inertial force to the damping force, i.e. $\Gamma_R\left(\omega_n^R\right)/\Gamma_I\left(\omega_n^R\right)$, while also being impacted by the dimensions and composition of the device and fluid, i.e. $\rho$, *A*, *b*, and $\rho_f$.

## 2.2. Energy Dissipation Mechanism: Q-Factors

One of the key performance measures for NEMS is their quality or Q-factor, which can be described either as the full-width at half maximum of the experimentally measured resonance peak, or the rate at which the NEMS loses energy per vibrational period due to interactions with its environment or due to intrinsic flaws or defects in the NEMS. Q-factors are critical to the sensing performance of NEMS because the ability of the NEMS to sense changes in its environment, i.e. adsorbed masses, changes in force or pressure, is strongly-dependent on its Q-factor. For example, the mass sensitivity of a NEMS resonator can be written as [102]

$$\Delta m \approx 2 m_0 \left(\frac{b}{Q\omega_0}\right)^{1/2} 10^{-DR/20} \qquad (20)$$

where *DR* is the dynamic range of the resonator, *b* is the bandwidth, or the available frequency range of detection, $\Delta m$ is the change in mass that is to be detected, $m_0$ is the mass of the resonator, and $\omega_0$ is the resonant frequency of the resonator with no attached mass. Eq. (20) clearly demonstrates that a higher Q-factor is necessary to detect ever smaller masses $\Delta m$; we note that similar inverse relationships between the Q-factor and the sensitivity of the NEMS can be found in detecting other physical quantities [34].

The factors that degrade the Q-factor can be categorized as intrinsic or extrinsic. Extrinsic damping mechanisms occur due to interactions of the NEMS with its surrounding environment, i.e. the air or gas molecules surrounding it, or the substrate on which it lies. Intrinsic damping mechanisms occur due to flaws or defects inherent to the NEMS, for example dislocations, grain boundaries, crystalline impurities, etc. There are essentially four major loss mechanisms for NEMS; these are surface losses [2, 103-107], clamping or support losses [104, 108-110], gas damping losses [95], and thermoelastic damping losses [2, 103-106, 111-114]. Surface and thermoelastic damping losses are intrinsic, while clamping and gas damping losses are extrinsic.

Surface losses arise in NEMS due to the fact that the atoms that lie at the surfaces of the NEMS have fewer bonding neighbors than atoms that lie within the bulk. Because of this, the surface atoms have a different vibrational frequency than do atoms that lie within the bulk. The importance of this on the Q-factor is that the resonant frequency of the surface atoms is also different from that of the overall NEMS; therefore, the NEMS resonance loses coherency due to the different vibrational frequency of the surface atoms, which leads to a decreased Q-factor with an increase in surface area to volume ratio (see Fig. 5). We note that research has shown that passivating surface atoms to make their bonding environments more bulk-like has resulted experimentally in higher Q-factors [115-118].

Clamping or support losses occur due to the fact that NEMS are generally fabricated on top of and clamped or fixed to substrates that are much larger than the operational NEMS device. The various analytical models that have been developed [104, 108-110] for Q-factor degradation due to clamping losses account for the fact that during the flexural motion that occurs during resonance, the waves that are generated in the NEMS device carry energy and leave the NEMS device by propagating into the substrate through the supports that fix the NEMS to the substrate. Judge et al. [108] developed analytical models that accounted for both infinite and finite thickness substrates on Q-factor degradation, while Cross and Lifshitz [110] also considered energy dissipation due to wave propagation into surrounding substrate. Wilson-Rae [109] used a different approach, that of phonon tunneling between beams and supports, and also support-induced modification of the density of states, to develop estimates for support-induced Q-factor degradation for a variety of geometries.

Gas damping losses also are significant for NEMS, and arise due to the perpetual interaction (collisions) between the oscillating NEMS and surrounding gas atoms or molecules. Gas damping effects tend to be more significant for NEMS than MEMS due to the fact that as the NEMS becomes smaller, the ratio of the mass between the NEMS and the surrounding gas molecules becomes non-negligible. In other words, while more massive MEMS can easily brush aside surrounding gas molecules during oscillation, NEMS can lose a significant amount of their energy via the collisions with the surrounding gas atoms or molecules.

Finally, NEMS can lose energy through so-called thermoelastic dissipation (TED), which is presented in Fig. 6. TED works via a mechanism in which a flexurally oscillating NEMS, due to being bent, has one surface that is in tension while the opposite surface is in compression. Due to thermomechanical coupling, the surface that is in compression becomes slightly warmer, while the surface that is in tension becomes slightly cooler; the resulting heat flow between the hotter and cooler surfaces is the source of TED as a non-recoverable loss of energy. Mathematically, TED has been accounted for by adding a thermal term to the classical Bernoulli-Euler beam equation of motion, i.e.:

$$\rho A \frac{\partial^2 u}{\partial t^2} + \frac{\partial^2}{\partial x^2}\left(EI \frac{\partial^2 u}{\partial x^2} + E\alpha I_T\right) = 0 \qquad (21)$$

where $\alpha$ is the coefficient of thermal expansion and $I_T$ is the thermal contribution to the beam's moment of inertia. It is likely that TED will be altered in NEMS due to the presence of surface stresses, which may enhance the tensile and compressive stresses that result at the surfaces of the NEMS due to the flexural mode of deformation [119].

### 2.3. Mechanical Modulation of Resonance Behavior

*2.3.1. Resonant Frequency*

As has been discussed, the resonant frequency is fundamentally important to not only characterize MEMS/NEMS devices, but also because it plays a critical role on detection sensitivity (for more details, see Sections 3.2.2, 4.2.1, and 4.2.2). This indicates that manipulation of the resonant frequency will enable the development of not only high-frequency devices but also mass sensors with increased detection sensitivity. One possible approach to tuning the resonant frequency is to control the actuation force driven by bias voltage as described in Refs. [81, 84, 87, 120] and Fig. 3. Another possible method is to apply mechanical stress (or strain) to the nanobeam, which leads to changes in the resonance behavior [121, 122]. The vibrational motion of a nanobeam, which is operated in vacuum or air under applied mechanical stress $\sigma_0$, is described as [68, 69, 71, 121]

$$EI\frac{\partial^4 w(x,t)}{\partial x^4} - \sigma_0 A\frac{\partial^2 w(x,t)}{\partial x^2} + \rho A\frac{\partial^2 w(x,t)}{\partial t^2} = 0 \quad (22)$$

where $E$, $I$, $A$, and $\rho$ represent the elastic (Young's) modulus, cross-sectional moment of inertia, cross-sectional area, and mass density of a beam, respectively, and gas damping effects are ignored for convenience. The mathematical solution to the eigenvalue problem that results from Eq. (22) (as described in Section 2.1.1) provides the resonant frequency in the form of $f_n = f_n^0\sqrt{1+\Gamma_n}$, where $f_n^0$ is the resonant frequency without any application of mechanical stress, and $\Gamma$ is the normalized mechanical tension, defined as $\Gamma_n = \sigma_0 L^2/\alpha_n E t^2$ where $t$ and $L$ indicate the thickness and length of a beam, respectively, and $\alpha_n$ is a constant that depends on the mode index $n$, e.g. $\alpha_1 = 3.4$ for fundamental flexural resonance for a doubly-clamped beam. It should be noted that, for the sign convention, a positive value indicates a tensile stress applied to the nanobeam, while a compressive stress is represented by a negative value. Here, the dimensionless parameter $\Gamma_n$ provides the ratio of mechanical tension, i.e. $\sigma_0 A$, induced by applied stress to a critical load, i.e. $P_{cr} = \beta EI/L^2$, that induces the buckling of a nanobeam. If the mechanical tension induced by applied tensile stress is comparable to a critical load $P_{cr}$, then the resonance is significantly amplified. On the other hand, when a compressive stress is applied that is comparable to the critical load, the resonant frequency can be reduced considerably, which indicates that mechanical stress (or strain) is an important control parameter to modulate the resonance of MEMS/NEMS. This hypothesis has been validated by a recent experimental works [81, 121], which shows that resonant frequency of a doubly-clamped nanobeam can be increased by application of mechanical tension that affects the bending behavior.

*2.3.2. Q-factors*

Due to the importance of enhanced Q-factors for NEMS-based sensing applications, and due to the reduction in NEMS Q-factors due to surface damping, clamping losses, thermoelastic damping and gas damping, researchers have actively been looking for ways in which the Q-factors of NEMS can be enhanced. For example, researchers at Cornell have had significant success in using mechanical stress to tune and enhance the Q-factors of NEMS made of various materials. In these experiments, tensile stress was applied to both silicon nitride and single crystal silicon NEMS resonators by placing the NEMS resonators on a silicon wafer substrate [121, 123]. By bending and flexing the substrate, the researchers were able to induce controllable tensile stress in the NEMS resonators. The effect of the tensile stress is that the Q-factor was able to be tuned and increased by several hundred percent, while eliminating a significant amount of the inherent clamping losses (see Fig. 7). Similarly, other researchers [124] fabricated AlN and SiC NEMS resonators on silicon substrates, then induced tensile strain in the NEMS by utilizing the thermal expansion mismatch between the NEMS and substrate. Q-factor enhancements of about one order of magnitude were reported for 50-250 nm thick NEMS with strains of about 0.0026%.

In addition to experiments, other researchers have utilized classical atomistic [molecular dynamics (MD)] simulations to examine the effects of applied mechanical strain on the Q-factors of NEMS. For example, Park and Kim used classical MD to examine how the Q-factors of metal nanowires could be tuned using tensile strain [125]. In doing so, they also found that the Q-factors of 2 nm cross section copper nanowires could be increased by nearly an order of magnitude through the application of tensile strain. To demonstrate that tensile strain can be used for various nanomaterials, they also studied the effects of strain on the Q-factors of monolayer graphene NEMS [126]. Similar effects resulting in order of magnitude enhancements in the Q-factors of graphene were also found as a result of the tensile strain application.

**3. MEMS/NEMS-Based Molecular Detection**

The last decade has witnessed significant progress towards utilizing MEMS/NEMS sensors that enable *in vitro* molecular detection. Unlike molecular recognition, which uses labeling, MEMS/NEMS devices have enabled the fast, reliable, label-free detection of specific molecules related to specific diseases, which implies their tremendous potential in performing early diagnosis of specific diseases such as cancer [37, 38, 42, 127-133]. The detection principle is the direct transduction of molecular binding on the device surface into a change of the device's physical properties such as its electrical signal [130-132], mechanical bending deflection and/or mechanical resonance [37, 38, 127]. For instance, molecular detection using microcantilever is attributed to the measurement of bending deflection change induced by surface stress that originates from molecular binding on the cantilever surface [64, 134, 135]. Recently, resonant MEMS/NEMS devices have been highlighted for their capability to weigh a single molecule [9, 10], which implies the great potential of NEMS resonators for single-molecule detection, where such detection sensitivity is usually inaccessible with a micro/nanoscale field effect transistor [130-132].

Therefore, in this section, we focus our review on the MEMS/NEMS devices that use mechanical transduction for molecular detection. Specifically, our review is restricted to experimental attempts on *in vitro* molecular detection using micro/nano-cantilevers and/or nanostructures such as nanowires and/or nanotubes based on their resonance motion.

## 3.1. Detection Principles

*3.1.1. Molecular Detection via Flexural Deflection Motion*

For the past decade, microcantilevers have been used for studying molecular adsorption. Specifically, molecular adsorption onto a surface has been well understood based on measuring the flexural deflection of the cantilever that is induced by molecular adsorption. This is ascribed to the principle that the surface stress induced by molecular adsorption on the surface is measurable by estimating the cantilever's resulting flexural deflection. This principle is expressed through the well-known "Stoney's formula" [64, 134, 135] that provides the relationship between surface stress and flexural deflection change, i.e.

$$\tau = \frac{Et^2}{6R(1-v)} \qquad (23)$$

where $\tau$ is the surface stress, $R$ is the radius of curvature, and $E$, $t$, and $v$ represent the elastic modulus, the thickness, and the Poisson's ratio of the cantilever, respectively. Herein, from the assumption of pure bending motion of a cantilever, the radius of curvature is associated with the flexural deflection change through the relation $1/R = 2\Delta w/L^2$, where $\Delta w$ and $L$ represent the cantilever's bending deflection change due to surface stress and the cantilever's length, respectively.

A quantitative understanding of molecular adsorption on surfaces has become possible using the cantilever's flexural deflection motion driven by such adsorption. Gerber and coworkers [136] first employed microcantilevers to study the adsorption kinetics of alkane thiol chains onto the gold surface. In their work, the intermolecular interaction responsible for adsorption kinetics is related to a surface stress change, which can be straightforwardly measured from Stoney's formula in Eq. (18). It was remarkably shown that the relationship between surface stress and the length of alkane thiol chains is well described by Langmuir kinetics. Moreover, their study suggested that microcantilevers are able to sensitively detect the different conformations of alkane thiol chains.

The detection principle based on Stoney's formula has enabled not only the detection of molecular adsorption but also the label-free detection of specific molecules. For label-free recognition of specific biological/chemical species, the surface of the MEMS/NEMS cantilever has to be chemically modified by immobilization of receptor molecules that have a high binding affinity to specific biological/chemical species. Briefly, the surface can be chemically modified using (i) interactions between the thiol group of the receptor molecules and the gold thin film [137], (ii) chemical reaction of amine group of receptor molecule onto silicon surface [24], and/or (iii) self-assembled monolayer [138] that acts as a cross-linker between the receptor molecules and the surface. Here, we do not review all details of surface modifications, which are well summarized in Refs. [37, 38, 133]. Specific molecular binding on the cantilever surface results in the generation of surface stress, which leads to a flexural deflection of the cantilever (see Fig. 8). This detection scheme was applied for sensing the specific DNA sequences relevant to specific diseases. For example, Majumdar and coworkers [139, 140] studied the role of ionic strength on the DNA hybridization mechanism on a cantilever's surface using measurements of the cantilever deflection motion. Further, they studied the effect of DNA chain length on the cantilever deflection motion. Recently, Tamayo and coworkers [141] studied the role of hydration on the DNA hybridization mechanism using cantilever deflection motion. Here, it should be noticed that ionic strength as well as hydration are known to play a critical role on intermolecular interactions between DNA chains, which shows the potential of microcantilever for studying the intermolecular interactions based upon the measurement of bending deflection change. Recently, it has been reported that a microcantilever is capable of label-free detection of specific marker proteins [43], specific RNA sequence [142], enzymatic activity [143, 144], and/or drug resistance [145] (e.g. superbug).

However, detection schemes that are based upon the flexural deflection of the cantilever are subject to important restrictions. First, the length of the cantilever should be larger than at least ~100 μm for reliable detection. This can be clearly elucidated from Stoney's formula in Eq. (23), which shows that the deflection change $\Delta w$ due to the surface stress is proportional to the square of the length $L$, i.e. $\Delta w \sim L^2$. In other words, the shorter the cantilever, the smaller the deflection change, which imposes significant difficulties on the experimental resolution, and which indicates that detection schemes based upon the flexural deflection of cantilevers prevents the scaling down of microcantilevers or MEMS to NEMS device sizes. Second, it is not straightforward to make a connection between the deflection change due to specific molecular adsorption and the amount of the adsorbed molecules. This implies that detection schemes based on the flexural deflection are not appropriate for a quantitative study on molecular interactions. These two drawbacks can be resolved by instead considering detection schemes based upon the resonant frequency. The details of the resonance-based detection will be described in the next section.

*3.1.2. Molecular Detection via Resonance Motion*

As described in the previous section, resonance-based detection allows the continuous scaling down of a cantilever to nanometer length scales. This can be attributed to the fact that scaling down leads to an increase of the resonant frequency of a device, which increases the detection sensitivity. In other words, the smaller the resonator, the more sensitivity it possesses. In the case of detecting minute amounts of molecules, the resonant frequency shift due to added molecules can be written as [30, 146]

$$\frac{\Delta f_n}{f_n} = -\frac{1}{2}\frac{\Delta m}{m} \qquad (24)$$

where $f_n$ and $\Delta f_n$ indicate the resonant frequency and the resonant frequency shift for the *n*-th flexural mode, respectively, and $m$ and $\Delta m$ represent the effective mass of a resonator and the mass of added molecules, respectively. Here, it is assumed that $\Delta m \ll m$, that is, the amount of adsorbed mass is much smaller than the mass of the cantilever. Clearly, Eq. (24) demonstrates that if the resonant frequency of the cantilever $f_n$ increases due to a scaling down of the cantilever towards nanometer length scales, it is able to detect even smaller changes in mass $\Delta m$, eventually reaching the ultimate limit of a single molecule or a single atom.

At first glance, the detection principle described by Eq. (24) provides straightforward insights into how to optimize the design of nanomechanical resonators for ultra sensitive detection. However, this relationship cannot be blindly applied to NEMS resonators because of unexpected small-scale effects such as surface effects, which cause the resonant frequency to deviate from values predicted using standard continuum beam theories. These unanticipated small-scale effects, which also play a key role in understanding the detection principle for NEMS, are discussed in Sections 4.3 and 4.4.

## 3.2. Resonator-Based Chemical Detection

For the past decade, the changes in resonant frequency of MEMS/NEMS in response to chemical adsorption have been used to validate its potential for sensor applications.
Chun et al. [146] have utilized resonant microcantilevers for sensing atomic adsorption, i.e. a thin gold layer. In their work, it was shown that the resonant microcantilever is able to detect a gold thin layer with detection sensitivity up to ~3 pico ($10^{-12}$) gram, equivalent to molecular weight of ~$10^6$ gold atoms. They also scrutinized the mass sensitivity with respect to the cantilever's dimension such as length and/or width, and found that the relationship between mass sensitivity and the cantilever's geometry is given by

$$\frac{\Delta f_n}{\Delta m} = -\frac{\lambda_n^2}{4\pi^2 bL^3}\sqrt{\frac{E}{\rho^3}} \qquad (25)$$

This indicates that the detection sensitivity can be enhanced by miniaturization of the resonator with a scaling of $(bL^3)^{-1}$, where $b$ and $L$ represent the resonator's width and length, respectively. However, we note that as the resonator thickness is scaled down to sub-micrometer length scales, the mass sensitivity predicted by Eq. (25) does not hold because the resonant frequency shift is determined by not only the mass but also the elastic properties of the adsorbed atoms (for details, see Section 4.2.3).
As mentioned above, Roukes and coworkers [6] have remarkably reported mass sensing with ultrahigh sensitivity up to zepto-gram resolution using nanomechanical resonators (see Fig. 9). They used a nanoresonator that has a length of ~2 μm and that is operated under extreme environmental conditions such as ultrahigh vacuum and cryogenic temperature; these idealized conditions were utilized because they reduce the energy dissipation that usually leads to the degradation of resonance, Q-factor, and thus the mass sensitivity.

More recently, Roukes and coworkers [7] have further developed the chemical detection technique using NEMS resonators. In their work [7], they utilized cantilevers which have various length scales ranging from ~500 nm to ~30 μm, and showed that the smaller cantilevers possess higher Q-factors in air at room temperature; this was attributed to the reduction of dissipated energy through gas damping that occurs due to the corresponding reduction in cantilever dimensions. For specific detection of 1,1-difluoroethane ($C_2H_4F_2$), the surface of the NEMS resonator was chemically modified using a polymer thin film, i.e. polymethylmethacrylate (PMMA). Their work showed that NEMS resonators enable the label-free, specific detection of 1,1-difluoroethane gas molecules with a detection limit up to 1 atto ($10^{-18}$) gram.

Recently, Zettl and coworkers [8] employed carbon nanotube (CNT) resonators as a mass sensor with sensitivity up to atomic resolution. They used a CNT resonator that operates under ultrahigh vacuum and room temperature. They experimentally studied the detection limit which arises from thermal noise using statistical analysis on the resonant frequency shift due to atomic adsorption. Their work suggests that the noise-level of CNT resonator is 0.13zg/Hz$^{1/2}$, and that a single gold atom can be detected using CNT resonators operated in ultrahigh vacuum at room temperature.

However, most chemical detection experiments [6, 8, 146] that we have discussed here have been implemented using physisorption and/or chemisorption such that a resonator is exposed to an environment that possesses a single chemical species rather than multiple species. This indicates that most current works [6, 8, 146] are restricted to sensing specific chemicals such as trinitrotoluene vapor [147] and/or toxic chemicals [148] among the mixture of various chemicals in air. To our best knowledge, the multiplexed specific chemical detection among various chemical species in an ambient environment has not been realized using nanomechanical resonators. The realization of the resonator-based detection of multiple, specific chemicals among various chemical species will eventually lead to the development of novel gas sensors that enable the fast, label-free detection of chemicals with high specificity as well as high sensitivity.

**3.3. Resonator-Based Biological Detection**

*3.3.1. Cell Detection*

The ability to detect specific cells such as cancerous cells is critical as it would enable the early, rapid diagnosis of specific diseases that can infect or kill humans. Moreover, the change of mass and/or density of a cell during its functional cycle (e.g. cell growth) can be quantitatively measured based on mass sensing using NEMS resonators.

The application of resonators to cell detection was first implemented by Craighead and coworkers [39, 149], who employed a resonant microcantilever to sense *Escherichia coli* (*E. coli*) cells. The cantilever surface was chemically modified using antibodies that enable the specific detection of *E. coli* cells. They systematically studied the relationship between the number of adsorbed cells and the resonant frequency shift. Here, the resonant frequency shift due to cell adsorption was measured under ambient conditions, where it was shown that resonant microcantilevers are able to detect a single cell. Later, Campbell and Mutharasan [150] utilized piezoelectric microcantilevers to detect *E. coli* O157:H7 in an aqueous environment (see Fig. 10). Their work showed that resonant microcantilevers operated in an aqueous environment allow a quantitative description of the kinetics of cell adsorption onto an antibody-modified surface based on the measured resonant frequency shift. Despite *in situ* detection, the piezoelectric microcantilever exhibited a mass sensitivity of $7 \times 10^2$ cells/mL, which indicates that it is challenging to detect cells with sensitivity up to $1\text{-}10^2$ cells/mL (clinically relevant to early diagnostics of cancer) using piezoelectric microcantilevers.

Hegner and coworkers studied the active growth of *E. coli* [151] and also reported the label-free detection of two major fungal forms [152] (i.e. *Aspergillus niger* and *Saccharomyces cerevisiae*) using resonant microcantilevers. For studying bacterial growth using microcantilevers, the *E. coli* was chemically attached onto the surface of a microcantilever. Their detection scheme was based on the fact that the added mass resulting from cell growth should decrease the resonant frequency of a microcantilever. The resonant frequency shift driven by cell growth was measured under ambient condition with relatively high humidity. They showed that the detection sensitivity for detection of *E. coli* using microcantilever is ~140 pg/Hz, which suggests that the theoretical detection limit of a microcantilever is ~200 *E. coli* cells. This is ascribed to the low resonant frequency (i.e. ~30 kHz) that is related to the detection limit, e.g. see Eq. (24). For detecting the growth of fungal spore, they measured the real-time resonant frequency shift induced by the growth of fungal spores. The mass sensitivity for detecting fungal spores was found to be 1.9 pg/Hz when the fundamental resonance frequency (~130 kHz) is used. It was emphasized that the detection based on resonant microcantilever allows the fast measurement of the active growth of cells in a couple of hours when compared to the conventional plating method which requires at least 24 hours. On the other hand, it should be noted that their detection is implemented in humid air, which indicates the restriction in understanding the growth behavior of cells under physiological conditions (e.g. aqueous environment under different pH).

Manalis and coworkers [25, 153] developed a suspended microchannel resonator (SMR), where a microchannel is embedded in a microcantilever. The advantage of SMR over the conventional microcantilever is the detection of specific molecules and/or cells in aqueous environment while maintaining a high Q-factor [25]. This is ascribed to the detection scheme in which specific molecules or cells are detected inside a channel while the microcantilever vibrates in air with a high Q-factor. They showed that SMR is able to sense a single cell in aqueous environment [25]. Specifically, the masses of *E. coli* and *Bacillus subtilis* (*B. subtilis*) were measured as ~110 fg and ~150 fg, respectively, using SMR. More remarkably, Manalis and coworkers [47] studied the cell cycle of yeast cells based on SMR. Specifically, the density change during the cell cycle was measured from an SMR in which a yeast cell undergoing a cell cycle is deposited. They found that cell density is increased prior to bud formation at the G1/S transition, consistent with previous studies.

Recently, Low and coworkers [154] considered the detection of *B. subtilis* spores using resonant and static microcantilevers. For label-free detection of *B. subtilis*, the cantilever surface was chemically modified using a specific peptide sequence. It was found that the detection based on microcantilever deflection motion rather than its resonance motion provides the better detection sensitivity. This is ascribed to the following detection principle, that is, the interactions between peptide sequence and *B. subtilis* spores generates a larger surface stress than that between antibody and spores, which results in a greater bending deflection, and consequently, the better detection sensitivity. However, the resonant frequency shift due to bacteria chemisorptions is not dependent on the interaction between probe molecules (functionalized on a microcantilever surface) and bacteria but instead is dependent on the mass of the chemically adsorbed bacteria. This indicates that surface chemistry plays a key role in increasing the detection sensitivity when microcantilever bending deflection is used for label-free detection [155], while the detection sensitivity in case of detection using resonant microcantilevers is not highly correlated with surface chemistry. On the other hand, the effect of surface chemistry on the resonant frequency shift may appear when the length scale is decreased to nanometer length scales [118].

Most current works [25, 39, 47, 149-152] presented here utilized the detection principle where the resonant frequency shift is correlated with the change of effective mass of the cantilever due to cell attachment and/or the change of the cell's density during its functional cycles. However, it should be noted that, during the cell's functions such as apoptosis, the functional cycle induces changes to not only the mass density but also the stiffness of the cell [156, 157]. This indicates that it is also critical to measure the stiffness change for the cell using the micro/nanomechanical resonators in order to quantitatively understand the cell's function. The fundamental relationship between the resonant frequency shift and the stiffness and/or the mass of adsorbed cells and/or biomolecules will be delineated later in Section 4.2.3. Moreover, the measurement of the cell stiffness also enables the sensitive detection of cancerous cells, since it has been recently found that cancerous cells are more flexible than normal cells [158, 159]. These issues are important as it will be desirable to design the resonator-based detection scheme to measure in real-time the changes of the stiffness of a cell for gaining insight into its function and/or fast diagnosis of cancer. It should be noted that the detection principle to measure the stiffness of adsorbed molecules using nano/micromechanical resonators is summarized in Section 4.2.3.

*3.3.2. Virus Detection*

Bashir and coworkers [160] have reported the label-free detection of virus particles such as the vaccinia virus using resonant microcantilevers (see Fig. 11). Their detection scheme compares the resonant frequencies of two prepared cantilevers, i.e. an *unloaded* cantilever and a virus-*loaded* cantilever. Here, the unloaded cantilever is prepared in such a way that a bare cantilever is cleansed with piranha solution ($H_2O_2:H_2SO_4 = 1:1$) for removal of any possible physisorption, while the loaded cantilever is obtained such that the cleansed cantilever is immersed into a virus dissolved solution and then dried in air. Based on these cantilevers, a single virus particle was detected and its mass was measured as ~9.5 fg. Similarly, Craighead and coworkers [161] utilized resonant cantilevers to detect virus particles. Their resonant cantilevers were shown to exhibit the detection sensitivity ranging from $10^5$ to $10^7$ pfu/ml. Here, the unit of "pfu (plaque forming unit)" is defined as the number of virus particles that are able to form plaques per unit volume. Further, they studied the role of the cantilever length on the resonant frequency shift due to virus chemisorption. They showed that the shorter the resonant cantilever, the more sensitivity it possesses. This is attributed to the fact that decreasing the cantilever length increases the resonant frequency, and consequently the resonant frequency shift due to mass adsorption. This is quite different from case of deflection-based detection scheme, where the cantilever length should be at least ~100 μm for reliable detection (for detail, see Section 3.1.1).

*3.3.3. Protein Detection*

Protein-protein interactions, protein-DNA interactions, and/or protein-peptide interactions have been widely considered as model systems for the label-free detection of specific marker proteins relevant to specific diseases such as cancer. So far, many research works on the label-free detection of marker proteins using the bending deflection of cantilevers have been thoroughly reviewed in the literature [38, 127, 162, 163]. The detection scheme based upon bending deflection change has the significant drawback such that it does not enable sensing the marker proteins in very low concentration < ~1 ng/ml. This shows that the detection scheme using deflection change is not suitable for early diagnostics of diseases such as cancer. On the other hand, the resonance-based detection enables the early diagnostics, because resonance-based detection is able to sense the marker proteins even in low concentration < $10^2$ pg/ml. Therefore, we will restrict our review to resonance-based detection of proteins.

Lee et al. [164, 165] reported the label-free detection of marker proteins using a resonant microcantilever vibrating in air. Their detection scheme was to measure the frequency difference in air between a clean cantilever (without protein adsorption) and a protein-adsorbed cantilever. Based on their microcantilevers, the marker proteins in the concentration of ~10 pg/ml (in buffer solution) can be detected. In their study [166], it is shown that the resonant frequency shift due to specific protein binding on the cantilever surface is larger than expected because the resonant frequency shift is determined by not only the mass of the adsorbed proteins but also the surface stress induced by protein adsorption. Their argument is consistent with the experimental study by Thundat et al. [167], who showed that the mass of added molecules is insufficient to describe the frequency shift due to molecular adsorption. However, this argument is still controversial since the resonant frequencies of the one-dimensional beams discussed in Section 2.1 are unaffected by surface stress, as was first shown by Gurtin and coworkers [168] (for details, see Section 4.3.1).

Bashir and coworkers [62] recently reported the label-free detection of proteins using resonant cantilevers whose thickness is comparable to the molecular size of the proteins. It was interestingly shown that protein adsorption onto the cantilever increases the resonant frequency, which contradicts the hypothesis shown clearly through Equation (19) that an increase in cantilever mass should necessarily result in a decrease of its resonant frequency. In their work [62], it was shown that the elastic properties of the adsorbed protein monolayer plays a critical role on the observed resonant frequency shift. Specifically, for ultra-thin cantilevers, the protein adsorption induces a significant change in flexural rigidity, which dominates the frequency shift (for details, see Section 4.3.3).

Recently, Roukes and coworkers [10] reported the ultra sensitive detection of proteins at single-molecule resolution using a nanomechanical resonator. It was remarkably reported that resonant NEMS devices allow the measurement of the molecular weight of specific proteins, indicating that NEMS resonators can be employed as nanomechanical mass spectrometers. Nevertheless, their detection strategy encounters the restrictions such that the biological sample has to be prepared by pre-concentration of one species of protein, and that the resonant frequency shift has to be measured under ultrahigh vacuum and low temperature. Although their detection scheme is implemented based on physisportion rather than chemisorption, it was shown that NEMS resonators are able to distinguish different protein oligomers based on the analysis of frequency jump statistics. For biosensing applications for future diagnostics, NEMS resonators must be capable of detecting a specific biological species among various biological species with high selectivity and specificity, which is typically achieved by chemical modification of the surface of the NEMS resonator.

Craighead and coworkers [169-171] have provided the label-free detection of specific marker proteins using resonant MEMS devices. Their detection scheme is based on mass sensing such that the frequency shift arises only from the mass of chemically adsorbed protein molecules. In their studies, the label-free detection of marker proteins such as prostate specific antigen (PSA) [170] and prion protein [169, 171] has been reported. They have also suggested a scheme to improve the detection sensitivity such that the frequency shift due to added mass is amplified using nanoparticles and/or secondary antibodies [169].

Most experimental studies of protein detection using resonators have been performed in either ambient conditions [62, 164-166], vacuum conditions [170], or cryogenic, high-vacuum conditions [10, 169, 171]. Until now, with few exceptions [22, 23, 172], there have been very few studies on *in situ* detection of proteins in aqueous environments. Here, *in situ* detection of proteins in solution is of high significance, because *in situ* detection allows the real-time monitoring of protein-protein interactions, which will enable novel insights into the kinetics of protein-protein interactions. Recently, Kwon and coworkers [22, 23] utilized a piezoelectric thick film microcantilever [79], which exhibits a relatively high Q-factor even in a viscous liquid, for the real-time monitoring of biomolecular interactions. In their work [22, 23], it is shown that the frequency shift due to biomolecular interactions on the cantilever surface is governed by not only the mass of target biomolecules but also the hydrodynamic loading change driven by hydrophilicity change during the biomolecular interactions (see Fig. 12). Furthermore, it is shown that the resonant frequency shift, measured in buffer solution, enables the quantitative descriptions on the binding kinetics of protein antigen-antibody interactions and/or DNA hybridization. Specifically, the resonant frequency shift measured in buffer solution suggests that the binding kinetics for protein antigen-antibody interactions and/or DNA hybridization can be modeled using Langmuir kinetic models.

### 3.3.4. DNA Detection

Detection of DNA sequences that are related to specific diseases has been extensively reported using cantilever bending deflection motion [139, 173, 174]. Furthermore, intermolecular interactions between DNA molecules have been thoroughly studied based on cantilever bending deflection motion [140]. However, the detection scheme based upon bending deflection exhibits the significant restrictions for ultra sensitive detection of DNA molecules in very low concentration < 1 pM for long DNA/RNA chains and/or < 1 nM for short DNA/RNA fragment chains.

In 2003, Dravid and coworkers [175] first reported the label-free detection of DNA molecules with concentrations as low as 0.05 nM using resonant microcantilevers. Their detection scheme is based on the measurement of frequency shifts due to the mass of hybridized DNA molecules. For sensitive detection, they considered mass amplification using nanoparticle-conjugated DNA. Specifically, short probe DNA molecules are functionalized on the cantilever surface in order to capture the target DNA molecules that are longer than the probe DNA. Subsequently, they introduced the nanoparticle-conjugated DNA that can be hybridized to some nucleotide sequence of target DNA that was already reacted to probe DNA molecules on the cantilever surface. Based on this scheme, they were able to detect the specific DNA chain even in low concentrations such as 0.05 nM, implying the possibility of clinically relevant early diagnostics. Nonetheless, it is difficult to measure the amount of target DNA molecules bound to the functionalized cantilever because the resonant frequency shift is attributed to the mass of both nanoparticles and DNA molecules.1

In 2005, Craighead and coworkers [9] suggested a remarkable detection scheme using NEMS resonators that possess ultrahigh resonant frequency up to ~100 MHz. In their work, it was shown that the resonant frequency shift due to DNA adsorption even at single-molecule resolution is measurable using a high-frequency resonator vibrating under high-vacuum conditions. In particular, they optically measured the resonant frequencies of an unloaded resonator and a DNA-loaded resonator, respectively, then estimated the loaded mass for adsorbed DNA based on the resonant frequency shift from Eq. (19). Based on the measurement of the frequency shift due to DNA loading onto a cantilever, they were able to measure the molecular weight of a single double-stranded DNA with a length of ~$10^3$ base-pairs. This indicates that their detection scheme using high-frequency resonators allows the enumeration of target DNA molecules adsorbed onto a resonator.

Recently, Eom and coworkers [176] theoretically studied the resonance behavior of CNT resonator in response to DNA adsorption onto a CNT due to π-π interactions between the CNT and DNA. In their work, it was shown that the resonant frequency shift measured in air due to DNA adsorption enables the identification of specific DNA sequences, while the frequency shift estimated in solution is difficult to be measured in addition to difficulties in identifying the specific sequences. This implies that NEMS resonator is unacceptable for *in situ* detection of DNA molecules in buffer solution.

In recent years, Kwon and coworkers [23] have suggested the *in situ*, label-free detection of specific DNA sequences such as HIV DNA fragments using resonant microcantilevers that exhibit a high Q-factor even in buffer solution (see Fig. 13). It was shown that resonant microcantilevers operated in buffer solution enable the real-time monitoring of DNA hybridization, making it possible to gain insight into the kinetics of DNA hybridization. They have shown that the time constants to describe the Langmuir kinetics for DNA immobilization and DNA hybridization using resonant cantilever are consistent with a previous study [177] based on Surface Plasmon Resonance (SPR).

*3.3.5. Detection of Enzymatic Activity*

Enzyme-protein interactions have played a vital role in signal transduction, which results in disease expression inside a cell [178, 179]. Specifically, enzyme-protein interactions cleave a protein into substrates, which drives the cellular response [178]. For therapeutics at a molecular level, insights into enzymatic cleavage and how to inhibit such cleavage are the key to developing *de novo* drug molecules.

Over the last decade, the insight into enzyme-protein (or enzyme-peptide) interactions has been gained using cantilever bending deflection motion. Here, specific enzymes or proteins (peptides) were immobilized on the cantilever surface. The details of label-free detection of enzymatic activity have been well summarized in Refs. [38, 143, 144, 180]. However, the detection of enzymatic activity using cantilever bending deflection motion is restricted such that it is difficult to quantify the enzymatic cleavage of proteins (or peptides), since the detection using deflection motion does not enable the measurement of the total mass of cleft peptides due to enzyme-peptide interactions in real-time, which is relevant to quantitative descriptions of the kinetics of enzymatic cleavage.

Recently, Kwon et al. [24, 44] experimentally studied the enzymatic activity as well as the kinetics of enzymatic cleavage using a resonant microcantilever operated in buffer solution (see Fig. 14). In their work [24], a cantilever was functionalized with specific peptide sequences linked to polyethylene glycol (PEG) chains. When such a cantilever is exposed to a specific enzyme, the occurrence of enzymatic cleavage leads to the decrease of the effective mass of the cantilever, resulting in an increase of the resonant frequency. Their detection scheme enables the quantification of the total mass of enzymatically cleft peptides due to enzymes based on the relationship between the frequency shift and the total mass of cleft peptides. In particular, they are able to evaluate the proteolysis efficacy with respect to enzyme concentrations ranging from ~0.2 μM to ~1 μM from the measured resonant frequency shift. Further, they have estimated the rate constant for enzymatic cleavage based on the resonant frequency shift measured in buffer solution such that the frequency shift is well fitted to a Langmuir kinetic model. Nevertheless, their detection using resonant microcantilever is limited in the viewpoint of detection sensitivity, that is, their microcantilever may be unable to sense and detect the proteolysis of peptide due to enzymes in extremely low concentrations (e.g. < ~1 nM).

To our best knowledge, except a recent study by Kwon et al. [24], there have not been previous attempts to employ micro/nanomechanical resonators to study the enzymatic activity and/or the inhibition of such activity. A quantitative understanding of the enzymatic activity and/or the drug-induced inhibition of enzymatic activity is of high significance [178, 179], since such understanding provide insights into novel drug design and/or drug efficacy related to drug screening. The recent study by Kwon et al. [24] can be regarded as a first step towards the development of a nanomechanical drug screening system that enables deep insights into the function of enzymes, which can provide the novel concept of drug design. Therefore, it will be important to consider the nanoresonators as a nanomechanical drug screening system for sensitive, label-free detection of enzymatic activity and/or drug-induced inhibition of enzymatic activity.

### 3.4. Perspectives and Challenges

As described in previous Sections 3.2 and 3.3, resonant MEMS/NEMS devices enable the label-free detection of specific chemicals and/or biological species related to specific diseases, which implies the early diagnostics of specific diseases such as cancer, even at single-molecule resolution. Until recently, most of detection has been conducted using resonant MEMS devices, whereas except for recent experimental efforts [6-8, 10], NEMS resonators have been rarely employed for chemical/biological detection. We anticipate that resonant NEMS devices will receive more attention for label-free detection in the near future because of their unprecedented detection sensitivity. Nevertheless, the following technological obstacles must be overcome for NEMS resonator-based detection strategies to reach their ultimate potential.

Until recently, biological/chemical detection using NEMS resonators has been implemented using physisorption rather than chemisorption [6, 8, 10], except a recent work by Roukes and coworkers [7]. This indicates that pretreated sample has to be prepared such that only a single chemical/biological species exists in the sample. In other words, if there are several species of chemical/biological molecules, the bare NEMS surface is unable to selectively capture a specific species of molecules. For lab-on-a-chip sensing applications, the NEMS should be able to selectively detect a specific species among various, different species in the sample. Therefore, current NEMS resonator-based detection scheme using physisorption is inappropriate for further biological/chemical sensing applications at lab-on-a-chip level.

As described above, chemical modification of the NEMS surface is required for sensing applications. In general, the vibrational frequencies of NEMS are dependent on surface effects [181-183]. For example, the surface modification may result in an undesirable decrease in the Q-factor [181] that is directly related to the detection limit [182, 184] as shown in Eq. (14). This indicates that, for specific chemical detection using nanoresonators, it is essential to systematically study the role of surface chemistry on the resonance behavior and Q-factors, and thus the resulting detection sensitivity of NEMS.

Unlike other MEMS/NEMS sensors, NEMS resonators have significant difficulties to sense and detect in aqueous environments because of large energy dissipation due to both damping and hydrodynamic loading effects [22, 88, 99]. It is still challenging to improve the Q-factors of NEMS resonators operated in aqueous environment, which indicates that NEMS resonators are unable to detect the toxic metal ions (e.g. $Ca^{2+}$, $Cr^{2+}$, $Cs^{+}$, and $CrO_4^{2-}$ ; see Ref. [38]) in water, implying that NEMS resonators are currently restricted from being used as toxicity detectors in an aqueous environment. Moreover, the large energy dissipation (i.e. low Q-factor) for NEMS resonators that are operated in an aqueous environment is currently preventing scientists from gaining insight into the binding/unbinding kinetics of various biomolecular interactions such as protein-protein interactions, protein-DNA interactions, DNA hybridization, and protein-enzyme interactions. This insight is of high importance, since a quantitative description of the binding/unbinding kinetics will provide the significant information that could be further used for novel drug design and/or drug screening. Therefore, we believe that it is of a great importance to improve the *Q*-factors in aqueous environment for bio-sensing applications such as *in situ*, real-time detection. Recent progress towards understanding Q-factor degradation due to surrounding fluidic environments has occurred due to work by Ekinci et al. [185, 186], who studied doubly clamped silicon beam resonators in a gaseous nitrogen environment. They found that fluidic dissipation appears to saturate at high frequencies that are related to the relaxation times of the fluid; this finding enabled them to quantitatively state how the NEMS geometry can be utilized to minimize fluidic dissipation. While such studies are fundamentally important, other approaches that can directly lead to enhanced Q-factors in liquid or gaseous environments are clearly needed.

Finally, to our best knowledge, currently proposed studies provide a single NEMS resonator for the label-free detection. Multiplexed detection based on resonant NEMS has not been attempted, though nanomechanical arrays [33, 187] have been recently suggested. Multiplexed detection is of a great importance because it allows the diagnosis of distinct, different diseases. Multiplexed detection may allow one to use the more realistic sample (that is not pretreated) such as blood serum, which consequently enables the fast, reliable, label-free detection, leading to early diagnostics.

## 4. Continuum Modeling Approaches

As shown above in Section 2.1, the harmonic resonance motion of MEMS/NEMS devices is well described by simple continuum elastic models such as the Euler-Bernoulli beam model. Nonetheless, theoretical and computational characterization of the resonance behavior of MEMS/NEMS has to be scrutinized, since such devices can exhibit the dynamic motion far from the simple harmonic oscillations. For instance, nonlinear oscillations [28, 50, 82, 84, 188] and/or coupled resonance [19-21] have been recently observed in the resonance motion of NEMS. Moreover, the detection principle of mass sensing based on continuum elasticity is insufficient to describe the generic detection principle for NEMS-based sensing. This is attributed to the hypothesis that the resonant frequency shift due to molecular adsorption is insufficiently described only by mass addition due to adsorbed molecules. However, there have been recent reports that the resonant frequency shift due to molecular adsorption can be dominated by various effects such as the stiffness of the adsorbed molecules [61, 62], surface stress effects [63, 189-191], nonlinear vibrations [120], and/or mechanically modulated vibrations [120]. Such effects have not been thoroughly studied, though some recent theoretical, computational, and/or experimental studies have been reported.

Modeling approaches enable not only the quantitative understanding of experimentally observed phenomena such as resonance behaviors, but also the design concept of *de novo* NEMS resonators for their specific functions such as actuation and sensing such that the performance (e.g. resonant frequencies, and/or detection sensitivity) of NEMS resonators can be anticipated from modeling-based simulations with respect to specific design parameters (e.g. surface-to-volume ratio related to surface effects, driving forces involved in nonlinear oscillation, externally applied force associated with modulation of frequencies, etc.).

In this Section, we overview the current attempts and perspectives on studying the resonance behavior of NEMS in response to aforementioned effects based on continuum elasticity theory and its related models.

## 4.1. Continuum Elastic Models: Beyond the Harmonic Vibrations

### 4.1.1. Nonlinear Oscillations

As described in Section 2.1.3, the equation of motion for a vibrating doubly-clamped nanowire is given by Eq. (7). In general, a doubly-clamped nanowire can easily reach the nonlinear oscillation regime because of geometric nonlinear effects due to its boundary conditions (see Fig. 15).

Specifically, the geometric nonlinearity [68, 69, 71], which is dictated by the term $(EA/2L)\int_0^L \{\partial w(x,t)/\partial x\}^2 dx$ in Eq. (7), plays a dominant role on the vibration motion. Using Galerkin's approximation [192-194], the bending deflection $w(x, t)$ is represented in the form $w(x, t) = \psi_k(x) \cdot z_k(t)$, where $\psi_k(x)$ and $z_k(t)$ represent the $k$-th deflection eigenmode and its corresponding amplitude, respectively, and repeated index indicates the Einstein summation convention. For convenience, we restrict our discussion to the fundamental resonance, i.e. $w(x, t) = \psi_1(x) \cdot z_1(t)$. For a doubly-clamped nanobeam, the fundamental deflection eigenmode $\psi_1(x)$ is assumed to be in the form of $\psi_1(x) = (2/3)^{1/2}[1 - \cos(2\pi x/L)]$, which satisfies the essential boundary conditions. By multiplying the deflection eigenmode $\psi_1(x)$ with the equation of motion, i.e. Eq. (7), and subsequent use of integration by parts, the equation of motion becomes the Duffing equation [48, 49, 194] given by

$$\ddot{z}_1(t) + \omega_0^2 z_1(t) + \alpha [z_1(t)]^3 = F(t) \qquad (26)$$

where $\omega_0 = (4\pi^2/L^2)[(EI/3\rho A)(1 + L^2 T_0/4\pi^2 EI)]^{1/2}$, and $\alpha = (9\rho L^4)^{-1} \times (8\pi^4 E)$ for the case of geometric nonlinearity-induced nonlinear oscillation. To consider damping effects on the resonance, the damping term $(\omega_0/Q)\dot{z}(t)$ is added to the Duffing equation, where $Q$ is the quality factor which describes the energy dissipation during the vibration due to damping. Then, the critical amplitude for the onset of geometric nonlinearity is given as $a_{cr} = (\omega_0 L^2/\pi^2)(3^{1/2}\rho/EQ)^{1/2}$ [48-50]. Nonlinear oscillation has been found in the case of vibration of double-clamped nanowire that is actuated by external field, where fields such as Lorentz force from magnetomotive technique [82], electrostatic force [84], and/or piezoelectric effect [195] have all been utilized to induce the nonlinear vibration. It should be noted that the nonlinear oscillation due to geometric nonlinear effect, dictated as a term $\alpha$ in Eq. (26), leads to the increase of the resonant frequency (i.e. *stiffening effect*).

Moreover, it should be noted that, in case of the in-plane vibration of a nanoresonator with d.c. bias voltage applied to a gate (see Fig. 16b), the nonlinear oscillation behavior of nanoresonators can be attributed to not only the geometric nonlinear effect due to the doubly-clamped boundary condition but also the electrostatic interaction between the gate electrode and the nanoresonator [50]. In particular, when vibrating in the plane of the gate, a nanoresonator is electrostatically attracted to the gate electrode, which leads to a decrease of the effective spring constant of a nanoresonator, thus resulting in the decrease of the resonant frequency (i.e. *softening effect*) [50]. In order to gain a fundamental insight into tuning of nonlinear oscillation (and also frequency) using electrostatic attraction, we consider the eqution of motion for a doubly-clamped beam, which is given by Eq. (7). Here, the actuating force per unit length, $f(x, t)$, is represented in the form $f(x, t) = (1/2)[\partial c(w)/\partial w] \cdot V^2$ [50, 196, 197], where $c(w)$ is the capacitance per unit length as a function of the deflection of a beam, $w(x, t)$, and $V$ is a d.c. bias voltage. When a nanoresonator is approximated as an infinite wire with diameter $d$ and it vibrates near the semi-infinite plane of the gate, the capacitance per unit length is given by $c(w) = 1/[2\ln(2(R - w)/d)]$ [197], where $R$ is the distance between the nanoresonator and the gate electrode. Using a Taylor series expansion, the capacitance per unit length can be approximated as $c(w) \approx c_0 + c_1(w/R) + c_2(w/R)^2 + c_3(w/R)^3 + c_4(w/R)^4 + O((w/R)^5)$. Furthermore, the bending deflection $w(x, t)$ of a nanoresonator can be decomposed into two terms such as a static d.c. displacement $u_{dc}(x)$ and a time-dependent a.c. displacement $u_{ac}(x, t)$, i.e. $w(x, t) = u_{dc}(x) + u_{ac}(x, t)$. Then, with substitution of $w(x, t) = u_{dc}(x) + u_{ac}(x, t)$ and $c(w) \approx c_0 + c_1(w/R) + c_2(w/R)^2 + c_3(w/R)^3 + c_4(w/R)^4$ into Eq. (7), the equation of motion for a nanoresonator vibrating in the plane of a gate becomes

$$EI\frac{\partial^4 u_{ac}}{\partial x^4} + \left[\frac{EA}{2L}\int_0^L \left(\frac{\partial u_{ac}}{\partial x} + \frac{du_{dc}}{dx}\right)^2 dx\right]\frac{\partial^2 u_{ac}}{\partial x^2} + \rho A \frac{\partial^2 u_{ac}}{\partial t^2} = -EI\frac{d^4 u_{dc}}{dx^4} - \left[\frac{EA}{2L}\int_0^L \left(\frac{\partial u_{ac}}{\partial x} + \frac{du_{dc}}{dx}\right)^2 dx\right]$$

$$\times \frac{d^2 u_{dc}}{dx^2} + \frac{V^2}{2R}\left[c_1 + 2c_2\left(\frac{u_{dc}+u_{ac}}{R}\right) + 3c_3\left(\frac{u_{dc}+u_{ac}}{R}\right)^2 + 4c_4\left(\frac{u_{dc}+u_{ac}}{R}\right)^3\right] \qquad (27)$$

Using Galerkin's approximation [192-194], the displacements are assumed as $u_{ac}(x, t) = z_1(t) \cdot \psi_1(x)$ and $u_{dc}(x) = A_{dc}\psi_1(x)$, where $z_1(t)$ and $A_{dc}$ represent the time-dependent a.c. amplitude for a fundamental resonance and the static d.c. amplitude, respectively, and $\psi_1(x)$ is the assumed deflection eigenmode given as $\psi_1(x) = (2/3)^{1/2}[1 - \cos(2\pi x/L)]$ for the fundamental resonance. Therefore, by multiplication of $\psi_1(x)$ with Eq. (27) and subsequent use of integration by parts, the equation of motion can be transformed into the form [50]

$$\ddot{z}_1(t) + \omega_0^2 z_1(t) + \alpha_2 [z_1(t)]^2 + \alpha_3 [z_1(t)]^3 = 0 \tag{28}$$

Here, the parameters are well summarized in Ref. [50]. It should be noted that the quadratic term $\alpha_2$ arises from the electrostatic attraction between a nanoresonator and a gate electrode, while the cubic term $\alpha_3$ is attributed to the geometric nonlinear effect due to boundary condition. The effective nonlinearity for this system can be described as $\alpha_{eff} = \alpha_3 - [10/(9\omega_0)]\alpha_2^2$ [198], which indicates that a d.c. bias voltage applied to the gate increases the quadratic nonlinearity $\alpha_2$ [50], which eventually decreases the effective nonlinearity. In other words, the nonlinear oscillation behavior of a nanoresonator that undergoes the in-plane vibration can be tuned by using electrostatic interactions.

Furthermore, Karabalin et al. [188] showed that by using nonlinear mode coupling they were able to tune the nonlinear dynamic behavior and to observe the chaotic behavior in nonlinear oscillation. The details of mode coupling are presented in Section 4.1.2.

*4.1.2. Coupled Resonance*

In recent years, there have been a few attempts [19-21] to develop coupled micro/nanomechanical resonators in order to understand the anamolous oscillation behavior, that is, the coupled resonance for future applications such as synchronization, which has been widely studied in various disciplines from physics to mathematics [199-202] for studying various natural systems such as neural networks [203], activity of pacemaker cells for heart beats [204], spin-orbit resonance of Mercury [205], social networks [202] and the internet [200]. For instance, Craighead and coworkers [20] have showed the intrinsic localized modes for coupled micromechanical resonators. Mohanty and coworkers [19] have remarkably shown that coupled resonators are able to capture the synchronization behavior that is consistent with theoretical predictions. Raman and coworkers [21] employed the coupled resonance for mass sensing applications, and showed that the coupled resonance enhances the detection sensitivity. It should be noted that the coupled resonance behavior shown in Refs. [19-21] is slightly different from synchronized oscillation in that the coupled resonators reported in Refs. [19-21] do not include the feedback system that makes the mechanical modes undergo self-sustained oscillation. In other words, the development of a feedback loop is essential for establishing a nanomechanical synchronization system (see the recent review of Rhoads et al. [206]). For instance, Cross et al. [207, 208] suggested a feedback constructed from a reactive coupling due to elastic or electrostatic interactions between nanomechanical resonators, and nonlinear frequency pulling.

For brief elucidation of coupled resonator, let us consider the case where two nanomechanical resonators are coupled by mechanical constraints. The dynamic motion for such a coupled resonator system is given as [188]

$$\begin{bmatrix} M_1 & 0 \\ 0 & M_2 \end{bmatrix} \begin{Bmatrix} \ddot{z}_1(t) \\ \ddot{z}_2(t) \end{Bmatrix} + \begin{bmatrix} K_1 + C & -C \\ -C & K_2 + C \end{bmatrix} \begin{Bmatrix} z_1(t) \\ z_2(t) \end{Bmatrix} + \begin{bmatrix} \alpha_1 & 0 \\ 0 & \alpha_2 \end{bmatrix}$$
$$\times \begin{Bmatrix} [z_1(t)]^3 \\ [z_2(t)]^3 \end{Bmatrix} = \begin{Bmatrix} f_1(t) \\ f_2(t) \end{Bmatrix} \tag{29}$$

Here, $M_1$ and $M_2$ are the effective masses of the two resonators, respectively, $K_1$ and $K_2$ are the effective force constants for each resonator, respectively, $z_1(t)$ and $z_2(t)$ are the variables representing the motion of the each resonator, respectively, $\alpha_1$ and $\alpha_2$ are constants that represent the nonlinearity, $f_1$ and $f_2$ are the excitation forces acting on each resonator, respectively, and $C$ indicates the spring constant for coupling between two resonators. The governing equation can be also written in the matrix form $\mathbf{M}\ddot{\mathbf{z}} + \mathbf{K}\mathbf{z} = \mathbf{F}$, where we note that the governing equation assumes that each resonator vibrates in the harmonic oscillation regime. That is, the nonlinear effect such as geometric nonlinearity has not been considered in the model. The localized mode can be easily found from the eigenvalue problem **Mu** = λ**Ku**, where λ is an eigenvalue of the system and **u** is its corresponding normal mode. In the recent study by Mohanty and coworkers [19], the coupled resonance has been observed by using an excitation force that has the driving frequencies of Ω = $(m/n)\omega_0$, where $m$ and $n$ are integer winding numbers of the two resonators and $\omega_0$ is the resonant frequency of the coupled system. This is essentially identical to the theoretical treatment of Arnold's tongue [209, 210], whose governing equation is obtained from Eq. (29) with excitation force **F** represented in the form of **F** = **q**cos(Ω$t$). The details of theoretical treatment of Arnold's tongue are provided in Ref. [209]. Furthermore, the coupled resonator has recently been highlighted as a sensitive mass-sensing tool [21], which is described in Section 4.2.2.

*4.1.3. Perspectives*

As demonstrated above, nanoscale resonant systems can be used to perform fundamental studies related to nonlinear dynamics. Importantly, this indicates that theories in nonlinear dynamics can be tested and validated using NEMS devices. Moreover, it is implied that researchers may be able to develop nanoscale nonlinear systems that can perform desired functions. For example, nanoscale functional devices may provide the opportunity to develop a nanoscale neural network that may replace the damaged neural network in human physiology [211]. Furthermore, there have been few experimental studies on nonlinear dynamics such as synchronization [19], stochastic resonance [18] and chaotic resonance [212]. We believe that it is noteworthy to consider small-scale nonlinear dynamics with various modeling techniques ranging from continuum mechanics models to atomistic models, since most NEMS device exhibits the high surface-to-volume ratio which leads to unique physical behaviors such as surface effects on the resonance motion. So far, most theoretical interpretations [18, 19, 212] of nanoscale nonlinear dynamics have been made using continuum mechanics model excluding the surface effect. These theoretical models cannot be utilized to study the nonlinear dynamics of nanowires where surface effect plays a dominant role. It is therefore evident that the nonlinear dynamics of nanoscale devices has to be carefully treated with novel theoretical models (to be discussed in Section 4.3) that take into account the surface effect (e.g. surface stress and surface elasticity).

**4.2. Mass Sensing: Beyond the Conventional Detection Principle for Mass Sensing**

We have already described the most commonly used principles for mass detection using MEMS/NEMS resonators in Section 3.1.2. For such a principle, we have presumed that the nanomechanical resonator vibrates within the harmonic oscillation regime, and that the stiffness of the adsorbed molecules is negligible. In this Section, we consider the detection principle for cases where the resonance motion is dominated by more than just the added mass, which is typically assumed in the conventional detection principle. Specifically, we first consider the mass detection using nonlinear oscillations. It is shown that the frequency shift due to mass adsorption can lead to quite unique behavior using nonlinear oscillations, and that the detection sensitivity can be increased by nonlinear oscillation. The detection using coupled resonators is also reviewed, and it is shown that the detection limit can be improved by coupled resonators when the change in normal modes due to mass adsorption is taken into account. We finally review the detection principle for the case where not only the mass but also the stiffness of adsorbed molecules plays a critical role on the observed resonant frequency shift.

The detection principles based upon continuum elastic models, which we reviewed in Sections 2 and 3, have been rarely considered for experiments on nanomechanical resonator-based detections. As mechanical resonators are scaled down to nanoscale in order to improve the detection sensitivity, they encounter the unanticipated, aforementioned surface effects. For instance, the resonant frequency shift due to chemisorption of proteins onto resonant microcantilevers is much larger than the anticipated frequency shift, which is attributed to not only the added mass but also the surface stress change due to mass adsorption [166]. This indicates that the conventional detection principle is unable to explain the experimentally observed phenomena for resonator-based detection. Therefore, our reviews on the theoretical framework for novel detection principles will provide not only the fundamental physics on experimentally observed phenomena in resonator-based sensing, but also further guidelines to design *de novo* detection schemes using nanoresonators.

*4.2.1. Mass Detection Using Nonlinear Oscillations*

As shown in Section 4.1.1, nonlinear oscillations due to geometric nonlinearity can lead to an increase of the resonant frequency of a nanoscale device, i.e. stiffening effect driven by geometric nonlinearity. This implies that the detection sensitivity could be improved by geometric nonlinearity-induced nonlinear oscillation, since the detection sensitivity is highly correlated with the resonant frequency, which is increased by nonlinear oscillation. To our best knowledge, mass detection based on nonlinear oscillation has been rarely studied except recent theoretical studies by Buks et al. [213] and Eom et al. [120]. In that work [120], they considered the vibrational motion of a nanobeam (e.g. CNT, nanowire, etc.) with molecular adsorption given by Eq. (7) and with effective mass per unit length defined as $\mu(x, t) = \rho_{NB}A + \Delta\mu(x) \cdot H(t - t_0)$, where $\rho_{NB}$ and $A$ indicate the mass density and the cross-sectional area of a nanobeam, respectively, $\Delta\mu(x)$ is the position-dependent mass density for adsorbed molecules onto a nanobeam, $H(t)$ is the Heaviside unite step function, and $t_0$ is the time at which molecular adsorption is initiated. For convenience, $t_0$ is set to $t_0 = 0$. In the case of homogeneous molecular adsorption over the entire length of a nanobeam, $\Delta\mu(x)$ becomes $\Delta\mu(x) = \Delta m/L$, where $\Delta m$ is the total mass of adsorbed molecules. For the case of molecular adsorption at the location $x = a$ (where $0 < a < L$), $\Delta\mu(x)$ is given as $\Delta\mu(x) = \Delta m \cdot \delta(x - a)$, where $\delta(x)$ is a Dirac delta function. Using the Rayleigh-Ritz method [71], which is essentially identical to Galerkin's method [192-194], the equation of motion for a nanobeam with molecular adsorption becomes the Duffing equation [48, 49, 194] given by

$$[\varphi + \psi T]z(t) + \kappa[z(t)]^3 + \sigma\ddot{z}(t) = p_0 \cos\Omega t \tag{30}$$

Here, the parameters $\varphi$, $\psi$, $\kappa$, and $\sigma$ for the Duffing equation are given as [120]

$$\varphi = \frac{16\pi^4 EI}{3L^2} \tag{31.a}$$

$$\psi = \frac{4\pi^2}{3L} \tag{31.b}$$

$$\kappa = \frac{8\pi^4 EA}{9L^3} \tag{31.c}$$

$$\sigma = \sqrt{2/3}\rho_{NB}AL + \Delta M \tag{31.d}$$

where $E$, $I$, $L$, and $A$ indicate the elastic modulus, cross-sectional moment of inertia, length, and cross-sectional area of a resonator, respectively, and $\Delta M$ is an effective added mass. In the case of uniform molecular adsorption over the entire length of the resonator, the effective added mass is given by $\Delta M = (2/3)^{1/2}\Delta m$, where $\Delta m$ is the total mass of adsorbed molecules. For the case of localized adsorption at $x = a$ (where $0 < a < L$), the effective added mass is written as $\Delta M = (2/3)^{1/2}\Delta m[1 - \cos(2\pi a/L)]$. The numerical simulation using the Duffing equation shows that the resonant frequency is increased due to mass adsorption when the excitation force is large enough to induce nonlinear oscillations (see Fig. 17). This phenomenon is quite unique, and is very different from the conventional mass sensing where the resonant frequency is usually decreased due to mass adsorption in the harmonic oscillation regime. Moreover, for a relatively long nanobeam, the magnitude of the frequency shift due to mass adsorption in the nonlinear oscillation regime is larger than that in the harmonic oscillation regime (see Fig. 17d), which indicates that the detection sensitivity can be noticeably enhanced using geometric nonlinearity-induced nonlinear oscillation.

In a recent study by Eom et al. [120], the role of mechanical modulation on the detection sensitivity based on nonlinear oscillation has been considered, since the frequency shift in the harmonic oscillation regime due to mass adsorption can be amplified using mechanical tension. This is attributed to the fact that mechanical extension of a nanobeam increases the resonant frequency [121, 122] (see Section 2.3.1), which leads to an increase in the detection sensitivity. Numerical simulations based on Eq. (30) showed that the mechanical extension reduces the magnitude of resonant frequency shift in nonlinear oscillation regime due to mass adsorption, which is contradictory to the frequency shift in harmonic oscillation regime driven by mass adsorption. Moreover, Buks et al. [213] have theoretically found the minimum detectable mass for nonlinear oscillation-based mass sensing. In addition, they have also theoretically showed that nonlinear oscillation leads to a slowing down of the resonator's response to mass adsorption [213], which is one key limitation of nonlinear oscillation for the real-time mass detection.

It should be noted that, in the recent study by Eom and coworkers [120], the numerical simulation is only valid for vibrating CNTs rather than nanowires, since the theoretical model does not take into account the surface effect which plays a critical role on the deflection motion and/or the resonance motion of a solid nanowire. The nonlinear oscillation-based mass detection using a solid nanowire is currently under consideration for the future work. We believe that the surface effect such as surface elasticity and/or surface stress plays a vital role on not only the resonance motion itself but also the detection sensitivity using vibrating nanowires in both the harmonic and nonlinear oscillation regimes.

*4.2.2. Mass Sensing by Coupled Resonators*

Recently, Raman and coworkers [21] first reported ultra sensitive mass detection using coupled micromechanical resonators. Specifically, they showed that the detection sensitivity based on the measurement of the frequency shift due to mass adsorption is not improved by coupling two resonators, whereas the detection limit based on change in deflection eigenmode due to mass adsorption is remarkably enhanced in coupled resonators. Here, the deflection eigenmode was experimentally determined by measurement of resonance amplitudes and phases. Until recently, except for the work by Raman and coworkers [21], the coupled resonators have not been employed for mass sensing, though the detection sensitivity could be augmented by coupled resonator.

For a quantitative understanding of mass sensing using coupled micro/nanomechanical resonators, we consider the coupled resonator modeled as 2 harmonic oscillators that are coupled to each other via a mechanical constraint. The equation of motion for this coupled resonator upon mass adsorption onto one of the coupled resonators is written as

$$\left\{ \begin{array}{c} \ddot{x}_1(t) \\ \ddot{x}_2(t) \end{array} \right\} + \omega_0^2 \begin{bmatrix} 1+\xi & -\xi \\ -\xi & (1+\xi)/(1+\varepsilon) \end{bmatrix} \left\{ \begin{array}{c} x_1(t) \\ x_2(t) \end{array} \right\} = \mathbf{0}, \text{ or } \ddot{\mathbf{x}} + \mathbf{K}^* \mathbf{x} = \mathbf{0} \qquad (32)$$

where $x_j$ indicates the deflection for the $j$-th resonator, $\omega_0$ is the resonant frequency for each resonator (without any coupling), i.e. $\omega_0 = (K/M)^{1/2}$ with $K$ and $M$ being the stiffness and the mass for a resonator, respectively, $\xi$ is the dimensionless force constant for coupling, i.e. $\xi = C/K$ with $C$ being the spring constant for the mechanical constraint, and $\mathbf{K}^*$ is an effective stiffness matrix (i.e. stiffness matrix normalized by mass $M$) for mass-adsorbed coupled resonator, and $\varepsilon$ is the normalized mass of adsorbed molecules, i.e. $\varepsilon = \Delta m/M$ with $\Delta m$ being the total mass of adsorbed molecules.

In order to understand the shifts in resonant frequency (equivalent to shifts in eigenvalue) and/or the changes in eigenmode due to mass adsorption, the perturbation theory [214] is utilized while denoting $\lambda^k$ and $\mathbf{u}^k$ as the $k$-th eigenvalue and eigenmode, respectively, for effective stiffness matrix $\mathbf{K}$ for a coupled resonator without mass adsorption, i.e. $\mathbf{K} = \begin{bmatrix} 1+\xi & -\xi \\ -\xi & 1+\xi \end{bmatrix}$. In particular, upon the molecular adsorption, the eigenvalue and eigenmode for effective stiffness matrix $\mathbf{K}^*$ for a mass-adsorbed coupled resonator given by Eq. (25) are assumed as $\bar{\lambda}^k = \lambda^k + \varepsilon \cdot \mu^k + O(\varepsilon^2)$ and $\bar{\mathbf{u}}^k = \mathbf{u}^k + \varepsilon \mathbf{v}^k + O(\varepsilon^2)$, respectively. In addition, with an assumption of $\varepsilon \ll \xi \ll 1$, the effective stiffness matrix given by Eq. (25) can be approximated as $\mathbf{K}^* = \mathbf{K} + \varepsilon\mathbf{G}$, where $\mathbf{G} = \begin{bmatrix} 0 & 0 \\ 0 & -(1+\xi) \end{bmatrix}$, from a Taylor series expansion while neglecting the higher order terms, i.e. $O(\varepsilon^2)$. As a consequence, perturbation theory [214] provides the relative shifts in the eigenvalue and eigenmode, respectively, given by

$$\frac{\bar{\lambda}^k - \lambda^k}{\lambda^k} = \frac{(\mathbf{u}^k)^T \mathbf{G} \mathbf{u}^k}{\lambda^k (\mathbf{u}^k)^T (\mathbf{u}^k)} \approx -\frac{\varepsilon}{2} \quad (33.a)$$

$$\frac{\|\bar{\mathbf{u}}^k - \mathbf{u}^k\|}{\|\mathbf{u}^k\|} = \frac{\left\|\sum_{j \neq k}(\lambda^j - \lambda^k)^{-1} \mathbf{G} \mathbf{u}^k\right\|}{\|\mathbf{u}^k\|} \approx \frac{1}{4}\left(1 + \frac{1}{\xi}\right)\varepsilon \quad (33.b)$$

This indicates that the normalized frequency shift is determined only by the normalized added mass $\varepsilon$, whereas the normalized eigenmode shift is linearly proportional to the normalized added mass $\varepsilon$ and inversely proportional to normalized coupling strength $\xi$. That is, for mass sensing using a coupled resonator, the relative shift in eigenmode induced by mass adsorption can be manipulated by controlling the coupling strength [21]. This suggests that the measurement of eigenmode shift due to mass adsorption would be a useful route to improvement of the detection sensitivity. For instance, Raman and coworkers [21] showed that the relative shift in eigenmode (e.g. amplitude) due to mass adsorption for two coupled microcantilevers is larger by two orders than the relative frequency shift induced by mass adsorption.

Nanomechanical detection based on coupled resonance (i.e. localized eigenmode) may exhibit a few attractive features. First, the detection sensitivity for localized mode-based sensing is independent of Q-factor, while the detection limit for conventional resonance-based sensing is sensitively governed by Q-factor. Second, as shown in Eq. (33.b), the detection sensitivity for coupled resonance-based sensing can be enhanced by controlling the coupling strength through fabrication. Third, localized mode-based detection using the multiple coupled resonators may enable the sensing paradigm to be utilized to sense and detect multiple analytes with ultrahigh detection sensitivity. Finally, the detection sensitivity for localized mode-based sensing could be enhanced by a particular attempt at optimal design of coupled resonator. For instance, while our review is restricted to the coupling between identical resonators, the relative shift in eigenmode for localized mode-based detection could be tailored by optimal design of coupled resonator, e.g. coupling of geometrically different resonators.

*4.2.3. Mass Effect vs. Stiffness Effect*

The detection principle for mass sensing that was discussed in Section 3.1.2 assumed that added mass plays a critical role on the frequency shift. This assumption is only valid when bending rigidity change due to added molecules is much smaller than the bending rigidity of a bare resonator. For instance, this detection principle is appropriate for sensing biological molecules using resonant piezoelectric microcantilevers [22], since the size and stiffness of the added biomolecule are usually given as ~10 nm and ~1 GPa [215, 216], respectively, and consequently the bending rigidity change is ~8 × 10$^{-13}$ Nm$^2$, which is much smaller than the bending rigidity of a bare microcantilever which typically has the units of ~1 × 10$^{-7}$ Nm$^2$.

However, as the resonator size (e.g. thickness) becomes comparable to that of the added molecules, the frequency shift due to molecular adsorption is dominated by not only the mass of added molecules but also the stiffness of the added molecules [61, 62]. Bashir and coworkers [62] first showed that biomolecular adsorption increases the resonant frequency of an ultra-thin cantilever, which had a thickness of less than 100 nm. This finding contradicts the conventional detection principle [6, 22, 30, 146] for mass sensing, and is attributed to the fact that the resonant frequency shift is controlled by the elastic stiffness of adsorbed biomolecular monolayer rather than the added mass. In general, the resonant frequency shift, $\Delta\omega$, due to molecular adsorption can be represented in the form $\Delta\omega/\omega_0 = \Delta k/2k - \Delta m/2m$ [22, 62], where $\omega_0$, $k$, and $m$ are the resonant frequency, the stiffness, and the mass of a resonator, respectively, and $\Delta k$ and $\Delta m$ are the stiffness, and the mass of adsorbed molecules, respectively. This indicates that the resonant frequency can be increased due to molecular adsorption if the effect of the stiffness of the adsorbed molecules dominates the resonance motion.

Tamayo and coworkers [61] have systematically studied how to modulate the frequency shift due to molecular adsorption such that either the stiffness effect or the mass effect can control the frequency shift. In their study [61], the adsorption of a single virus particle, a single cell, or protein monolayer onto a thin cantilever is considered as a model system (see Fig. 18a). The equation of motion for such a model system, where a single virus particle or a single cell (or even protein monolayer) is adsorbed at the location of $a < x < b$ is given by

$$\mu(x)\frac{\partial^2 w(x,t)}{\partial t^2} + \frac{\partial^2}{\partial x^2}\left[D(x)\frac{\partial^2 w(x,t)}{\partial x^2}\right] = 0 \tag{34}$$

Here, $\mu(x) = \rho_c b h_c + \rho_a b h_a(x)[H(x-a) - H(x-b)]$ and $D(x) = E_c I_c^0[(1 - H(x-a)) + H(x-b)] + [E_c I_c(x) + E_a I_a(x)][H(x-a) - H(x-b)]$, where $\rho_c$, $h_c$, $b$, $E_c$, and $I_c^0$ are the density, the thickness, the width, the elastic modulus, and the moment of inertia (i.e. $I_c^0 = bh_c^3/12$) for an unloaded cantilever, respectively, $I_c(x)$ is the cross-sectional moment of inertia for a virus (or cell)-loaded cantilever, $\rho_a$, $h_a(x)$, $E_a$, and $I_a(x)$ are the mass density, the thickness, the elastic modulus, and the moment of inertia of an adsorbed virus particle or cell, respectively, and $H(x)$ is the Heaviside unit step function. Here, upon localized adsorption, the moment inertias for virus-loaded cantilever and adsorbed virus, i.e. $I_c(x)$ and $I_a(x)$, are evaluated as

$$I_c = \frac{b(h_c/h_a)\left[h_c^3\left\{2(E_c/E_a) + (E_c/E_a)^2(h_c/h_a)\right\} + h_a\left\{3h_a^2 + 6h_c h_a + 4h_c^2\right\}\right]}{12\left[1 + (E_c/E_a)(h_c/h_a)\right]^2} \tag{35.a}$$

$$I_a = \frac{b\left[h_a^2\left\{h_a + (E_c/E_a)h_c\right\} + h_c(E_c/E_a)^2(h_c/h_a)\left(4h_a^2 + 6h_c h_a + 3h_c^2\right)\right]}{12\left[1 + (E_c/E_a)(h_c/h_a)\right]^2} \tag{35.b}$$

Consequently, one can easily compute the effective bending rigidity $D(x)$ by using moment of inertias given by Eq. (35). The explicit form of effective bending rigidity $D(x)$ is summarized in Ref. [61]. Assuming that $w(x, t) = \psi(x)\cdot\exp[i\omega t]$, where $\omega$ and $\psi(x)$ are the resonant frequency and its corresponding deflection eigenmode, the Rayleigh-Ritz method [71] provides the relation

$$\omega^2 = \frac{\int_0^L D(x)[\psi''(x)]^2 dx}{\int_0^L \mu(x)[\psi(x)]^2 dx} \tag{36}$$

where $\psi(x)$ is the deflection eigenmode for an unloaded cantilever, i.e. $\psi(x) = \sin(\lambda x/L) - \sinh(\lambda x/L) + (\sin\lambda + \sinh\lambda)/(\cos\lambda + \cosh\lambda)[\cosh(\lambda x/L) - \cos(\lambda x/L)]$ with $\lambda$ being a constant given as $\lambda = 1.86$. It is interestingly shown that the frequency shift due to the adsorption is dependent on either the adsorbant mass effect or the adsorbant stiffness effect (see Fig. 18b). It is found that, when the adsorption occurs at the clamped end, the resonant frequency shift is determined by the stiffness of adsorbed molecules (or virus or a single cell), whereas the mass effect is dominant in the frequency shift for the adsorption at the free end. This is consistent with theoretical model dictated by Eq. (36) such that at the clamped end, $\psi|_{x \to 0} \to 0$ indicating that frequency shift due to the mass effect of the adsorbant is negligible. For molecular adsorption at the free end, we have $\psi''|_{x \to L} \to 0$, which shows that the effect of the adsorbant stiffness on the frequency shift is negligible. Further, for the molecular adsorption over the entire length of a cantilever, one can easily find that $\Delta\omega/\omega_0 = E_a I_a/2EI - \rho_a bh/2\rho A$, which is consistent with the finding by Bashir and coworkers [62].

As discussed in Section 3.3.1, the detection principle that accounts for variations in the resonant frequency of NEMS cantilevers due to both mass and stiffness effects of the adsorbates can be employed to monitor the functionality of individual cells because the stiffness of a cell is significantly changed during the cell's function such as apoptosis [156]. Furthermore, the detection principle to estimate the cell stiffness enables the sensitive detection of cancerous cells, since cancerous cells are more flexible than normal cells [158, 159].

**4.3. Effect of Surface Stress on the Resonance**

In the 1970s, the role of surface stress on the resonance behavior of a structure was studied experimentally and theoretically [217, 218]. More recently, this issue has been reconsidered for the label-free detection using nanomechanical resonators. The intrigue in understanding the effect of surface stress on the vibration dynamics of nanomechanical resonators is attributed to the hypothesis that molecular adsorption and/or specific molecular binding on the functionalized surface results in the generation of surface stress that originates from the intermolecular interactions. This hypothesis has been understood using cantilever bending deflection motion in response to such adsorption and/or molecular binding [139]. Moreover, as a resonator is miniaturized to nanometer size scales, it can be characterized by an increasing surface-to-volume ratio which results in an increase in the surface energy [52-56], or the energetic cost to create a free surface. This increased surface energy is also dependent on the mechanical deformation, which leads to the generation of surface stress. In addition, the surface stress is also inherent to nanostructures, which is attributed to the fact that surface atoms have fewer bonding neighbors than do bulk atoms; because the surface atoms are not at equilibrium, the nanostructure is subject to surface stress [57]. Therefore, for the label-free detection using nanocantilevers, the intrinsic surface stress may have a significant effect on the resonant frequencies of nanocantilevers, and consequently the frequency shift due to adsorption [63]. However, the role of surface stress driven by molecular interactions on the resonance behavior of a nanoscale device is still a controversial issue [63]. Furthermore, there is still lacking a general theoretical model that is able to resolve this controversial issue, though a variety of different models have recently been proposed.

*4.3.1. One-Dimensional Beam Model: Gurtin's Argument*

Recently, Thundat and coworkers [167] realized that the mass of added molecules onto a microcantilever is insufficient to describe the resonant frequency shift due to such adsorption. They therefore conjectured that the surface stress induced by intermolecular interactions between adsorbed molecules may also play a central role on the frequency shift of a microcantilever. They provided the equation of motion for a resonant cantilever on which the surface stress resides.

$$\rho A \frac{\partial^2 w(x,t)}{\partial t^2} - S \frac{\partial^2 w(x,t)}{\partial x^2} + EI \frac{\partial^4 w(x,t)}{\partial x^4} = 0 \tag{37}$$

Here, $\rho$, $A$, $E$, and $I$ are the density, the cross-sectional area, the elastic modulus, and the moment of inertia for a resonant cantilever, and $S$ is the effective force due to surface stress where $S = \tau_0 A$ and where $\tau_0$ is a constant surface stress. This is essentially identical to the work by Lagowski et al. [218], who insisted that the resonant frequency of a structure depends on the surface stress. Later, the hypothesis on the role of surface stress in the resonance has been employed in recent works by McFarland et al. [219], Hwang et al. [189], and Dorignac et al. [220] in order to understand the frequency behavior of microcantilever in response to (bio)molecular adsorption. The theoretical solution to the governing equation provides that the resonant frequency shift $\Delta\omega$ due to the constant surface stress is given by $\Delta\omega/\omega_0 = \tau_0 L^3/\pi^2 EI$, where $L$ is the cantilever's length.

However, the model given by Eq. (37) was contradicted by Gurtin and coworkers [168] and later by Lu et al. [63], who showed that the resonant frequency is unaffected by the constant surface stress. This is ascribed to Newton's third law, which was not considered in the model depicted by Eq. (37). Specifically, the surface stress $\tau_0$ induces a residual stress $\sigma_r$ in order to satisfy the force equilibrium, i.e. $\int \sigma_r(y)dy = \tau_0$, where $y$ is the coordinate along the thickness. Consequently, the bending moment exerted by a cantilever with surface stress is

$$M(x) = EI\kappa(x) - \tau_0 h^* + \int_0^h y\sigma_r(y)dy \tag{38}$$

where $\kappa(x)$ is the bending curvature for a cantilever, and $h^*$ is the distance between the surface and the neutral axis, i.e. $h^* = \int y\sigma_r(y)dy / \int \sigma_r(y)dy$. In Eq. (38), it is straightforwardly shown that the effective bending moment is independent of constant surface stress, i.e. $M(x) = EI\kappa(x)$, which leads to the important result that the resonant frequency is independent of the constant surface stress. This implies that the one-dimensional beam model cannot be utilized to analyze the role of surface stress on the resonance behavior of micro and/or nanomechanical resonators.

*4.3.2. Three-Dimensional Elastic Model: Plate Model*

Recently, Sader and Lachut [66, 190] have revisited the theoretical model for studying the role of surface stress on the resonant properties of nanocantilevers in order to resolve Gurtin's argument [168]. In their work, instead of the one-dimensional beam model, a three-dimensional elastic model (i.e. plate model) was used to find the relationship between the constant surface stress and the resonant frequency shift. Because of difficulty in the analytical calculations, they obtained a scaling behavior for this relationship and also utilized the finite element analysis to validate the relationship.

First, the plate exerting the surface stress without any constraints (i.e. without essential boundary conditions) was taken into account. The force equilibrium with assumption of no-traction boundary conditions at all edges provides the in-plane stress state ($\sigma_x$, $\sigma_y$) given by $\sigma_x = \sigma_y = \tau_0/h$, where $\tau_0$ and $h$ are the total constant-surface stress and the thickness of a plate, respectively. Here, $x$ is the coordinate along the cantilever's length while $y$ is the transverse coordinate. With such in-plane stresses, continuum elasticity suggests that the displacement field ($u_x$, $u_y$) for a plate is given as $u_i = -(1 - \nu)\tau_0\xi_i/Eh$, where $E$ and $\nu$ indicate the elastic modulus and Poisson's ratio of a plate, respectively, the subscript $i$ indicates the coordinate index, i.e. $i = x$ or $y$, and $\xi_i$ is the coordinates defined as $\xi_i = x$ or $y$.

In the case of a cantilevered plate with surface stress, the in-plane stresses are no longer uniform due to the essential boundary condition. Specifically, for $x < O(b)$ where $b$ is the width of a cantilevered plate, the in-plane stresses exist, otherwise the in-plane stresses are zero. Sader and Lachut [66] assumed that, for $x < O(b/L)$, the in-plane stresses are uniform. With the plate equation [68, 69, 221], i.e. $D\nabla^4 u_z - \mathbf{N} \cdot \nabla u_z = q$, where $\mathbf{N}$ is the in-plane stress tensor, the effective bending rigidity, $D_{eff}$, for a plate with surface stress is given by the relationship

$$\frac{D_{eff} - D}{D} \approx \frac{(1-\nu)\tau_0}{Eh}\left(\frac{b}{L}\right)\left(\frac{b}{h}\right)^2 \tag{39}$$

where $D$ is the bending rigidity for a plate without surface stress, i.e. $D = Eh^3/12(1 - \nu^2)$, and $h$ is the thickness of a plate. Consequently, the resonant frequency shift due to surface stress is given as

$$\frac{\Delta\omega}{\omega_0} = f(\nu)\frac{(1-\nu)\tau_0}{Eh}\left(\frac{b}{L}\right)\left(\frac{b}{h}\right)^2 \tag{40}$$

This does not contradict the Gurtin's argument such that, for $b \ll L$ (i.e. one-dimensional beam), the resonant frequency shift due to surface stress approaches zero. Sader and Lachut [66] validated the relationship between surface stress and resonant frequency shift obtained from the scaling law using finite element analysis (see Fig. 19).

Based on the theoretical model given by Eq. (40), Sader and Lachut [66] have revisited the previous experimental works by Lagowski et al. [218], McFarland et al. [219] and Hwang et al. [189]. In their work [66], it is shown that the conjecture on the relationship between surface stress and resonance behavior of GaAs cantilever by Lagowski et al. [218] may be incorrect, since the frequency shift due to the constant surface stress using the three-dimensional elastic model is negligible for a GaAs cantilever, which implies that the frequency response of the GaAs cantilever is not impacted by the constant surface stress. In a similar manner, Sader and Lachut [66] reconsidered the experimental work by McFarland et al. [219], who took into account the resonance behavior of cantilever in response to molecular adsorption. They showed that the resonant frequency shift for microcantilevers due to a constant surface stress (estimated from Stoney's formula) is negligible, implying that the resonant frequency shift due to molecular adsorption may be attributed to other factors than a constant surface stress, and that the resonant frequency shift due to surface stress driven by molecular adsorption is almost negligible.

*4.3.3. Surface Elasticity*

Most theoretical and experimental works have concentrated on the effect of a constant surface stress on the resonant frequencies of nanomechanical structures (e.g. cantilevers, etc.). However, as scaled down, the nanomechanical resonator can be characterized by an increasing surface-to-volume ratio, implying that the nanomechanical motion of nano-resonators will be dominated by surface energy rather than strain energy [222, 223]. In other words, unlike the bulk material, the surface elastic energy (in the context of surface elasticity) plays a significant role in the mechanical deformation of nanostructures. The importance of surface effect on the mechanical behavior of nanostructures has been supported by recent experimental and theoretical studies, which show that the effective elastic modulus of nanowires is different from that of bulk material when the nanowire diameter becomes smaller than about 100 nm (for details, see the recent review of Park et al. [67]). As summarized in Ref. [67], there is a discrepancy not only between different experimental predictions on the elastic modulus of nanowires over various size ranges for both metals [224-230] and semiconductors [231-235], but also discrepancies between experiment and theory for both metals [236-242] and semiconductors [231-235].

While there is significant scatter in the experimental measurements of the elastic moduli of nanowires, there are notable specific trends in nanowire's elastic moduli with respect to their cross-sectional sizes. First, there appears to be an experimental trend that for nanowire cross-sectional sizes that are about 100 nm or smaller, there is a distinct deviation in Young's modulus, either stiffening or softening, as compared to the bulk value. Second, it is interesting that the molecular dynamics (MD) simulation results are, as a group, generally self-consistent in that either stiffening or softening is predicted for a specific material; these results are in constrast to the experimental scatter that has been observed. Third, there is a distinct discrepancy in the nanowire sizes at which non-bulk elastic properties are observed between MD simulations and experiments. Specifically, the MD simulations tend to find that the elastic properties approach the bulk value when the nanowire cross-sectional size becomes about 10 nm, which is about one order of magnitude smaller than what is predicted experimentally.

There are many likely and conceivable reasons that individually or collectively may explain the difference in experimental and theoretical predictions of surface effects on the nanowire elastic properties. For example, experimental studies of the nanowire elastic properties have utilized different loading modes, i.e. axial loading (tension), as well as bending or flexural (resonance) type of loading. This may be critical for nanowires because surface effects are expected to contribute most strongly to flexural deformation, since it is known that the stresses are largest at the surfaces in bending deformation as compared to axial loading. Second, the theoretical modeling should account for the nanowires in their actual experimental condition (i.e. pre-existing defects, native oxide layers, possible polycrystalline texture, etc.) rather than perfectly single crystal nanowires that are studied theoretically. Finally, there are likely experimental difficulties and uncertainties to be overcome in terms of instrument calibration, fixing and mounting of samples, boundary and loading conditions, and measurements of nanowire diameters, strains, or cross-sectional areas.

In order to understand surface effects on the mechanical properties of nanobeams, Miller and Shenoy [241] first suggested a theoretical approach to treating and accounting for surface elasticity (i.e. the stiffness of the surface of the nanocantilevers, which is different from the stiffness of the bulk, and which has a larger effect on the overall stiffness of the nanocantilever as the nanocantilever size decreases, or equivalently the surface area to volume ratio increases). In recent years, Lu et al. [63] have considered the effect of surface elasticity on the resonant frequencies of nanomechanical cantilevers. In their work [63], it is remarkably shown that the surface elastic stiffness dominates the resonant frequencies of cantilevers, whereas the constant surface stress does not affect the resonance motion. Here, the surface elastic stiffness is defined as the second-derivative of surface elastic energy with respect to surface mechanical strain.

Recent studies by Wang et al. [243, 244] and Lilley et al. [245, 246] suggested a theoretical model based upon the Laplace-Young equation [247, 248], which takes into account the surface elasticity as well as the constant surface stress in order to gain insight into how nanoscale surface effects impact the resonant frequencies [243, 246], the bending deformation [245], and/or buckling [244] of nanocantilevers. It should be noticed that the Laplace-Young equation [247, 248] to be employed for description of the effect of constant surface stress on the mechanical behavior of a nanostructure violates the Newton's third law, as described in Section 4.3.1, where it is demonstrated that constant surface stress cannot affect the resonance of a nanostructure. Specifically, Newton's third law states that a constant surface stress on the beam induces a residual stress in order to satisfy the force equilibrium [66, 168, 222]. Recently, Sadeghian et al. [249] developed the finite element formulation for a three-dimensional, elastic solid in the presence of a constant surface stress and the surface elastic stiffness in order to understand the size-dependence of elastic modulus of a nanostructure. Gavan et al. [250] experimentally studied how the resonant frequencies of silicon nitride nanocantilevers are determined by either of inherent surface elastic stiffness or the constant surface stress (see Fig. 20); they determined that the resonance frequencies of nanocantilevers are governed by the inherent surface elastic stiffness rather than a constant surface stress [250]. Furthermore, Park et al. [242, 251] computationally reported that both the constant surface stress and the surface elastic stiffness make a significant contribution to the resonant frequencies of nanocantilevers when nonlinear, finite deformation kinematics is taken into account.

We consider the continuum mechanics theory in order to gain fundamental insights into the effect of surface elasticity on the resonance motion of a nanomechanical device. There is strong evidence that continuum mechanics can capture this behavior even down to very small scales, e.g. sub-5 nm size scales. While various continuum-based theories [57, 222, 240, 241, 252-255] that incorporate surface effects have been proposed, we note the representative work by Dingreville et al. [240], who used thermodynamic concepts from Gibbs to develop a continuum model that is able to predict surface effects on the elastic properties of nanostructures down to size scales under 5 nm as compared to benchmark atomistic calculations. This suggests that continuum mechanics, if appropriately modified to account for nanoscale surface effects, can be utilized to study the elastic properties, and therefore the resonant behavior of nanobeams and NEMS.

In order to account for surface effects on nanostructures within a continuum mechanics framework, we note that the surface stress acting on a beam is given by $\tau = \tau_0 + S\varepsilon$ [63, 222, 223, 256], where $\tau_0$, $S$, and $\varepsilon$ represent the constant surface stress, the surface elastic stiffness, and the surface mechanical strain, respectively. Here, strain is given as $\varepsilon = \kappa y$, where $\kappa$ and $y$ are bending curvature and the coordinates along the thickness, respectively. With the surface stresses $\tau_u = \tau_0 + S\kappa h/2$ (on the top surface) and $\tau_u = \tau_0 - S\kappa h/2$ (on the bottom surface), the equilibrium equation provides the bending moment in the form

$$M(x) = \left[EI + Sbh^2/2\right]\left\{\partial^2 w(x,t)/\partial x^2\right\} \tag{41}$$

where $b$ and $h$ are the width and the thickness of the nanomechanical resonator. Eq. (41) indicates that surface elasticity changes the effective bending modulus, which would lead to changes in the resonant frequencies. In general, the equation of motion for a vibrating doubly-clamped nanowire, where the surface elasticity plays a role, is given by

$$\left(EI + \frac{Sbh^2}{2}\right)\frac{\partial^4 w(x,t)}{\partial x^4} - \left[T_0 + \int_0^L (EA/2L)\{\partial w(x,t)/\partial x\}^2 dx\right]\frac{\partial^2 w(x,t)}{\partial x^2} \qquad (42)$$
$$+\rho A \frac{\partial^2 w(x,t)}{\partial t^2} = f(x,t)$$

Consequently, in the harmonic oscillation regime, the resonant frequency shift due to surface elasticity is given by $\Delta\omega/\omega_0 = Sbh^2/4EI$, where it is assumed that $Sbh^2/2 \ll EI$. Lu et al. [63] have systematically studied the role of surface elasticity on the resonance motion. Specifically, the resonance behavior of a cantilever with respect to its size effect (i.e. surface-to-volume ratio) has been studied. More importantly, they have also investigated the resonance behavior in response to the change of surface stress due to molecular adsorption. It is shown that surface stress changes due to molecular adsorption play a minor role on the frequency shift for a typical microcantilever, while such surface stress effects play a critical role on the resonance of ultrathin (nano)cantilevers. Furthermore, the dynamic characteristics of nonlinearly oscillating nanocantilevers will be significantly influenced by surface stress effects, which have not been considered previously.

### 4.4. Perspectives on Molecular Detections

As presented in Sections 4.2 and 4.3, we have overviewed the novel detection schemes using nanomechanical resonators. To our best knowledge, except for the attempts reviewed here, there has been little discussion in the literature how to improve the detection sensitivity. Here, we suggest the further considerations based on theoretical and/or computational modeling, which can provide insights into enhancement of detection sensitivity for resonator-based detections. Furthermore, the insights may be useful for experimentalists to design the novel detection schemes that can be employed in resonator-based sensing experiments.

*Nonlinear Oscillations with Surface Effects:* As suggested, nonlinear oscillations enable the unique, sensitive molecular detections using nanomechanical resonators. Moreover, Buks et al. [213] have theoretically found the minimum detectable mass for nonlinear oscillation-based mass sensing. In addition, they have also theoretically showed that nonlinear oscillation leads to a slowing down of the resonator's response to mass adsorption [213], which is one key limitation of nonlinear oscillation for the real-time mass detection. In addition, Eom and coworkers [120] have theoretically reported that, in case of CNT resonator-based detection, nonlinear oscillations can improve the detection sensitivity. We conjecture that nonlinear oscillation-based detection principles may be employed for nanowire resonator-based detection in order to increase the detection sensitivity. For such a case, the surface effect due to the large surface-to-volume ratio for a nanowire has to be carefully considered with nonlinear oscillations for molecular detection. We anticipate that the effect of surface elasticity and nonlinear oscillation may provide the unprecedented opportunity to improve the detection sensitivity. Our conjecture is based on the fact that goemetic nonlinearity-driven nonlinear oscillations increase the resonant frequencies [28, 48-51, 120], and that surface effects (e.g. surface elastic stiffness) increases the effective bending rigidity leading to the increase of the resonant frequencies for those materials that have the positive surface elastic stiffness [245, 246].

*Coupled Resonance with Nonlinear Effect:* Molecular detection using coupled resonance has not been implemented except a recent work by Raman and coworkers [21]. In their work, they used a coupled, harmonically oscillating cantilever array in order to improve the detection limit. However, in general, coupled nanomechanical systems (e.g. doubly clamped nanowire arrays) experience the nonlinear oscillations [19]; furthermore, nonlinear coupling between nanoresonators will become important due to the importance of developing large area arrays of nanoresonators for bio/chemical sensing applications. It will be worthwhile to consider the nonlinear oscillations of a coupled nanomechanical system in response to mass adsorptions because not only nonlinear oscillations [28, 48-51, 120] but also coupled resonance [19-21] can increase the resonant frequency, and consequently, the detection sensitivity.

*Mass Effect vs. Stiffness Effect in Nonlinear Oscillation:* The analysis of mass sensing using resonators has traditionally been done assuming harmonic oscillations of the resonators [61, 62]. However, it is also worthwhile to consider the mass effect and/or the stiffness effect for adsorbed molecules on the frequency shift in both harmonic and nonlinear oscillations. This detection principle can be applied to characterizing the cell's function, since a cell undergoes the changes of stiffness [156] and mass density [47] during the cell's function such as apoptosis and/or cell cycle. Furthermore, the detection principle enables the separation of cancerous cells from normal cells, since a cancerous cell is more flexible than a normal cell [158, 159].

*Biomolecular Adsorption-Induced Change in Surface Elasticity:* As stated earlier, the constant surface stress due to molecular adsorption does not play any role on the frequency shift [63, 168]. However, a recent study by Lu *et al*. [63] found that the adsorption-induced surface elasticity change may play a role on the frequency shift for a nanomechanical device. The change in surface elasticity has been reported for the case of atomic adsorptions [64, 65] (e.g. oxygen [65], hydrogen [257] and/or metals [258] such as Ga, As, Ge, etc.). On the other hand, the change of surface elasticity has not been suggested for biomolecular adsorption (e.g. protein adsorption). The origin of surface elasticity change due to biomolecular adsorption may be attributed to the intermolecular interactions between the adsorbed biomolecules [139]. This may be studied using multiscale model that couples the atomic model for intermolecular interactions to the continuum model for the flexural, or resonance motion of a nanomechanical system. This multiscale modeling concept will be demonstrated in Section 5.3.1.

## 5. Multiscale/Molecular Modeling-Based Simulations

As described in Section 4, the continuum modeling approaches enable the understanding of the underlying mechanisms in nonlinear dynamics and/or molecular detections. However, as resonators are scaled down to nanometer length scales, the continuum modeling approaches that were discussed in Sections 2 and 3 may not, in their current form, provide fundamental insights into the unexpected nanoscale behavior that impacts the ability to analyze and predict the response of nanoscale resonant sensors such as novel energy dissipation mechanisms, or why the resonant frequencies of nanocantilevers differ from that expected from classical beam theory. Furthermore, the continuum modeling approaches may not, in their current form, enable one to understand the role of intermolecular interactions on the resonant frequency shift for nanoresonators due to (bio)molecular adsorptions. This Section is dedicated to the current state-of-art in molecular modeling approaches such as atomistic simulations, coarse-grained, and/or multiscale simulations that provide alternative approaches to studying nanomechanics, and thus elucidating the unique behavior and properties of nanomaterial-based nanoresonators that have not been resolved by currently available continuum models.

### 5.1. Atomistic Models: Carbon Nanotube Resonators and Graphene Sheet Resonators

The fundamental principle of atomistic simulations such as molecular dynamics (MD) is to numerically solve Newton's equation of motion for all atoms of a system [259], where the interactions, i.e. the forces between atoms are prescribed using an empirical potential field. In general, an empirical potential field prescribed to atoms consists of energies for covalent bond stretch, bending of bond angle, twist of dihedral angle, and non-bonded interactions such as van der Waal's interaction and electrostatic interaction, respectively. Atomistic simulations have been broadly employed to study the dynamics of molecular structures ranging from nanomaterials [260] to biological structures at atomistic scales [261-263]. However, our review is only restricted to atomistic simulations nanoscale resonators; of these, carbon nanotube (CNT) and graphene-based nanoresonators have attracted the most attention.

Energy dissipation in oscillating CNTs has been studied using classical MD simulations for both single-walled (SW) CNTs [264] and multi-walled (MW) CNTs [264-266]. Of these, Jiang et al. [264] studied the temperature dependence of Q-factor degradation in both fixed/free SWCNTs and MWCNTs, while the other works [265, 266] focused on frictional effects on energy dissipation in MWCNTs.

We note that these works [264-266] have focused exclusively on the effects of intrinsic mechanisms, i.e. temperature and interlayer friction between CNTs, while neglecting extrinsic mechanism such as gas damping and clamping losses. Despite this, insights into the Q-degrading intrinsic mechanisms have been found. Specifically, Jiang et al. [264] found that the Q-factors of cantilevered CNTs degrade with an increase in temperature according to a $1/T^{0.36}$ relationship. They also found that the Q-factors of MWCNTs at a given temperature can be nearly one order of magnitude lower than the corresponding SWCNTs. Similar results regarding the deleterious effects of internal friction between CNTs on the Q-factor were found by Zhao et al. [266] and Guo et al. [265]. Interestingly, a recent experimental study by Huttel and coworkers [267] on SWCNT resonators found the same relationship between Q-factor and temperature as did Jiang et al. [264], though it is worth noting that the experimental study was performed across a very small range of low temperatures (20 mK to 1K), while the boundary condition of the experiment (fixed/fixed) differed from the cantilever boundary condition used in the MD simulation. It is also interesting that similar Q-factor dependence on temperature has been found for non-CNT systems, i.e. GaAs/InGaP/GaAs [268] and also single crystal silicon [269].

Classical atomistic modeling has also been utilized to study the Q-factors of both mono- and multi-layer graphene NEMS by Kim and Park [126]. The motivation for the study was the low Q-factors, which typically range between 2 and 2000, that have consistently been experimentally reported for graphene NEMS despite their high structural purity [85, 86, 270]. The simulations of Kim and Park demonstrated that a key factor underlying the low Q-factors arise from so-called edge effects in graphene, where because graphene is a purely 2D material, the edge effects are analogous to surface effects on 3D nanostructures such as nanowires [271, 272]. Specifically, the edge atoms of the graphene NEMS were found to oscillate at different vibrational frequencies than the remainder of the graphene NEMS [126]. Because of this, the spurious edge modes were found in the simulations to quickly propagate into the graphene NEMS, thereby leading to mode mixing, vibrational incoherency, and a rapid loss in the Q-factors of graphene. Therefore, it was determined that experimentally synthesized free-standing graphene NEMS, which thus have two free edges, are not optimal for high-Q graphene NEMS.

Kim and Park [273] also studied the Q-factors of multilayer graphene NEMS, including both intrinsic effects (friction between the graphene monolayers), and extrinsic effects, or the effects of clamping strength between the graphene NEMS and the substrate. Interestingly, it was found that the quality of the attachment between the graphene NEMS and substrate had a strong effect on the Q-factor, where weaker attachment forces between the graphene NEMS and the substrate led to lower Q-factors. Similarly, it was determined that effective friction between graphene layers increased with a decrease in the relatively weak non-bonded van der Waal's interactions that govern the interactions between adjacent graphene layers; this friction between graphene layers also lead to a significant loss in energy, and thus a lower Q-factor.

## 5.2. Coarse-Grained Models: Carbon Nanotube Resonators and Metallic/Semiconducting Nanowire Resonators

In this Section, we review the current state-of-art in computational models of different nanomaterials that have been widely used as the basic building blocks of NEMS, or nanoresonators; these include carbon nanotubes (CNTs), and both metallic and semiconducting nanowires. In general, the mechanical behavior of nanomaterials such as CNTs and metallic/semiconducting nanowires can be understood from all-atom simulations, i.e. classical MD. However, all-atom simulations are computationally prohibitive for larger nanostructures, which can easily contain upwards of tens or hundreds of millions of atoms. This computational restriction has motivated the recent development of coarse-grained models, which are computationally favorable and appropriate for the simulation of mechanical motion of nanoresonators with their various length scales. Specifically, we review highly utilized coarse-grained models such as the atomic finite element method (inspired from quasi-continuum method) and the molecular structural mechanics method for carbon-based nanostructures, and the surface Cauchy-Born model for surface-dominated metallic and semiconducting nanostructures.

*5.2.1. Atomic Finite Element Method: Carbon Nanotube Resonators*

In recent years, Huang and coworkers [274-276] have developed the "atomic finite element method (aFEM)" that allows the fast computation for characterization of nanomechanical deformation of nanomaterials including CNTs. The aFEM is spiritually identical to "quasi-continuum method (QCM)" developed by Phillips, Ortiz, and coworkers [277-281]. The key idea of QCM (or aFEM) is to reduce the degrees of freedom by introduction of mesh (unit cell) which includes the set of atoms. QCM assumes that set of atoms in a mesh undergoes the homogeneous deformation. Moreover, the potential field for all atoms can be simplified by quasi-harmonic approximation. Despite its assumptions, QCM enables the analyses of mechanical deformation of nanomaterials, which are inaccessible with all-atom simulations. In general, aFEM inspired from QCM can be employed to simulate the mechanical deformation of nanomaterials, but we review current state-of-arts in aFEM to model and simulate CNT devices.

For modeling CNT using aFEM, short-range interactions such as covalent bonds or non-bonded interaction (e.g. van der Waal's interaction) dominates the deformation of CNT, which allows one to just consider the unit cell for constructing the element stiffness matrix. If long-range interaction plays a role, then the interaction between unit cells has to be considered when the element stiffness matrix is established. The element stiffness matrix $\mathbf{K}^{el}$ is composed of $3N \times 3N$ block matrices $\mathbf{K}_{ij}^{el}$ defined as $\mathbf{K}_{ij}^{el} = \partial^2 V(\mathbf{R})/\partial \mathbf{r}_i \partial \mathbf{r}_j |_{\mathbf{R} = \mathbf{R}^*}$, where $V(\mathbf{R})$ is the potential energy as a function of atomic coordinates $\mathbf{R} = [\mathbf{r}_1, \ldots, \mathbf{r}_N]$ with $N$ being the total number of atoms in a unit cell (see Fig. 21), $\mathbf{r}_i$ is the coordinates of $i$-th atom in a unit cell, and superscript * indicates the equilibrium state. Once the element stiffness matrix $\mathbf{K}^{el}$ is constructed, the assembly process used in finite element method [193, 282, 283] enables the establishment of stiffness matrix, $\mathbf{K}$, for a system (e.g. CNT). Quasi-harmonic approximation provides the total energy of a system given by

$$E_{tot} \equiv T_{tot} + V_{tot} \approx \frac{1}{2}\dot{\mathbf{u}}^t \mathbf{M} \dot{\mathbf{u}} + \frac{1}{2}\mathbf{u}^t \mathbf{K} \mathbf{u} - \mathbf{u}^t \mathbf{P} \tag{43}$$

where $T_{tot}$ and $V_{tot}$ represent the total kinetic energy and the total potential energy, respectively, $\mathbf{u}$ is the atomic displacement field defined as $\mathbf{u} = \mathbf{R} - \mathbf{R}^*$, $\mathbf{M}$ is the mass matrix (that is diagonal matrix whose component is the molecular weight of carbon atom), $\mathbf{P}$ is the external force field applied to a system, a superscript $t$ indicates the transpose of a vector, and dot represents the differentiation with respect to time. Accordingly, the equation of motion for mechanical motion of CNT is given by $\mathbf{M}\ddot{\mathbf{u}} + \mathbf{K}\mathbf{u} = \mathbf{P}$. For free vibration (i.e. $\mathbf{P} = \mathbf{0}$), the displacement field vector is assumed as $\mathbf{u}(\mathbf{R}, t) = \mathbf{v}(\mathbf{R}) \cdot \exp[i\omega t]$, where $\omega$ and $\mathbf{v}$ are the natural frequency and its corresponding displacement eigen-mode. Consequently, the equation of motion becomes the eigen-value problem such as $\mathbf{K}\mathbf{u} = \omega^2 \mathbf{M}\mathbf{u}$.

Recently, Shi et al. [284] have numerically studied the vibration mode of single-walled CNT (SWCNT) and multi-walled CNT (MWCNT) using aFEM scheme. In their study [284], the CNTs are under the free vibration without any boundary conditions. For such a case, the low-frequency motion corresponds to the breathing mode [285-287] of CNTs such that CNT experiences the vibration motion in radial direction. However, for application to CNT-based resonator, it is essential to prescribe the boundary condition (e.g. cantilevered boundary, or double-clamped boundary) to the CNT. By application of constraints to the system whose total energy is dictated by Eq. (43), one has to modify the stiffness matrix that includes the constraints [193, 282, 283]. Then, it is possible to numerically find the resonant frequencies and their corresponding deflection eigenmode for CNT-based resonators. Furthermore, aFEM is applicable to studying not only the harmonic vibration but also various issues ranging from mass sensing to nonlinear oscillations. We anticipate that aFEM is one of useful simulation tools that allow the computationally efficient analyses on CNT resonators and their related sensing applications.

*5.2.2. Molecular Structural Mechanics Method: Carbon Nanotube Resonators*

In recent years, Li and Chou [288-291] have developed the "molecular structural mechanics method" to model CNTs for characterization of their mechanical behavior. Their key concept is to replace the atomic potential field with the mechanical strain energy composed of energies for axial extension, bending, and torsion. In other words, CNT is modeled as frame network such that carbon atoms regarded as a node (in structural mechanics) are connected by structural frame elements. Unlike aFEM, the molecular structural mechanics model neglects the non-bonded interactions (e.g. Lennard-Jones potential for non-bonded carbon atoms). Nevertheless, the molecular structural mechanics method has been successfully utilized to analyze the mechanical deformation of CNTs, which may be attributed to the conjecture that deformation of covalent bonds (e.g. stretching, bending, and/or torsion) plays a key role on mechanical behavior of CNTs.

As shown in Fig. 22, the potential energy for CNT is associated with strain energies for covalent bond stretch, bending of bond angle, torsion of dihedral angle (and torsion for out-of-plane motion), and non-bonded interaction. The potential energy $V$ can be represented as $V = \Sigma V_s + \Sigma V_b + \Sigma V_t + \Sigma V_{nb}$, where $V_s$, $V_b$, $V_t$, and $V_{nb}$ indicate the strain energies for bond stretch, bending of bond angle, torsion, and non-bonded interaction, respectively. As stated earlier, non-bonded interaction (i.e. $V_{nb}$) can be neglected for mechanical deformation of CNT. With harmonic approximation, $V_r$, $V_b$, and $V_t$ can be expressed as $V_s = (k_r/2)(\Delta r)^2$, $V_b = (k_b/2)(\Delta \theta)^2$, and $V_t = (k_t/2)(\Delta \varphi)^2$, where $k_r$, $k_b$, and $k_t$ represent the force constants for bond stretch, bending of a bond angle, and torsion of a dihedral angle, respectively, $\Delta r$, $\Delta \theta$, and $\Delta \varphi$ are the changes in covalent bond length, bond angle, and dihedral angle, respectively. In the molecular structural mechanics model, the strain energies involved in deformation of a covalent bond are only taken into account. Furthermore, the covalent bond is equivalently replaced with a mechanical beam that can undergo the axial extension, the bending, and the torsion in order to mimic the deformation of a covalent bond. That is, $V_s$, $V_b$, and $V_t$ are regarded as strain energies for axial extension, bending motion, and torsional motion, respectively. Consequently, the structural mechanics parameters such as axial stiffness $EA$, bending modulus $EI$, and torsional stiffness $GJ$ are given by $EA/l = k_r$, $EI/l = k_b$, and $GJ/l = k_t$, where $l$ is the covalent bond length. With the structural mechanics parameters, CNT is modeled as network of mechanical beams with empirically determined stiffness, i.e. $EA$, $EI$, and $GJ$.

Li and Chou [289] have employed the molecular structural mechanics method to analyze the resonant frequencies of SWCNTs. With the structural model of SWCNT with stiffness parameters, the element stiffness matrix for an element consisting of carbon atoms $i$ and $j$ is given by

$$k_{ij}^{el} = \begin{bmatrix} [k]_{3\times 3} & -[k]_{3\times 3} \\ -[k]_{3\times 3} & [k]_{3\times 3} \end{bmatrix} \tag{44}$$

where $[k]_{3\times 3}$ is the 3 × 3 block stiffness matrix for a mechanical beam element. The stiffness matrix, **K**, for SWCNT can be obtained from assembly process usually used in finite element method [193, 282, 283]. For free vibration of SWCNT, the equation of motion becomes the eigenvalue problem such as $\mathbf{Kv} = \omega^2 \mathbf{Mv}$, where **M** is the mass matrix (that is diagonal matrix), $\omega$ and **v** are the resonant frequency and its related deflection eigenmode. Based on such a scheme, Li and Chou [289] have studied the resonance behavior of SWCNTs with respect to their aspect ratio as well as chirality. Furthermore, they have also investigated the resonance behavior of SWCNT in response to mass adsorption at the location far from the clamped end. They have showed that detection sensitivity is related to CNT's length, and that the detection limit is theoretically obtained as zeptogram ($10^{-21}$ g) [291]. Moreover, they have extended the molecular structural mechanics model to modeling the MWCNTs in such a way that the interaction between CNT layers is equivalently substituted with truss rod elements whose stiffness depends on the distance between CNT layers [290].

Coarse-grained models such as the molecular structural mechanics model and/or the atomic finite element method have been useful in elucidating the vibrational properties (e.g. resonant frequencies) and the detection sensitivity of CNT resonators with respect to their structural parameters such as chirality [284, 289], CNT diameter [284, 289], CNT length [289, 290], and structural defects for both SWCNTs and MWCNTs, and/or the number of CNT layers for MWCNTs [290]. Moreover, the detection sensitivity of CNT resonators can be theoretically understood with respect to the aforementioned CNT structural parameters [291], which indicates that coarse-grained models can provide fundamental insights into how to optimize the design of CNT resonators for their specific functions such as actuation and sensing.

*5.2.3 Surface Cauchy-Born Model: Nanowire Resonator*

It is important to note that the discrepancy between experiment and classical atomistic modeling for predictions on the elastic properties of nanowires (e.g. see the recent review of Park et al. [67]) has led to significant demand for multiscale models, which can operate at length scales bridging experiments and simulations, to shed new insights into surface effects on the mechanical properties of nanowires across a range of technologically relevant length scales. Preliminary work in this direction has been performed by Park et al. [242, 251, 292], who developed the surface Cauchy-Born model to capture surface stress effects on the mechanical behavior and properties of both metallic and semiconducting nanomaterials and NEMS [293-295]. The SCB model builds upon the standard Cauchy-Born (CB) model [274, 277, 296], which is a hierarchical multi-scale relationship that enables the calculation of continuum stress and moduli directly from atomistic principles. Because the CB model does not admit surface stress effects, the SCB model was developed by Park et al. [293-295] such that the energy density of a material would include contributions not only from the bulk, but also the material surfaces thus leading to the incorporation of atomistic-based surface stress effects into standard continuum stress measures. The SCB model thus enables, for the first time, the solution of 3D nanomechanical boundary value problems using standard nonlinear finite element (FEM) methods while fully accounting for the effects of atomistic surface stresses. Because of this, as compared to classical atomistic simulations such as molecular dynamics (MD), the SCB enables studies of significantly larger nanostructures at significantly longer time scales.

The SCB model is formulated by establishing a relationship between the continuum strain energy density and the total potential energy of the corresponding, defect-free atomistic system as

$$\sum_{i}^{natoms} U_i(\mathbf{r}) = \int_{\Omega_0^{bulk}} \Phi(\mathbf{C}) d\Omega + \int_{\Gamma_0} \gamma(\mathbf{C}) d\Gamma \qquad (45)$$

where $U_i(\mathbf{r})$ is the potential energy for atom i, $\mathbf{r}$ is the inter-atomic distance, $\phi(\mathbf{C})$ is the bulk strain energy density, $\mathbf{C}$ is the stretch tensor, defined as $\mathbf{C}=\mathbf{F}^T\mathbf{F}$, where $\mathbf{F}$ is the continuum deformation gradient, $\Omega_0^{bulk}$ represents the volume of the body in which all atoms are fully coordinated, $\gamma(\mathbf{C})$ is the surface strain energy density, *natoms* is the total number of atoms in the system and $\Gamma_0$ is the surface area.

From the bulk and surface energy densities, stress and stiffness, which are needed for nonlinear finite element calculations, can be obtained. Specifically, the bulk ($\mathbf{S_B}$) and surface ($\mathbf{S_S}$) second Piola-Kirchoff stresses can be found as

$$\mathbf{S}_B(\mathbf{C}) = 2\frac{\partial \Phi(\mathbf{C})}{\partial \mathbf{C}}, \mathbf{S}_S(\mathbf{C}) = 2\frac{\partial \gamma(\mathbf{C})}{\partial \mathbf{C}} \qquad (46)$$

while the bulk ($C_B$) and surface ($C_S$) tangent moduli can be calculated as

$$\mathcal{C}_B(\mathbf{C}) = 4\frac{\partial^2 \Phi(\mathbf{C})}{\partial \mathbf{C}^2}, \mathcal{C}_S(\mathbf{C}) = 4\frac{\partial^2 \gamma(\mathbf{C})}{\partial \mathbf{C}^2} \quad (47)$$

The potential energy in Equation (39) can be turned into a variational form suitable for solution by standard nonlinear FEM techniques as:

$$\frac{\partial \Pi}{\partial \mathbf{u}_I} = \int_{\Omega_0^{bulk}} \mathbf{B}^T \mathbf{S}_B \mathbf{F}^T d\Omega + \int_{\Gamma_0} \mathbf{B}^T \mathbf{S}_S \mathbf{F}^T d\Gamma - \int_{\Gamma_0} N_I \mathbf{T} d\Gamma \quad (48)$$

where **T** are applied external forces, and **B** is the FEM strain-displacement matrix.

The SCB model has been used extensively to examine, understand and predict how surface stresses impact the resonant frequencies of nanowire-based NEMS [242, 251, 292, 297, 298], resulting in several interesting and novel findings. First, Park et al. [242] found that surface stresses shift the resonant frequencies of NEMS differently depending on the material and also on the boundary condition (see Fig. 23). For example, fixed/fixed FCC metal NEMS were found to show a significant increase in resonant frequency, indicating a significant stiffening in elastic modulus. In contrast, the resonant frequencies of fixed/free FCC metal NEMS were found to show little deviation from the expected bulk values. Furthermore, due to the multiscale, FEM-based nature of the SCB model, stiffening and softening at size scales (greater than 30 nm) that are much larger than what is tractable using full-scale atomistic simulations and comparable to what has been studied experimentally have been predicted using the SCB model. It is worth noting that all of while these trends have been observed experimentally [67, 224, 225, 228, 229], other experiments have not observed the same size-dependence in elastic properties [226, 227, 299].

In contrast, silicon-based NEMS are predicted to show a significant decrease in resonant frequencies due to surface stresses if tested in the fixed/fixed configuration, which indicates substantial elastic softening [251, 292]. However, if fixed/free boundary conditions are used, the resonant frequencies of silicon nanowire-based NEMS are predicted to show little variation as compared to the bulk material. These computational observations are presented in Fig. 24. While these predictions have been observed in some experiments [67, 300-303], other experiments have found that regardless of boundary condition, the elastic properties of silicon nanowire-based NEMS show little difference from the elastic properties of the bulk materials [227, 304, 305].

Finally, the SCB model has also been utilized to quantify the effects of both the residual (strain-independent) and surface elastic (strain-dependent) parts of the surface stress on the resonant frequencies of both metallic and silicon-based NEMS if fully nonlinear, finite deformation kinematics is utilized. It was shown that if finite deformation kinematics is considered, which is in contrast to the linear surface elastic theory of Gurtin and Murdoch [222] described in Section 4.3, the strain-independent surface stress substantially alters the resonant frequencies of the nanowires. However, it was also found that the strain-dependent surface stress has a significant effect, one that can be comparable to or even larger than the effect of the strain-independent surface stress depending on the boundary condition, in shifting the resonant frequencies of the nanowire-based NEMS as compared to the bulk material [242, 251].

### 5.3. Multiscale Models on Biomolecular Detection

This Section is dedicated to a review of current efforts to model biomolecular adsorption onto nanomechanical resonators such as nanocantilevers and/or CNTs. As presented in Section 4.4, it is unclear how the frequency shift due to biomolecular adsorption is governed by surface stress that originates from biomolecular interactions [63, 66, 168]. Furthermore, it is still challenging to make connections between the biomolecular interactions, and the resulting surface stress and/or surface elastic stiffness. In this Section, we review the current state-of-art in multiscale models that have been utilized to elucidate the relationship between biomolecular interactions and the resulting resonant frequency shift of the nanocantilever.

*5.3.1. Cantilever-Based Molecular Recognition*

As stated earlier, micro- and/or nanocantilevers are considered to be a nanomechanical tool enabling the sensitive, label-free detection of biological molecules. As presented in Section 4.4, currently available continuum mechanics models have not been unable to capture the mechanical response of micro or nanocantilevers to biomolecular adsorptions. Recently, there have been theoretical attempts to understand the mechanical response of micro or nanocantilevers to the biomolecular adsorption through multiscale modeling, where the fundamental principle underlying multiscale modeling is to couple two models across multiple spatial/length scales. For example, an atomistic model is used to describe the biomolecular interactions, while a continuum mechanics model is utilized to describe the resulting bending motion of a cantilever. Hagan et al. [306] first suggested the theoretical model based on multiscale modeling concept in order to understand the cantilever's bending deflection motion in response to adsorption of polymer chains and/or DNA chains. Thundat and coworkers [307] developed the multiscale model for atomistic adsorption onto a microcantilever. They numerically studied the bending deflection change due to atomistic adsorption. Huang et al. [308] studied the resonance behavior of a microcantilever in response to atomistic adsorption using multiscale modeling. They showed that interatomic interactions play a key role on the frequency shift due to atomistic adsorption. Eom et al. [309] have provided the multiscale model for adsorption of DNA molecules onto nanocantilevers in order to study the resonant frequency shift induced by DNA adsorption. They reported that intermolecular interactions between adsorbed DNA molecules dominate the resonance behavior when the thickness of a cantilever is comparable to the length of DNA chain. Duan et al. [191] have also scrutinized the nanocantilever's motion (e.g. deflection change and/or frequency shift) in response to atomistic adsorption using multiscale model.

Instead of a thorough review of current attempts, we consider instead a multiscale modeling concept employed in recent studies [191, 306-309], since this modeling concept can provide basic insights into the role of intermolecular interactions between adsorbates on the detection sensitivity of cantilevers. Furthermore, this concept can also be employed to model the dynamic behavior of nanomechanical resonators in response to (bio)molecular and/or atomic adsorptions, as long as the empirical potential field to describe such intermolecular interactions is available. Let us consider molecular adsorption onto the cantilever's surface. The packing density of adsorbed molecules is denoted as $\xi = N/L$, where $N$ is the total number of adsorbed molecules and $L$ is the cantilever's length. As shown in Fig. 25a, the interspacing distance $d$ between adsorbed molecules under the cantilever's bending motion is given by $d(\kappa) = d_0[1 + \kappa c(1 + s/c)]$, where $d_0 = 1/\xi$, $\kappa$ is the bending curvature, $2c$ is the thickness of the cantilever, and $s$ is a distance from the cantilever's surface. Assuming that the potential energy $U(d)$ that governs the intermolecular interactions is known, the total potential energy for a cantilever with molecular adsorption is given by

$$V = \int_0^L (EI/2)\kappa^2 \, dx + \int_0^L \xi U(\kappa) \, dx \tag{49}$$

In a similar manner, the total kinetic energy is

$$T = \int_0^L (\rho A + \xi M) \left[ \frac{\partial w(x,t)}{\partial t} \right]^2 dx \tag{50}$$

where $EI$, $\rho$, and $A$ indicate the bending modulus, the density, and the cross-sectional area of a cantilever, respectively, $M$ is the molecular weight of adsorbed molecule, and $w(x, t)$ is the bending deflection of a cantilever. Using a Taylor's series expansion, the total potential energy $V$ becomes

$$V = \int_0^L \left[ v_0 + \alpha\kappa + (1/2)(EI + \beta)\kappa^2 + O(\kappa^3) \right] dx \tag{51}$$

Here, the parameters $v_0$, $\alpha$, and $\beta$ are defined as $v_0 = \xi U|_{\kappa=0}$, $\alpha = \partial(\xi U)/\partial\kappa|_{\kappa=0}$, and $\beta = \partial^2(\xi U)/\partial\kappa^2|_{\kappa=0}$. For vibrational motion, the bending deflection $w(x, t)$ can be represented in the form of $w(x, t) = u(x) \cdot \exp[i\omega t]$, where $\omega$ and $u(x)$ are the resonant frequency and its corresponding deflection eigenmode. The total energy for a vibrating cantilever during one oscillation cycle is

$$\langle E_{tot}\rangle = \langle T+V\rangle = -\frac{\omega^2}{2}\int_0^L (\rho A + \xi M)u^2\, dx + \int_0^L \left[v_0 + \alpha u'' + (1/2)(EI+\beta)(u'')^2\right]dx \quad (52)$$

where the angle bracket indicates the mean value per oscillation cycle. Knowing the total energy per oscillation cycle, the equation of motion can be found using a variational method (the principle of least action), i.e. $\delta \langle E_{tot}\rangle = (\partial \langle E_{tot}\rangle/\partial u)\delta u = 0$, where $\delta$ represents the variation. Here, $\delta u$ represents the variation of the deflection eigenmode, i.e. the virtual deflection eigenmode. Specifically, the variational method based on the principle of least action with integration by parts provides the weak form of the equation of motion as follows.

$$\delta\langle H\rangle = \int_0^L \left[-\omega^2(\rho A + \xi M)u + (EI+\beta)u^{IV}\right]dx\,\delta u$$
$$+ \left[\alpha + (EI+\beta)u''\right]\delta u\Big|_0^L - (EI+\beta)u'''\delta u\Big|_0^L = 0 \quad (53)$$

The weak form depicted in Eq. (53) provides the equation of motion given by $(EI+\beta)(d^4u/dx^4) - \omega^2(\rho A + \xi M)u = 0$, and consequently the resonant frequency becomes

$$\frac{\omega}{\omega_0} = \sqrt{\frac{1+\beta/EI}{1+\xi M/\rho A}} \quad (54)$$

Here, $\omega_0$ is the resonant frequency for a cantilever without any molecular adsorption. Consequently, the resonant frequency shift due to molecular adsorption is given by

$$\frac{\Delta\omega}{\omega_0} \equiv \frac{\omega-\omega_0}{\omega_0} \approx -\frac{\xi M}{2\rho A} + \frac{\beta}{2EI} \quad (55)$$

Eq. (55) indicates that intermolecular interactions between adsorbed molecules induce a change in the stiffness of the cantilever by the amount $\beta$, leading to the frequency shift. Eom et al. [309] have found that intermolecular interactions between adsorbed DNA molecules induce the change of the effective stiffness of a nanocantilevers, and consequently, the resonant frequency of a nanocantilever (see Fig. 25b – c). Moreover, Duan et al. [191] have showed that the term $\beta$ arising from intermolecular interactions is equivalent to the change of surface elastic stiffness (i.e. strain-dependent surface stress). This implies that, if the intermolecular interactions are explicitly known, the change of surface elastic constant due to molecular adsorption can be computed from the second-derivative of the intermolecular interaction energy with respect to bending curvature. It should be kept in mind that the multiscale model introduced here ignores the effect of binding energy between the adsorbed molecules and the cantilever's surface. In other words, the effect of binding energy for molecular adsorption is assumed to be insignificant in the cantilever's resonance motion.

In practice, this multiscale modeling can be validated by devising experiments that can measure the cantilever bending deflection change and/or resonant frequency shift in response to atomic adsorption, while the deflection change and frequency shift can be theoretically predicted from the above-described multiscale models. In particular, one can consider the several kinds of atomic/molecular adsorptions that result in different types of intermolecular interactions (for which the interatomic potentials are well-known; for details, see Ref. [184]), which leads to the cantilever bending deflection change and/or resonant frequency shift, while the deflection change and frequency shift can be evaluated from experiments. For instance, the atomic adsorption onto cantilever surface is mainly governed by van der Waals interactions [307]. This suggests that for the case of atomic adsorption, the multiscale modeling concept based on van der Waals interaction in the form of $U(d) = -(A/d)^6 + (B/d)^{12}$ is able to theoretically provide the atomic adsorption-induced deflection change (e.g. see Ref. [307]) and/or frequency shift that can be also experimentally measurable (e.g. see Ref. [307, 310, 311]). In the case of DNA adsorption-induced cantilever bending deflection change and/or frequency shift, the intermolecular interactions governing the DNA adsorption consist of electrostatic repulsion (with repulsion amplitude $\beta$ and $\lambda_D$) and hydration replusion (with repulsion amplitude $\alpha$ and screening length scale $\lambda_H$) [312, 313], i.e. $U(d)/L_0 = \alpha(d/\lambda_H)^{-1/2}exp(d/\lambda_H) + \beta(d/\lambda_D)^{-1/2}exp(d/\lambda_D)$, where $L_0$ is the DNA chain length. With this explicit form of intermolecular interactions between adsorbed DNA molecules, it is straightforward to calculate the bending deflection change (e.g. see Ref. [306]) and/or frequency shift (e.g. see Ref. [309]) due to DNA adsorption using multiscale models, while microcantilever-based experiments allow the measurement of DNA-adsorption-induced cantilever bending deflection (e.g. see Ref. [139]) and/or frequency shift.

*5.3.2. Carbon Nanotube-Based Molecular Detection*

In recent years, it has been found that CNTs possess the high binding affinity to DNA chains, and that DNA helically wraps around the CNT with specific configuration depending on the nucleotide sequences [314-317]. This is attributed to the π-π interactions between DNA and CNT surface, which can be modulated with electrostatic repulsion. This has led many researchers to consider CNT-DNA complex for various applications such as CNT sorting [315], CNT-based biosensing [318], and/or CNT separation [317]. Nevertheless, CNT resonators have not been utilized for biosensing applications, though the CNT exhibits a high affinity for the DNA chain.

In a recent study by Eom and coworkers [176], the multiscale modeling concept was developed in order to understand the characteristics of DNA adsorption onto CNT as well as the resonance behavior of CNT in response to DNA adsorption. The key concept is inspired from multiscale models described in previous section such that the atomistic model which is used to depict the electrostatic interactions between nucleotide sequences in DNA chain is coupled to a continuum model describing the bending motion of CNT. Unlike the previous multiscale model demonstrated in Section 5.4.1, the energetically favorable helical configuration of a DNA chain bound onto a CNT can be found. It is attributed to the fact that the atomistic model, which is supposed to be coupled to the continuum model, depends on the energetically favorable configuration. The procedure to find the energetically favorable configuration is given in Ref. [176]. Here, our review is restricted to the multiscale model enabling the description of resonant frequency shift for CNT due to DNA adsorption.

As shown in Fig. 26a, the interspacing distance between nucleotide sites $m$ and $n$ on DNA under the bending of CNT is given by $d_{mn} = d_{mn}^0 [1 - P_{mn}\kappa + Q_{mn}\kappa^2]^{1/2}$, where $d_{mn}^0$ is the interspacing distance in undeformed configuration (i.e. energetically favorable equilibrium configuration), and $\kappa$ is the bending curvature. The electrostatic interactions are represented in the form $V_{el} = \Sigma_m \Sigma_n V_{mn}$, where $V_{mn}$ is the electrostatic interaction between nucleotide sites $m$ and $n$ given as $V_{mn} = q^2/4\pi<e_{mn}>d_{mn}e_0$ with $q$, $<e_{mn}>$, and $e_0$ being the electrical charge, the average permittivity, and the vacuum permittivity, respectively. Consequently, the total potential energy $V$ and the kinetic energy $T$ for CNT resonator with DNA adsorption are given by

$$V = V_{CNT} + V_{el} \approx \int_0^L \left[ v_0 + \alpha\kappa + (1/2)(EI + \beta)\kappa^2 \right] dx \quad (56)$$

$$T = \int_0^L (1/2) \left[ \rho A + \left( \mu_{DNA}\sqrt{1 + \tan^2\theta} \right) G(x) \right] dx \quad (57)$$

Here, $EI$ and $\rho A$ indicate the flexural rigidity and the mass per unit length for CNT, respectively, $\mu_{DNA}$ is the mass per unit length for DNA, an angle $\theta$ represents the helical pattern of adsorbed DNA onto CNT, $G(x)$ is the function defined as $G(x) = 1$ if there exists DNA chain at location $x$; otherwise $G(x) = 0$, and parameters $v_0$, $\alpha$, and $\beta$ are given in Ref. [176]. With the given potential energy $V$ and kinetic energy $T$, one can obtain the resonant frequency for CNT with DNA adsorption using Rayleigh-Ritz method such as

$$\omega^2 = \frac{\int_0^L (EI + \beta)\left[\psi''(x)\right]^2 dx}{\int_0^L \left\{ \rho A + \left( \mu_{DNA}\sqrt{1 + \tan^2\theta} \right) G(x) \right\} \left[\psi(x)\right]^2 dx} \quad (58)$$

where $\psi(x)$ is the admissible function (e.g. deflection eigenmode for CNT without DNA adsorption) which satisfies the essential boundary condition. Fig. 26b shows the resonant frequency shift for cantilevered CNT resonator due to adsorption of specific DNA sequences, e.g. homopolymer such as poly(dT) or poly(dA). This theoretical study suggests that CNT resonator is able to not only detect the DNA chain with high sensitivity but also identify the specific DNA sequences, which implies that a CNT resonator can serve as a nanomechanical sensing toolkit, which can be further applied to genomic sequencing.

### 5.4. Perspectives and Challenges on Coarse-Grained/Multiscale Models

As presented in Sections 5.2 and 5.3, the multiscale/coarse-grained modeling concept has recently been considered to characterize the mechanical response of micro/nanocantilevers, nanowires, and/or CNTs to molecular adsorptions such as atomic adsorption, DNA adsorption, and/or polymer adsorption. However, with regards to studying sensing and detection using micro/nanocantilevers, the multiscale models have been most useful to understand and predict the resonant frequencies of nanoresonators and/or understand how molecular adsorptions impacts the resonant frequencies. For broader application of multiscale modeling to bring novel insights into chemical/biological sensing and detection principles, the following issues should be considered and addressed.

One key area where methodological improvements may yield significant reward is if the multiscale models are further developed to bridge disparate time scales. This issue arises because most multiscale methods have been used to bridge disparate length scales ranging from atomistics to continua, but not time scales. For example, the basic timescale of an atomistic simulation is governed by the vibration frequency of individual atoms or molecules, which is on the order of $10^{-15}$ seconds. However, to accurately study the principles, kinetics and mechanics of sensing using nanocantilevers, researchers must be able to access time scales on the order of micro-seconds to seconds, which is considerably longer than what is available using classical atomistic simulations such as MD. We emphasize that most current multiscale models do not address the time scale issue, and instead focus on the length scale issue.

A significant challenge for multiscale methods (e.g. SCB, aFEM, molecular structural mechanics, etc.) is that they have not been used to develop fundamental insights into the mechanisms underlying Q-factor degradation in NEMS, from both intrinsic and extrinsic sources. For example, they should be further developed to study the influence of gas/fluidic damping, clamping or substrate losses, thermoelastic damping, or surface losses over length and time scales that both are significantly longer than what is currently possible using MD.

Improvements in multiscale models are also necessary to fully understand the mechanisms of molecular adsorption. For example, most multiscale models have been developed by coupling the atomic model of intermolecular interactions to the continuum model of bending motion of nanostructures such as nanocantilevers and/or CNTs. For gaining deep insights into the resonant frequency shift due to (bio)molecular adsorptions, it will be desirable to develop the multiscale model established in such a way that the multiscale model (e.g. SCB, aFEM, etc.) to describe the bending motion of nanostructures is coupled to the atomistic model for adsorbed (bio)molecules. These types of multiscale models would enable the understanding of the underlying mechanism of the mechanical response of nanostructures to (bio)molecular adsorptions over large spatial/temporal scales, which are longer than those accessible with classical MD simulation.

### 6. Future Outlook

Because it is currently impossible to predict which applications and physical discoveries will drive sensing and detection research for the next decade, we instead prefer to discuss some intriguing issues that are of particular interest in the sense that they will enable an understanding of the fundamental physics and make further progress in the development of nanoresonators and their sensing applications. Because we anticipate that, in the next decade, there will be a significant effort to bridge experiments, theories, and computational simulations, our outlook will be based on breakthroughs that may be possible by exploiting this interplay between theories, computational simulations and experiments.

Among these, a particularly intriguing issue in the molecular/multiscale simulation-based design of nanoresonators is the surface effect-driven dynamic behavior of nanoresonator, which strongly impacts the fundamental physics for resonator-based atomic/molecular detections. For example, a recent study by Hines and coworkers [118] experimentally showed that surface chemistry of a nanoresonator plays a key role on the dissipation mechanism (Q-factors). Specifically, the chemical modification of the nanoresonator's silicon surface with alkanethiol induced a significant enhancement in the Q-factor. A recent work by Ru [183] provided the theoretical model (based on continuum model) for understanding the role of surface stress on Q-factors of nanoresonators. A recent study by Seoanez et al. [319] suggested the quantum mechanics approaches in order to gain insight into the relationship between surface roughness and Q-factors. Even though these experimental [118] and theoretical [183, 319] efforts have been made, there is still a gap between experimental observations and theoretical predictions in the role of surface effects (e.g. surface stress, and surface elastic stiffness) on the resonance behavior (e.g. Q-factor) of nanoresonators, and thus there remains a lack of fundamental understanding of surface effect-driven dynamic characteristics. We anticipate the molecular/multiscale simulation-based design may play a critical role in influencing not only our fundamental understanding of experimentally observed surface effects, but also in informing novel design concepts of nanoresonators for specific applications such as sensing and detection.

Another issue of particular interest is the modeling/simulation-based design of nanomechanical resonators for their specific functions such as actuations. For example, resonant MEMS devices have been suggested using continuum modeling (finite element simulation)-based design [320]. Specifically, for design of high-frequency MEMS devices, finite element simulations have been used to gain insight into the relationship between design parameters such as device geometry and resonant frequencies and/or resonant frequency shift due to mass adsorption [320]. However, as devices are scaled down to the nanoscale, the currently available continuum modeling-based design may not capture all of the essential physical effects, for example the resonant frequencies may differ from what is anticipated from currently available continuum models due to nanoscale-specific surface effects. This indicates that molecular and/or multiscale simulation-based design techniques, particularly those based on well-established finite element techniques in order to facilitate usage by existing design engineers and scientists, will receive significant attention for the further development of nanomechanical resonators based on nanostructures.

This interest from scientists and engineers to have well-established computational techniques for nanomaterial-based NEMS is likely to increase significantly due to the recent advances in nanotechnology that have enabled researchers to fabricate single-crystalline nanoscale structures such as nanowires, nanotubes, and/or nanobeams with controllable size and geometry, which will allow these nanostructures to be employed for experimental validation of the nanoscale dynamic behavior of nanoresonators that is predicted by theoretical models and/or computational simulations. For instance, recent studies [321, 322] reported the size-dependent elastic moduli of ZnO nanowires, which were experimentally obtained from measurement of the resonant frequencies of ZnO nanowires. They found that the elastic modulus of ZnO nanowires (nanoresonator) increases as the nanowire diameter decreases, which is contradictory to the results of other experimental work such as AFM bending test [323] and/or theoretical work based on density functional theory [324], which suggests that there may be still other unknown factors outside of the previously discussed surface effects that significantly affect the elastic properties and consequently the resonant frequencies of ZnO nanowires.

Another important factor which has not been studied extensively to-date is the coupling between mechanical deformation and other physical quantities, for example externally applied electric fields [325, 326], or the optics-driven mechanical deformation and resonance motion of nanostructures [327-329]. The work of Zheng et al. [325] demonstrated that applied electric fields can couple with the previously discussed mechanical surface effects to have a significant effect on the elastic properties of nanostructures; this finding is potentially significant as many NEMS are actuated using electrostatic forces or externally applied electric fields. Specifically, in the experiment by Chen et al. [321], the resonant frequencies of the ZnO nanowires were measured in scanning electron microscope (SEM) so that the electrons of the SEM may have an effect on the resonant frequencies and also the elastic properties. This effect was also found recently by Mao et al. [330], who found photoinduced stiffening of ZnO nanobelts under nanoindentation. In a different multiphysics coupling, a recent experiment by Lassagne et al. [331] found that coupling between mechanical oscillations and electrical charge transport strongly limited the potential Q-factors of CNT resonators, where the damping rate due to electromechanical coupling was found to be nearly 6 orders of magnitude larger than that expected in bulk materials. Therefore, it seems clear that new theoretical developments coupled to novel multiscale, coupled physics computational models will need to be developed in order to achieve a fundamental understanding of how different physical fields couple at nanometer length scales, and how these complex couplings impact the dynamic characteristics, i.e. mechanical stiffness and resonant frequencies, of NEMS for sensing-based applications.

Molecular/multiscale simulations which can elucidate the fundamental detection principles for nanoresonators is also of particular interest [332]. Conventional detection principles that are based on continuum elasticity theory are clearly unable to explain the experimental observation that the resonant frequency shift due to chemisorption of biomolecules differs from what is anticipated from theoretical models including surface stress effects [66]. As discussed in Section 4.3, the continuum mechanics models are insufficient to gain the fundamental understanding of chemisorption-driven resonant frequency shift. We expect that molecular/multiscale simulations will enable the fundamental understanding of the resonance behavior in response to chemisorption, and also the development of novel detection schemes for not only sensing the specific biomolecules but also quantifying the biomolecular interactions such as protein antigen-antibody interactions and DNA hybridization. Furthermore, since the surface stress is also dependent on the configuration of adsorbed biomolecules such as DNA [333], the resonance behavior of nanostructures upon biomolecular chemisorption may be governed by molecular conformation of such biomolecules. This has not been studied yet, but it can be quantitatively understood from molecular/multiscale modeling-based simulations.

Another important unresolved issue for NEMS-based sensing is that, even though the multiscale simulations described in Section 5.3.1 have provided some insights into the role of intermolecular (or interatomic) interactions between adsorbates on the deflection motion and/or the resonant frequencies of nanocantilevers [191, 306-309], there have been few attempts to experimentally validate the relationship between intermolecular interactions and the detection sensitivity for nanoresonators, which are well described by such multiscale simulations. We expect that, owing to recent advances in nanotechnology, it will be possible to fabricate a nanocantilever and/or nanoresonator which can be suitable for validating the relationship that is currently available from molecular/multiscale simulations described in Section 5.3.1. For example, there has recently been an experimental effort [334] to validate insights into the role of intermolecular interactions between adsorbates on the bending deflection motion of microcantilevers, which were obtained from multiscale simulations [335]. We anticipate that the atomic adsorption onto nanoresonators and/or nanocantilevers can be considered as a model system for experimental validation, because interatomic interactions for adsorbed atoms can be straightforwardly described by non-bonded interactions such as van der Waals interactions and/or electrostatic interactions [191, 307, 308] (see also discussions in Section 5.3.1). On the other hand, when polymers or biological molecules are adsorbed, there are several possible factors such as ionic strength [312, 313], conformational states of adsorbed molecules [312, 313, 336, 337], hydration [141], which can significantly affect the intermolecular interactions. This indicates that it is easier to experimentally validate the multiscale simulations on atomic adsorption onto nanocantilevers and/or nanoresonators than molecular adsorptions such as polymer/biopolymer adsorptions. Moreover, computational simulations based on multiscale/molecular models may also suggest the experimental design to study the interatomic (or intermolecular) interactions using nanoresonators and/or nanocantilevers.

Finally, we anticipate that modeling and simulation of nanoresonators will, in conjunction with ongoing efforts to modify classical elasticity theory to account for nanoscale physics and phenomena [338], lead to new governing equations that will, in place of the classical results described in Section 2 and 3, be used by experimentalists to interpret their experimental data on nanocantilevers. For example, all relationships and equations that are based upon the Young's modulus of cantilevers will need to be modified due to the fact that the elastic properties of nanocantilevers differ from their bulk counterparts because of surface effects, and because the surface effects impact the elastic properties of different nanomaterials in different, and often unexpected ways [67].

In summary, in a next decade, we anticipate that there will be significant effort to bridge the experiments, theories, and computational simulations to understanding the mechanisms underlying the dynamic behavior of nanoresonators with a particular emphasis on sensing applications. We anticipate that concepts which emerge from computational simulation-based design will provide useful information to experimentalists for not only gaining insight into experimentally observed phenomena, but also in generating novel designs of nanoresonators that optimize performance for their specific functions such as sensing and actuation.

## 7. Conclusion

In this review article, we have demonstrated the experimental, theoretical, and computational approaches that have recently been utilized to gain insights into the underlying mechanisms of nanomechanical resonators as well as their related applications, in particular chemical/biological sensing and detection. Recent experimental approaches to develop nanomechanical resonators for sensing and detection applications have been briefly reviewed along with simple theoretical models based on continuum elasticity that are used to predict and explain their behavior. We have extensively overviewed the current state-of-arts in nanomechanical resonator-based chemical/biological detection, and have elucidated the challenging and unresolved issues and/or perspectives in such detection using nanomechanical resonators. Specifically, we have shown that currently available continuum modeling approaches may need to be modified in order to shed further insights into the atomic-scale mechanisms that govern energy dissipation mechanisms, frequency behavior of nanoresonators in response to intermolecular interactions, and nanoscale surface effects on the resonant frequencies. In order to gain insights into such issues, we have described molecular modeling-based simulations such as atomistic simulations and coarse-grained (multi-scale) modeling, while simultaneously discussing the strengths and limitations to these approaches, and the important issues that must be resolved in order for these approaches to continue making contributions to the understanding of nanomaterial-based NEMS capabilities for sensing and detection. Overall, we hope that our review has elucidated the coupled experimental, theoretical and computational challenges that must be overcome in order to gain fundamental insights into nanomechanical resonators and also to design novel resonator-based applications for specific purposes such as single-molecule detection.

**Acknowledgements**

K.E. appreciates the financial support from National Research Foundation of Korea (NRF) under Grant No. NRF-2009-0071246 and NRF-2010-0026223. H.S.P. acknowledges the financial support from the US National Science Foundation (NSF) under Grant No. CMMI-0750395. D.S.Y. is grateful to the financial support from NRF under Grant No. NRF-2008-0059438 and NRF-521-2008-1-D00580. T.K. gratefully acknowledges the financial support from NRF under Grant No. NRF-2008-313-D00031 and NRF-2010-0009428. The authors collectively thank and acknowledge the anonymous referees for their detailed and insightful comments.

# Tables

**Table 1.** Parameters for PZT piezoelectric and non-piezoelectric layers in the PZT thick film cantilever. Table is reprinted from Ref. [79] under courtesy of D.S. Yoon.

| Material | | Thickness (μm) | Young's modulus (GPa) | Density (kg/m$^3$) |
|---|---|---|---|---|
| PZT | | 22 | 53.9 ± 3.85 | 5250 ± 325 |
| non-PZT | Au | 0.3 | 120 | 19300 |
| | Pt | 0.5 | 168 | 21400 |
| | SiN$_x$ | 1.5 | 290 | 3100 |
| | Si | 10 | 190 | 2330 |

**Table 2.** Experimentally measured and theoretically estimated resonant frequencies (driven at 15 V) of PZT thick film microcantilevers with their different dimensions. Table is reprinted from Ref. [79] under courtesy of D.S. Yoon.

| Cantilever dimension (width × length) [μm × μm] | Experimentally measured resonant frequencies (Hz) | Theoretically calculated resonant frequencies (Hz) |
|---|---|---|
| 400 × 380 | 154,857 | 154,770 – 155,130 |
| 400 × 480 | 99,250 | 99,052 – 99,281 |
| 400 × 580 | 68,875 | 68,786 – 68,945 |

# Figure Captions

**Fig. 1.** Resonant frequencies and their corresponding deflection eigenmodes for various atomic force microscope (AFM) microcantilevers: The low-frequency motions of various AFM microcantilevers are well described by harmonic bending (flexural) modes, whereas some high-frequency motions cannot be depicted by elastic beam model due to coupling between flexural and torsional eigenmodes. Figures are adopted with permission from Ref. [72] © (2008) Elsevier Limited.

**Fig. 2. a.** Scanning electron microscope (SEM) image of PZW-PZT thick film. **b.** SEM image of piezoelectric unimorph microcantilevers made of PZW-PZT thick film. **c.** Resonance curves for PZT microcantilevers with respect to peak-to-peak voltages. The resonant frequencies of PZT microcantilevers are well described by harmonic vibrational motion.

**Fig. 3. a.** Scanning electron microscope image of doubly-clamped silicon carbide nanobeam. A scale bar indicates 1 μm. **b.** Resonance curves for doubly-clamped nanobeam that is actuated by a transversely applied electric field, i.e. d.c. bias. **c.** Resonant frequency shifts for doubly-clamped beam due to d.c. bias. Here, the driving a.c. amplitude is fixed at 70 mV. Blue and red curves show the resonant frequency shifts for beams fabricated along the [110] and [-110] crystallographic directions, respectively, due to d.c. bias. Figures are adopted with permission from Ref. [81] © (2007) The American Association for the Advancement of Science.

**Fig. 4. a.** Experimental set up for measuring the resonant frequencies of microcantilevers immersed in a viscous medium. A fabricated piezoelectric unimorph microcantilever is mounted in a quartz liquid cell, and the vibrational motion is optically measured using laser Doppler vibrometer. **b.** Experimentally measured resonant frequency shifts and Q-factors for piezoelectric unimorph microcantilevers with respect to the viscosity of a medium. Figure 4b is adopted with permission from Ref. [22] © (2007) American Institute of Physics.

**Fig. 5.** Illustration of the degradation in Q-factor with a decrease in volume from nano to macro. Figure is adopted with permission from Ref. [102] © (2005) American Institute of Physics.

**Fig. 6.** Illustration of thermoelastic damping (TED), or a reduction in Q-factor for GaAs and Si cantilevers with an increase in temperature. Figure is adopted from Ref. [112].

**Fig. 7.** Illustration of tensile stress-induced Q-factor enhancement in low stresses SiN nanocantilevers. Figures are adopted with permission from Ref. [121] © (2007) American Chemical Society.

**Fig. 8. a.** Schematic illustration of molecular detection using a cantilever's bending deflection motion. The specific molecular binding on the cantilever surface induces the measurable bending deflection change originating from the surface stress change due to such molecular binding. **b.** The bending deflection change for a microcantilever is presented due to binding of free prostate specific antigen (fPSA) onto the cantilever that was functionalized by fPSA antibodies. Figures are adopted with permission from Ref. [43] © (2001) Nature Publishing Group, Macmillan Publisher Ltd.

**Fig. 9. a.** Experimental configuration of nanomechanical resonator-based mass sensing. Gas molecules flow through nozzle so as to enter into the chip, in which nanoresonator is embedded. **b.** Resonant frequency shifts for two nanoresonators (whose resonant frequencies are initially given as 133 MHz and 190 MHz, respectively) due to physisorption of Xe atoms. The resonant frequency shifts for the 190 MHz device at 37 K were measured with respect to increment of adsorbed mass of ~100 zg. The 133 MHz device at 46 K was used to measure the resonant frequency shift due to increment of adsorbed mass of ~200 zg. Figures are adopted with permission from Ref. [6] © (2006) American Chemical Society.

**Fig. 10. a.** Micrograph image of cells (marked as bright dots) attached to the microcantilever surface. The density of attached cells is $7 \times 10^2$ cells/ml. **b.** Resonant frequency shifts for microcantilevers due to cell adsorption with the cell densities of $7 \times 10^2$ cells/ml, $7 \times 10^4$ cells/ml, $7 \times 10^6$ cells/ml, and $7 \times 10^7$ cells/ml, respectively. Figures are adopted with permission from Ref. [150] © (2005) Elsevier Limited.

**Fig. 11. a.** Scanning electro microscope image of microcantilever, on which vaccinia virus particles are attached. **b.** Resonance curves for a bare microcantilever (colored as gray) and a cell-attached microcantilever (colored as black). The resonant frequency of a microcantilever decreases in response to cell attachment. Figures are adopted with permission from Ref. [160] © (2004) American Institute of Physics.

**Fig. 12. a.** Resonant frequency shift, measured in air, due to protein-protein binding on the cantilever surface. **b.** Real-time measurement of the frequency shift for a microcantilever immersed in buffer solution due to protein-protein binding. The measured frequency shift in air due to protein-protein binding is ~2.7 kHz, while the frequency shift is estimated as ~11 kHz in buffer solution. The discrepancy between frequency shifts evaluated in air and buffer solution is attributed to the change of hydrodynamic loading induced by hydrophilicity change during the protein-protein binding in buffer solution. Figures are adopted with permission from Ref. [22] © (2007) American Institute of Physics.

**Fig. 13. a.** Experimental set up for real-time measurement of biomolecular interactions using a resonant microcantilever immersed in liquid. A prepared solution including biological molecules such as proteins or DNA is injected into a liquid cell, in which a microcantilever is mounted. **b.** *In situ* resonant frequency shift for a microcantilever in response to immobilization of single-stranded DNA and DNA hybridization. The frequency shifts due to DNA immobilization and/or DNA hybridization are well depicted by Langmuir kinetics. Figures are adopted with permission from Ref. [23] © (2008) American Institute of Physics.

**Fig. 14.** Illustration of nanomechanical detection of enzymatic cleavage of peptide chain using resonant microcantilevers immersed in buffer solution. **a.** The resonant frequency of a microcantilever functionalized by peptide chains is inversely proportional to the square root of the effective mass of a functionalized cantilever. **b.** chemical structure of tetrapeptide that can be cleft by an enzyme (i.e. Cathepsin B). **c.** Enzymatic cleavage of tetrapeptide due to Cathepsin B reduces the effective mass of a functionalized cantilever, which eventually increases the resonant frequency of such a cantilever. **d.** The efficiency of enzymatic cleavage with respect to enzyme concentrations in buffer solution was measured from the resonant frequency shifts induced by enzymatic cleavage of tetrapeptide chains. Figures are adopted from Ref. [24] under Creative Commons Attribution License (CCAL).

**Fig. 15.** A family of resonant curves for Si doubly-clamped beam to a progressively increasing drive is presented. Inset shows a SiC doubly-clamped nanobeam actuated by electrostatic forces. Figure is adopted with permission from Ref. [50] © (2006) American Institute of Physics.

**Fig. 16.** Two vibration modes for a nanoresonator. **a.** Out-of-plane mode: a nanoresonator vibrates out of the plane of a gate. **b.** In-plane mode: the nanoresonator oscillates in the plane of a gate. A black arrow indicates the direction of the flexural motion of a nanoresonator. Figures are adopted with permission from Ref. [50] © (2006) American Institute of Physics.

**Fig. 17. a.** Schematic illustration of mass sensing using carbon nanotube (CNT) resonator. **b.** Resonant frequency shifts for a CNT resonator (whose length is 200 nm) due to mass adsorption with respect to amounts of added mass, $\Delta m$, and electrostatic forces, $p_0$. For $p_0 > 0.6$ fN, the resonant frequency of a CNT resonator is increased in response to mass adsorption, which is contrary to conventional mass sensing principle. **c.** Resonant frequency shifts for a CNT resonator due to added mass of 6 ag as a function of electrostatic force, $p_0$. The unique resonant frequency shift due ot mass adsorption in nonlinear oscillation regime is observed when $p_0 > 0.5$ fN. **d.** Resonant frequency shifts for CNT resonators due to added mass of 6 ag with respect to CNT length, $L$, and electrostatic force, $p_0$. Figures are adopted with permission from Ref. [120] © (2009) American Institute of Physics.

**Fig. 18. a.** Schematics of molecular adsorption onto a cantilever. Here, $E_c$, $\rho_c$, $T_c$, and L represent the elastic modulus, density, thickness, and length of a cantilever, respectively, and $E_a$, $\rho_a$, and $T_a(x)$ indicate the elastic modulus, density, and thickness of adsorbed molecules. **b.** Relative resonant frequency shifts for Si cantilever or polymer (photoresist SU-8) cantilever due to uniform adsorption of protein or alkathiol self assembled monolayer (SAM) as a function of thickness ratio $T_a/T_c$. The resonant frequencies for SU-8 cantilever (for protein or SAM adsorptions) and/or Si cantilever (for SAM adsorption) increase due to molecular adsorption. This is attributed to the fact that elastic stiffness of adsorbates critically increases the effective bending ridigity of a cantilever, and consequently the resonant frequencies of a cantilever. Figures are adopted with permission from Ref. [61] © (2006) American Institute of Physics.

**Fig. 19. a.** Relative resonant frequency shifts with respect to $(b/h)^2 \bar{\sigma}$ for a given $L/b$ and Poisson's ratio (e.g. $v = 0, 0.25, 0.49$), where $b$, $L$ and $h$ represent the width, length, and thickness of a nanocantilever, respectively, and $\bar{\sigma}$ is a normalized surface stress defined as $\bar{\sigma} = (1-v)\tau_0/Eh$ with $\tau_0$ and $E$ being the constant surface stress and the elastic modulus of a nanocantilever, respectively. **b.** Normalized resonant frequency shift $\Omega$ due to a constant surface stress $\tau_0$ with respect to $b/h$ and $b/L$ for a given Poisson's ratio $v = 0.25$, where $\Omega = |\Delta\omega/\omega_0|/|\bar{\sigma}(b/L)(b/h)^2|$. Figures are adopted from Ref. [66].

**Fig. 20. a.** Scanning electron microscope image of silicon nitride nanocantilever. **b.** Resonant frequencies of silicon nitride nanocantilevers. **c.** Effective elastic moduli of silicon nitride cantilevers computed from the fundamental resonant frequencies. Blue line shows the theoretical predictions on the effective elastic modulus due to surface elastic stiffness, while red solid line indicates the theoretical predictions from a constant surface stress (in Laplace-Young equation). **d.** Effective elastic moduli of silicon nitride cantilevers calculated from the resonant frequencies for the second lowest normal mode. It is shown that the effective elastic moduli are determined by the surface elastic stiffness rather than a constant surface stress. Figures are adopted with permission from Ref. [250] © (2009) American Institute of Physics.

**Fig. 21.** Carbon nanotube (CNT) structure (shown in left panel) and atomic finite element (aFEM) mesh (unit cell), which contains 10 carbon atoms. Figures are adopted from Ref. [275].

**Fig. 22.** Schematic illustration of interatomic interactions for carbon nanotube (CNT): **a.** Stretching of a covalent bond, **b.** Bending of a bond angle formed by two adjacent covalent bonds, and **c.** Torsion of dihedral angle, defined as an angle between a bond vector and a plane formed by other two adjacent covalent bond vectors.

**Fig. 23.** Illustration of surface effects on shifting the resonant frequency as compared to the bulk material for 16 nm cross section gold nanowires with different boundary conditions.

**Fig. 24.** Illustration of surface effects on shifting the resonant frequency as compared to the bulk material for 16 nm cross section silicon nanowires with different boundary conditions.

**Fig. 25. a.** Schematic illustration of bending of a nanocantilever with DNA adsorption. **b.** Bending stiffness change for a silicon nanocantilever due to DNA adsorption with a given ionic strength of buffer solution. **c.** Relative resonant frequency shift for a nanocantilever due to DNA adsorption with respect to DNA packing density and the thickness of a nanocantilever. Here, the normalized DNA packing density $\eta$ is deinfed as $\eta = \theta/10^9$, where $\theta = N/L$ with $N$ and $L$ being the number of adsorbed DNA molecules and the length of a nanocantilever, respectively. Figures are adopted with permission from Ref. [309] © American Physical Society.

**Fig. 26. a.** Schematic illustration of DNA adsorption onto carbon nanotube (CNT): single-stranded DNA (ssDNA) chain helically wraps the CNT (show in top). The bending motion of the CNT with ssDNA adsorption is presented in the bottom panel. **b.** Relative resonant frequency shift for CNT resonator (whose length is 500 nm) due to adsorption of ssDNA such as poly(dA) and/or poly(dT), respectively. Figures are adopted from Ref. [176].

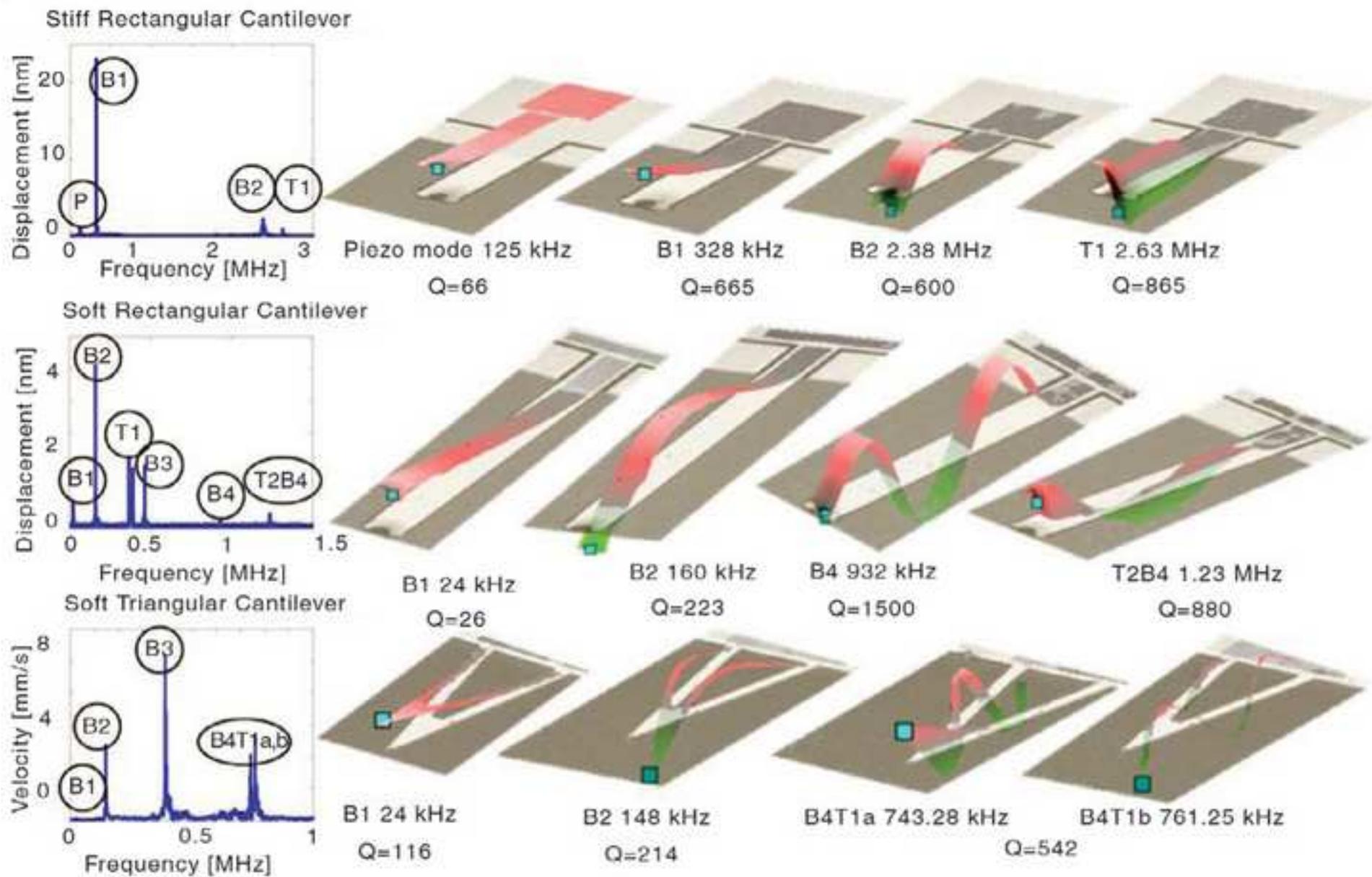

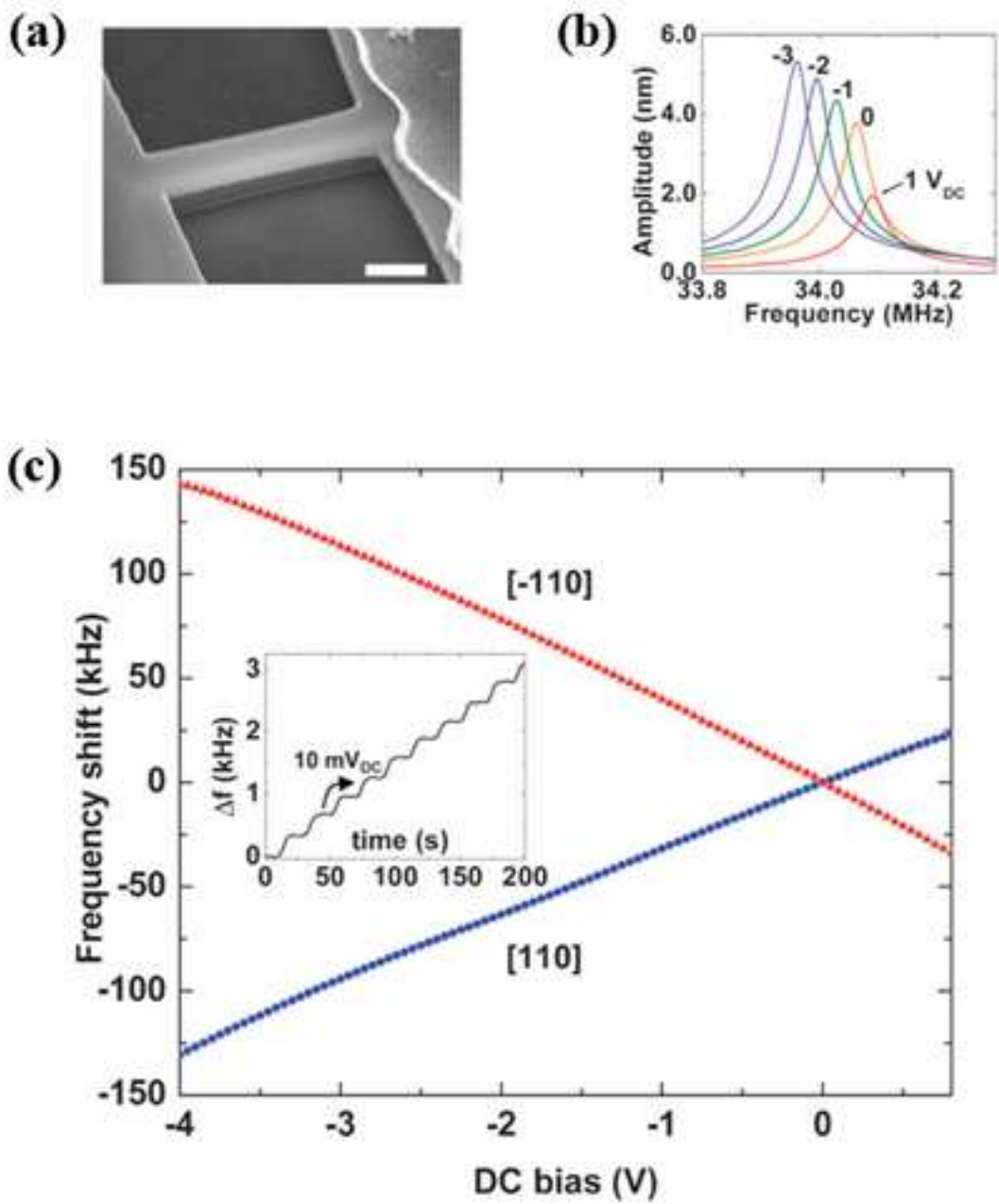

**Figure 4**
Click here to download high resolution image

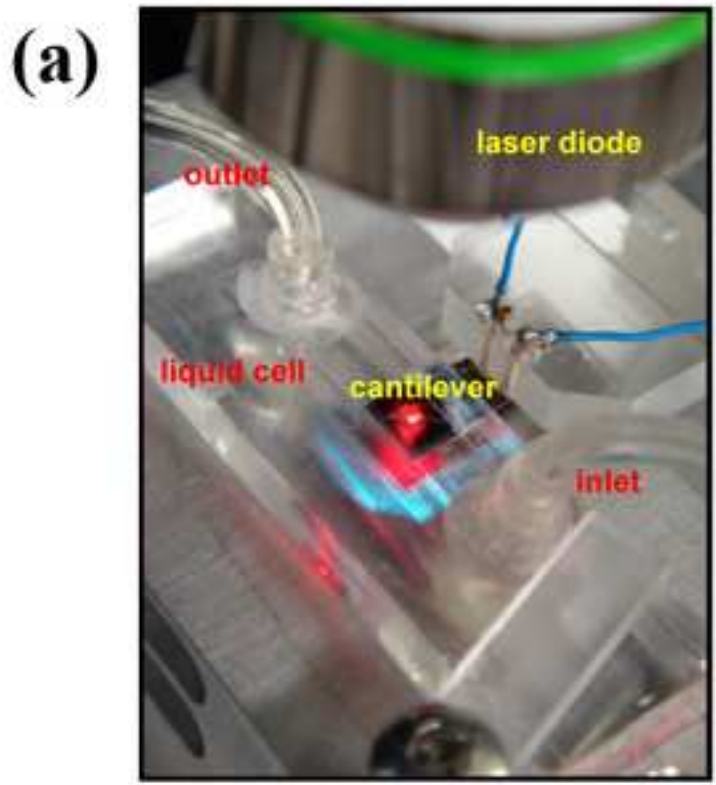
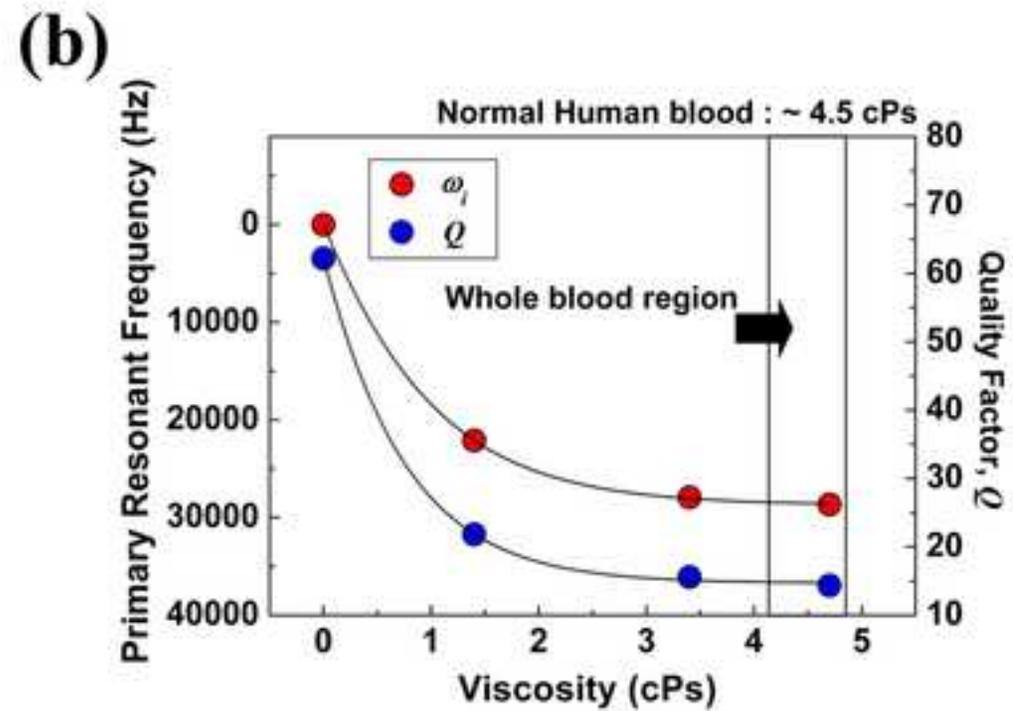

**Figure 5**
**Click here to download high resolution image**

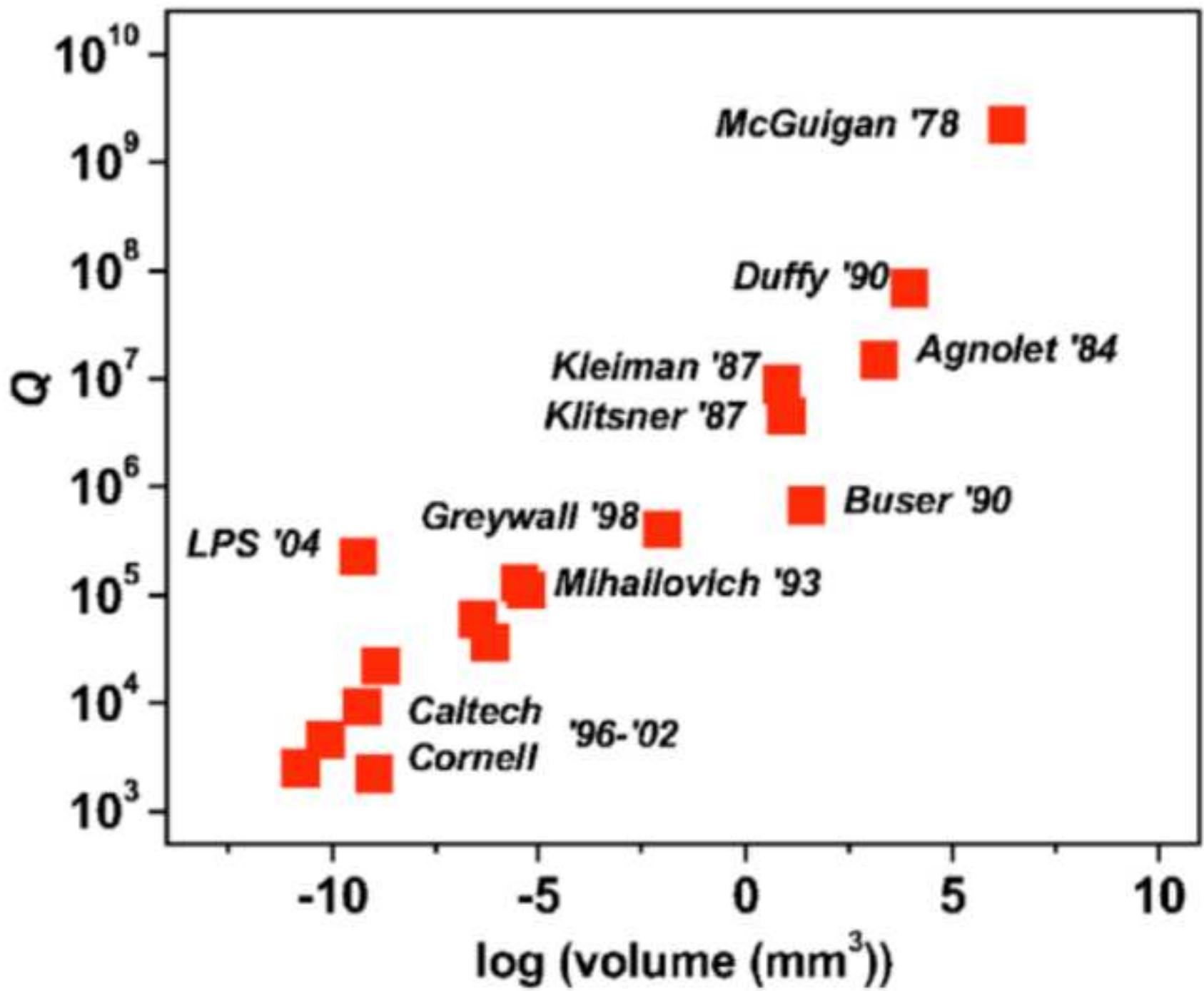

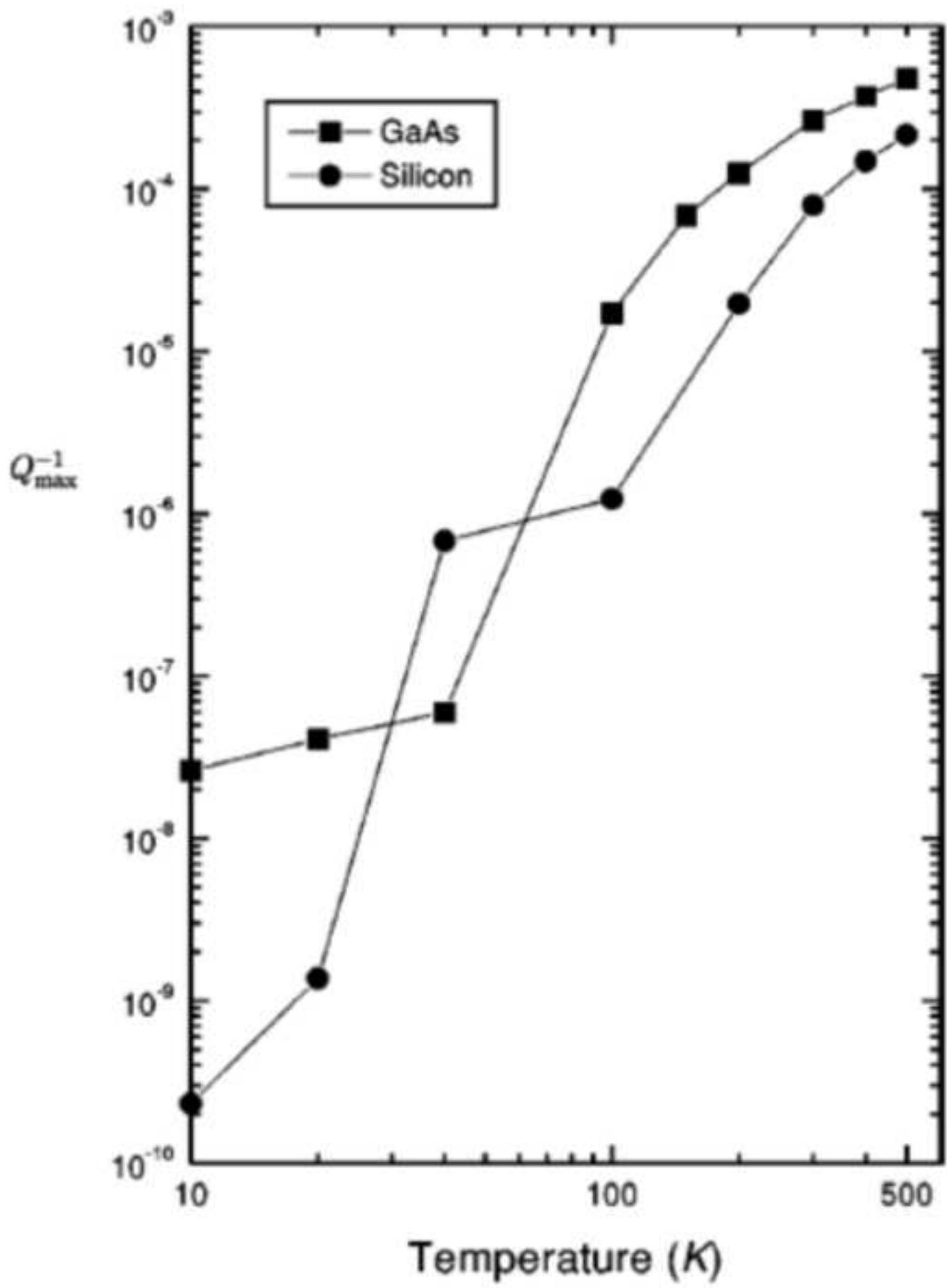

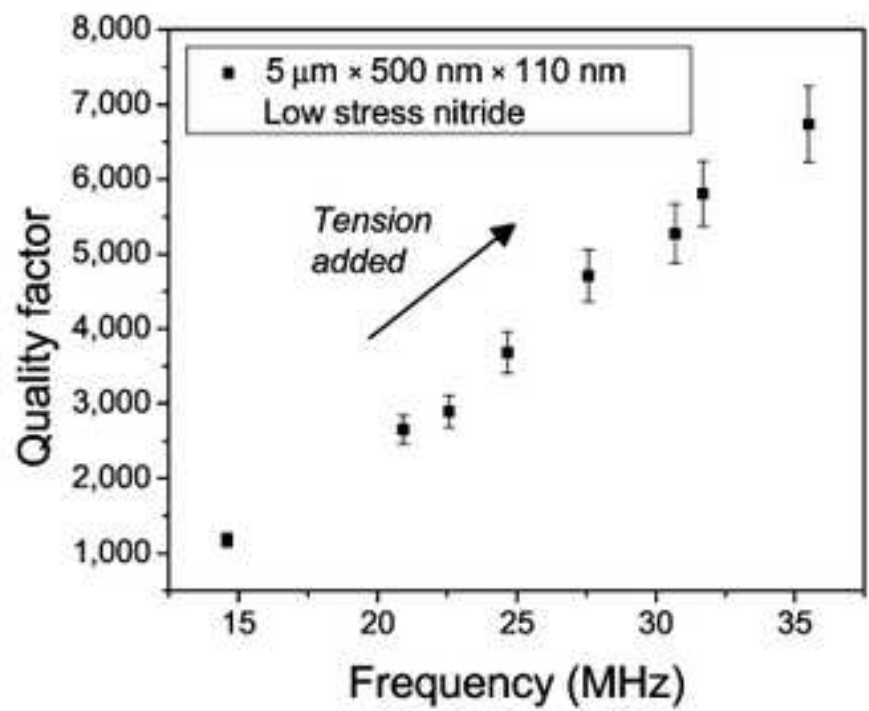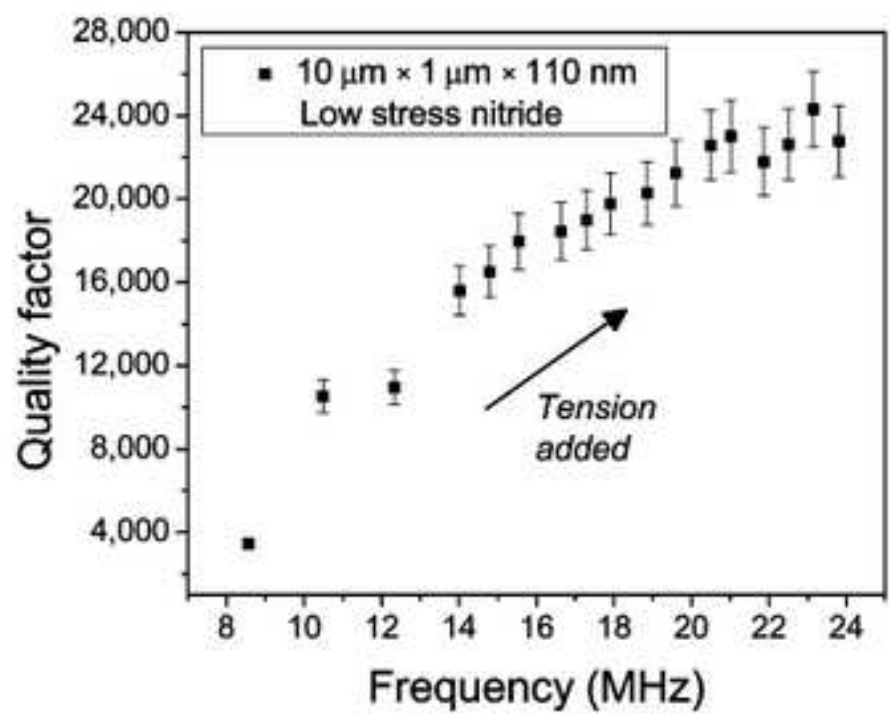

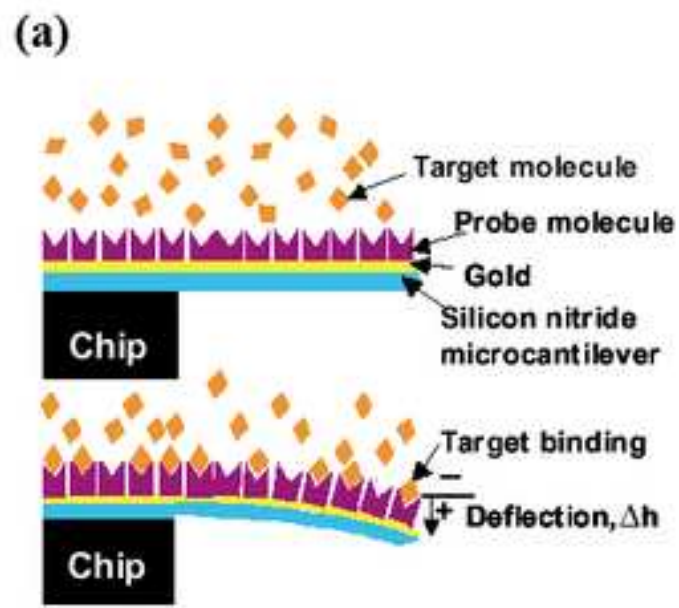
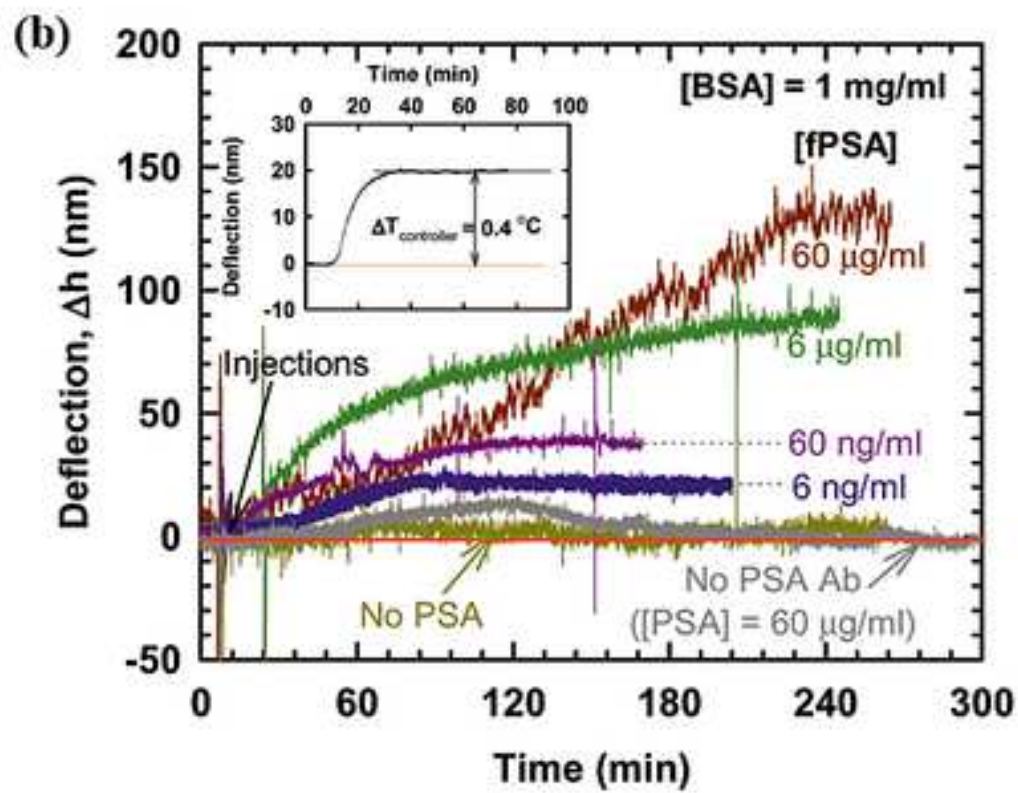

**Figure 9**
**Click here to download high resolution image**

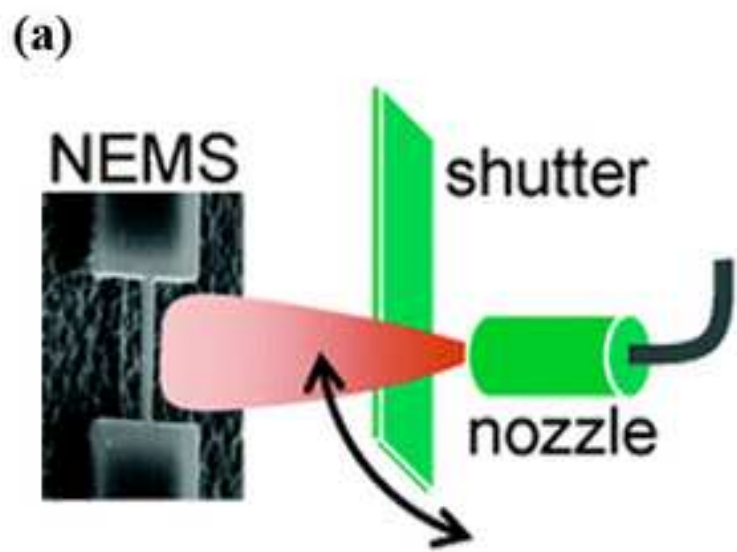
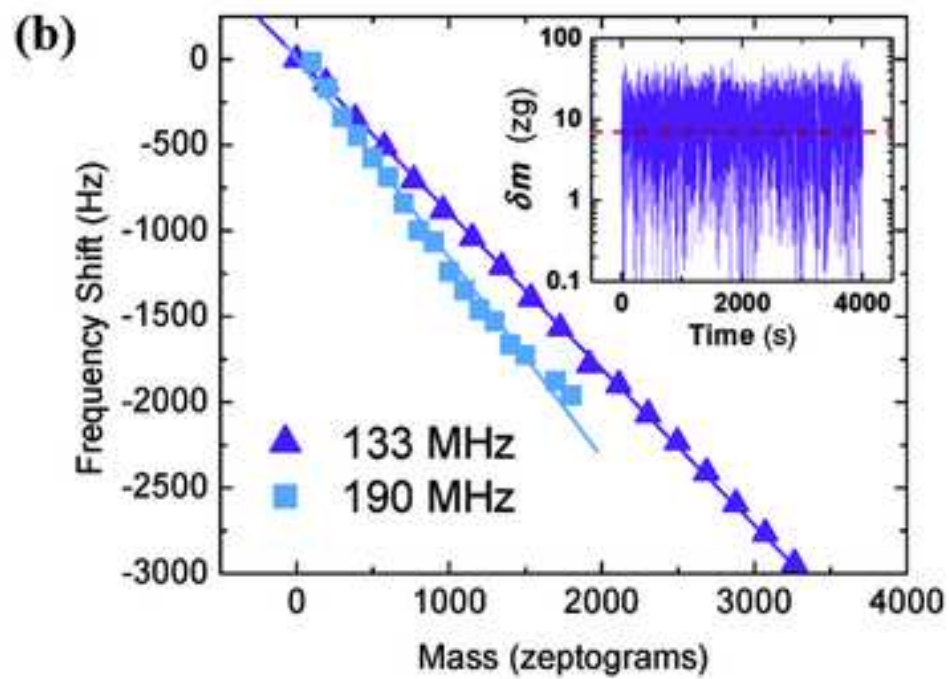

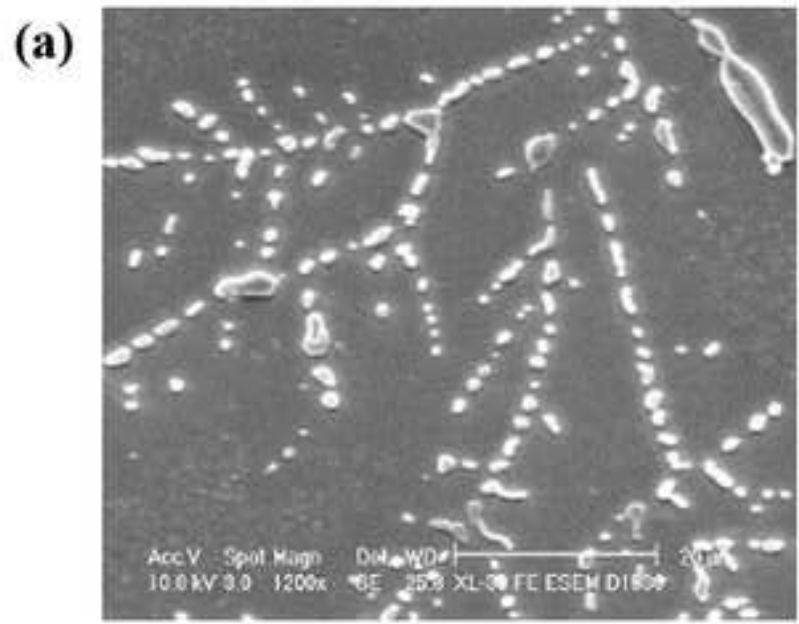 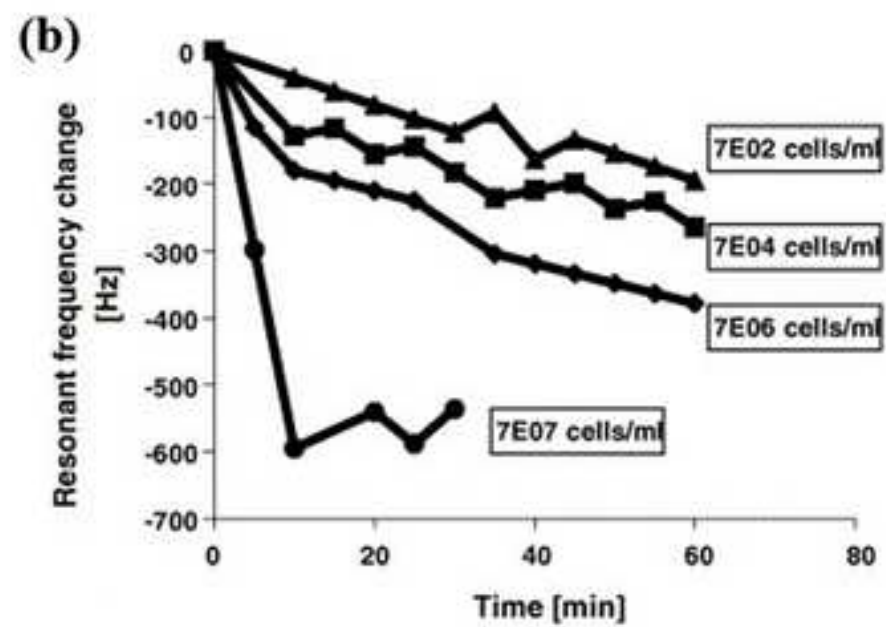

**Figure 11**
**Click here to download high resolution image**

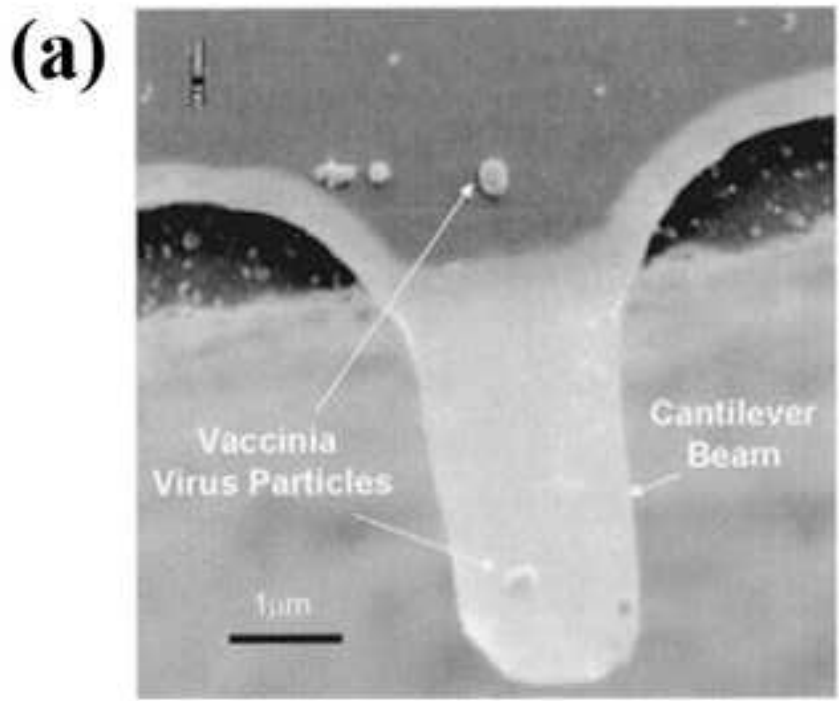 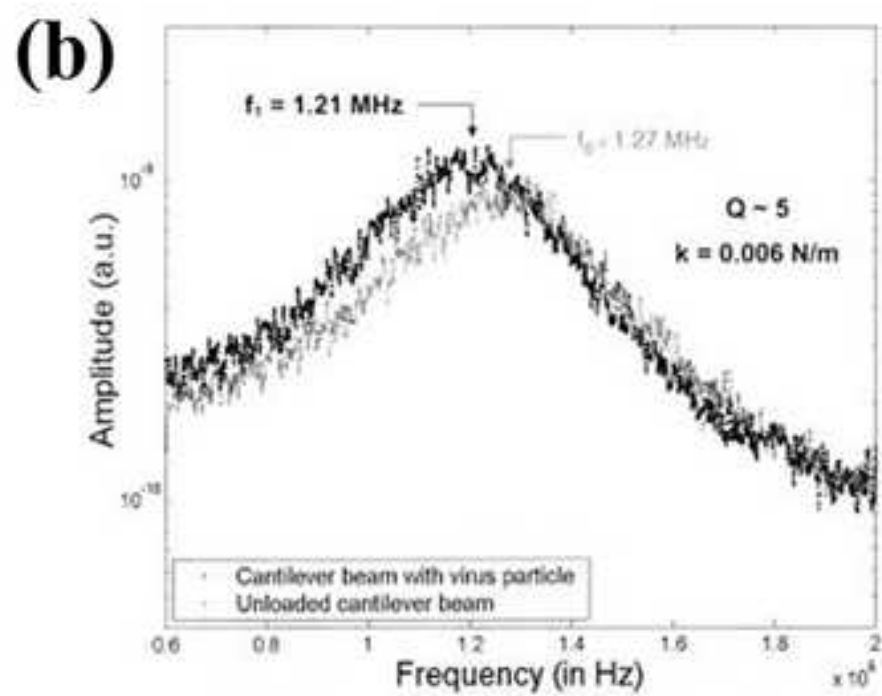

**Figure 12**
**Click here to download high resolution image**

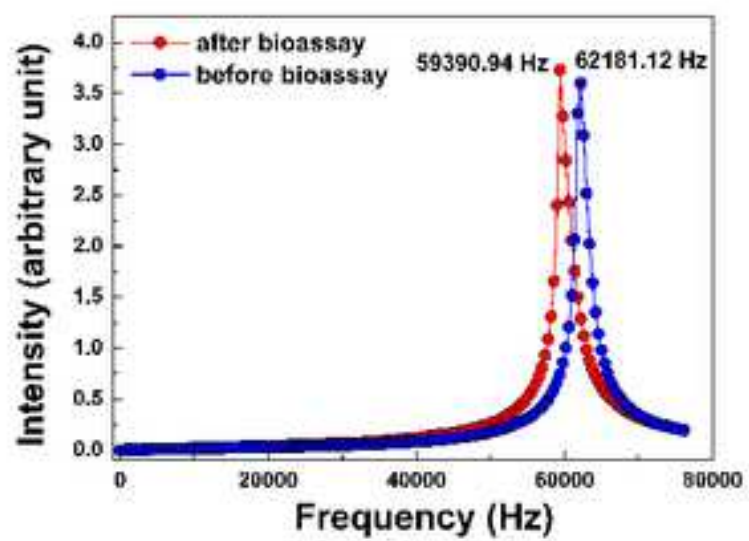 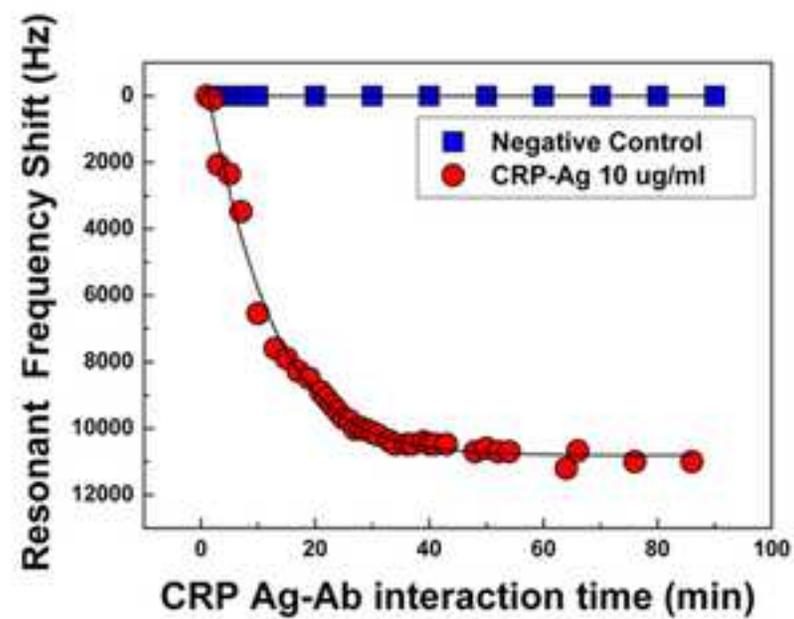

**Figure 13**
Click here to download high resolution image

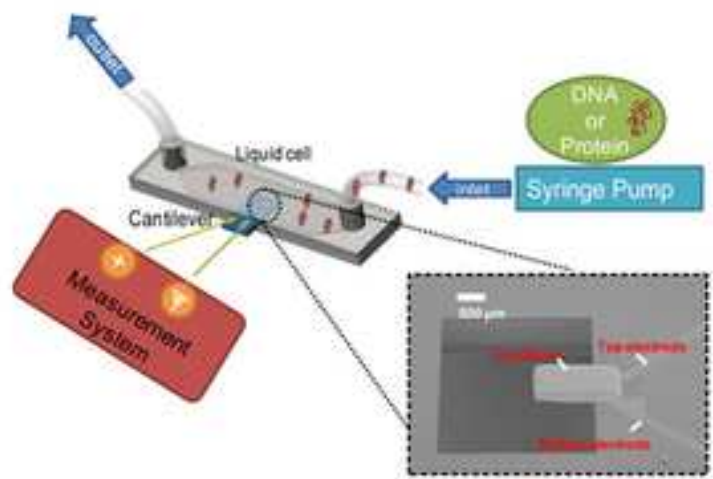 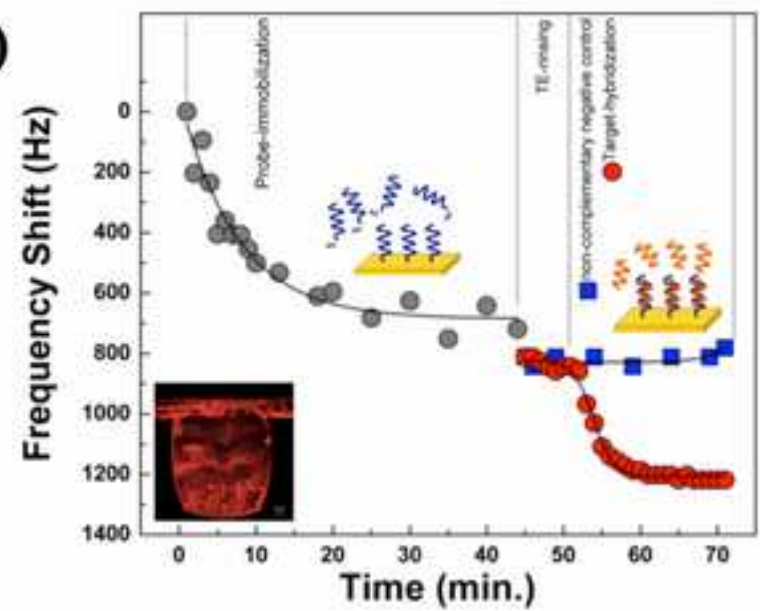

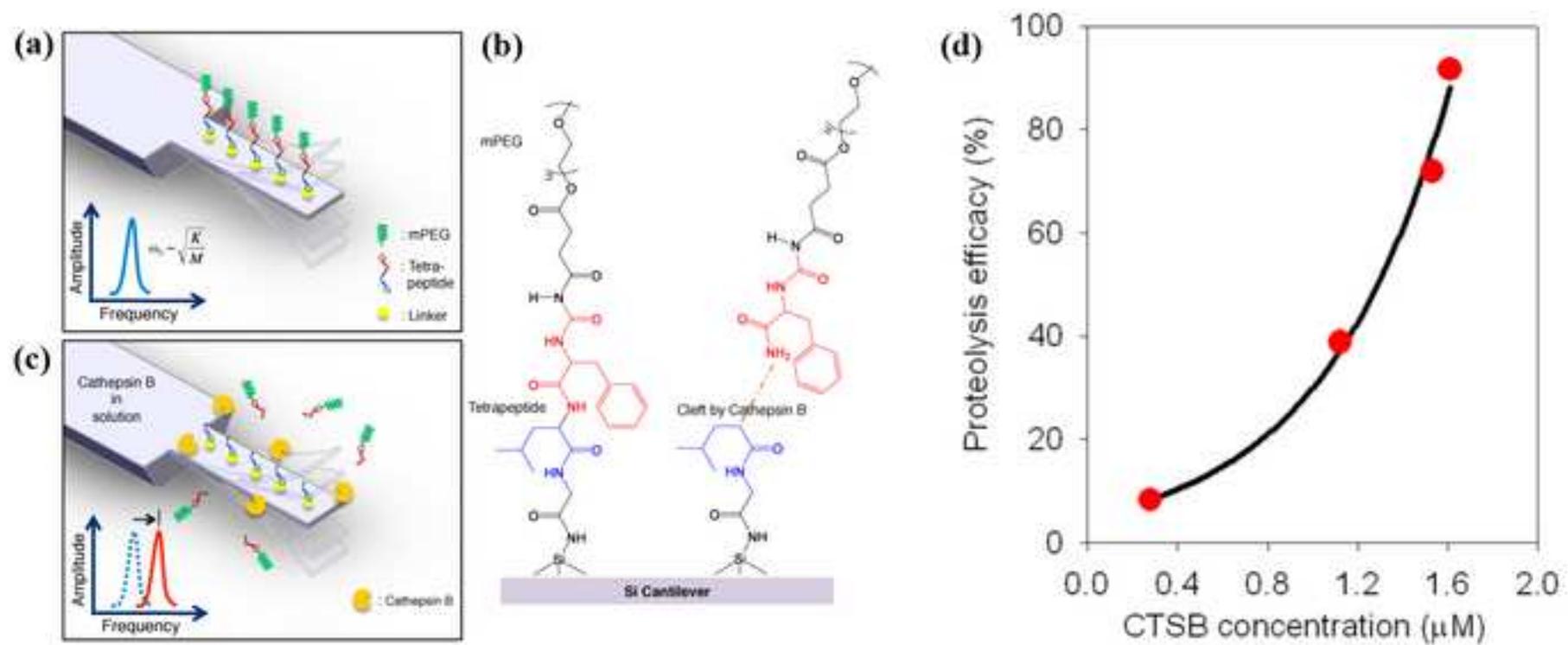

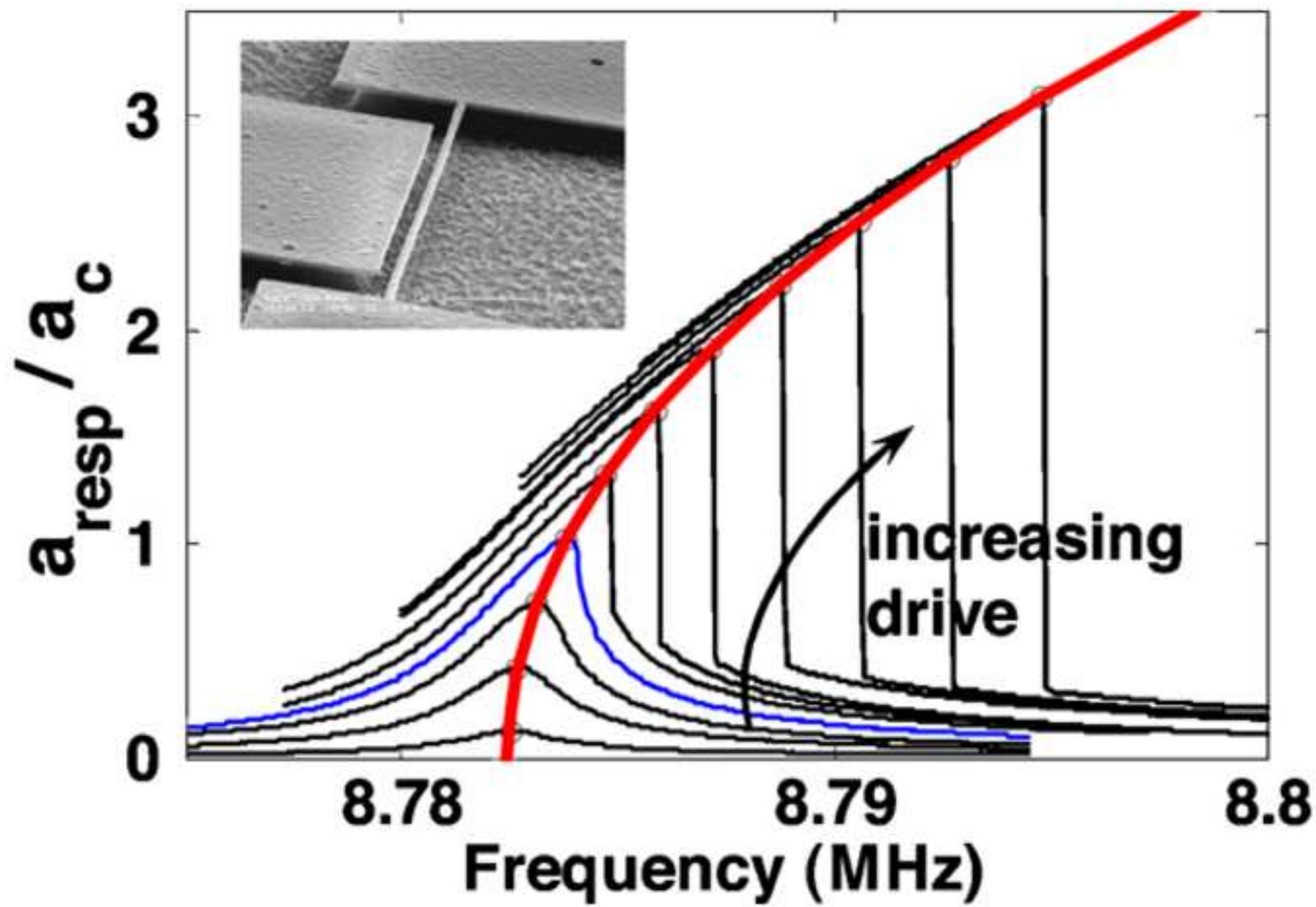

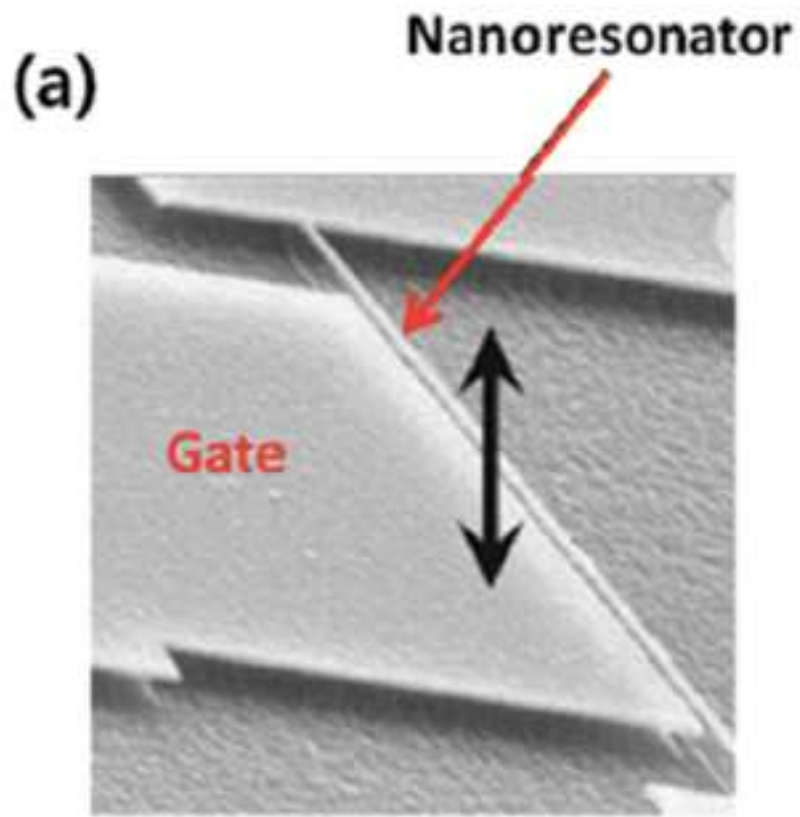 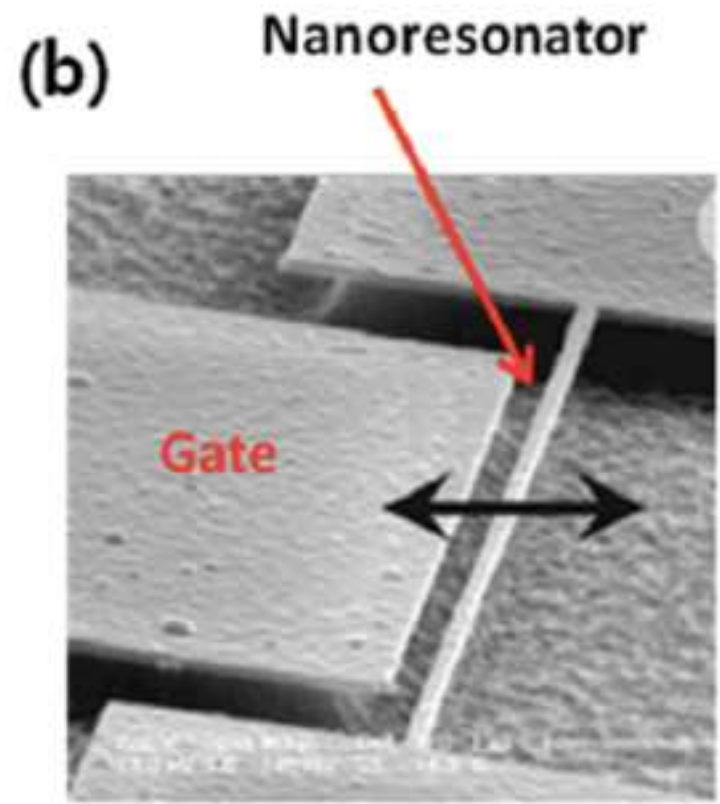

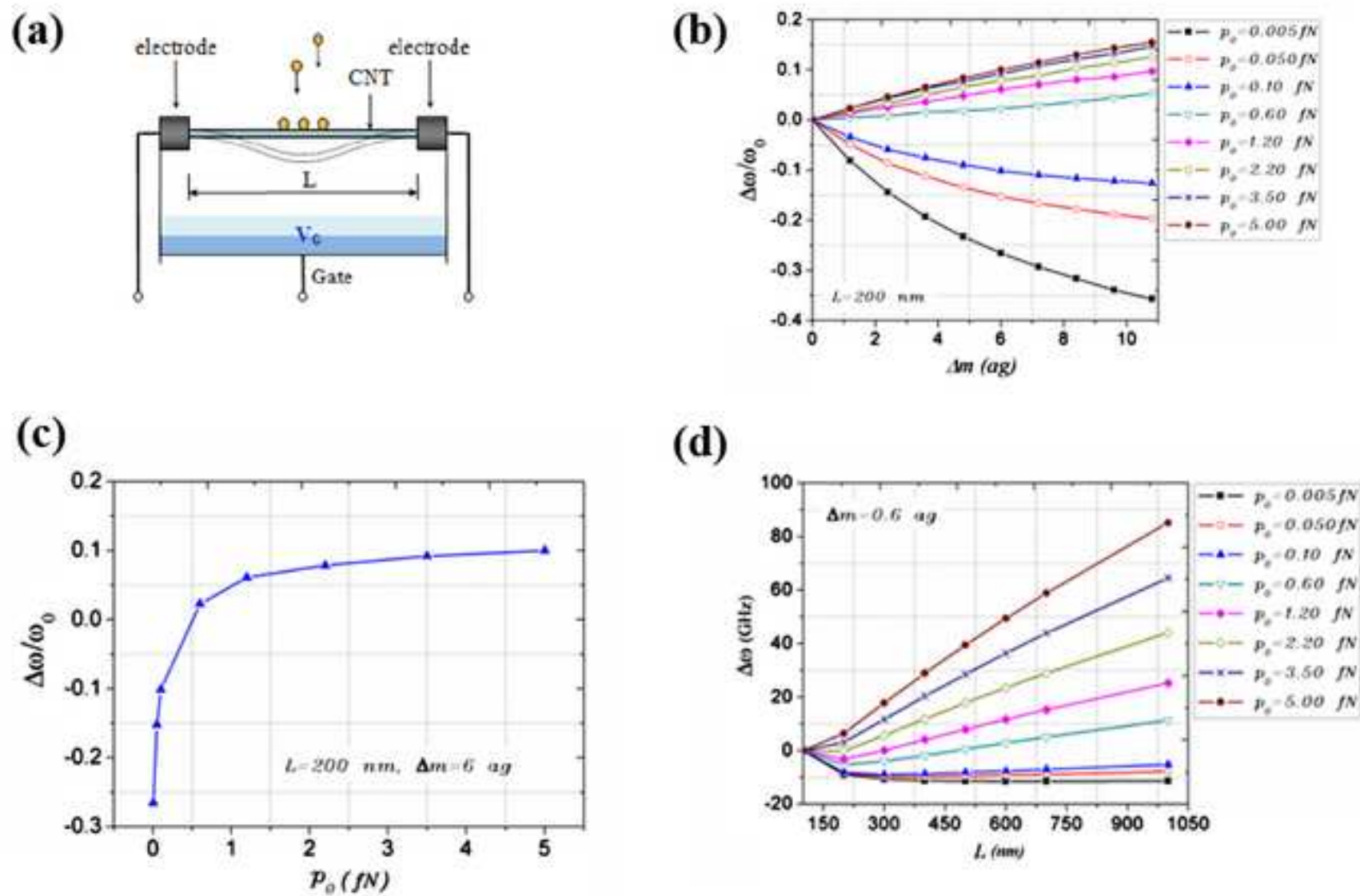

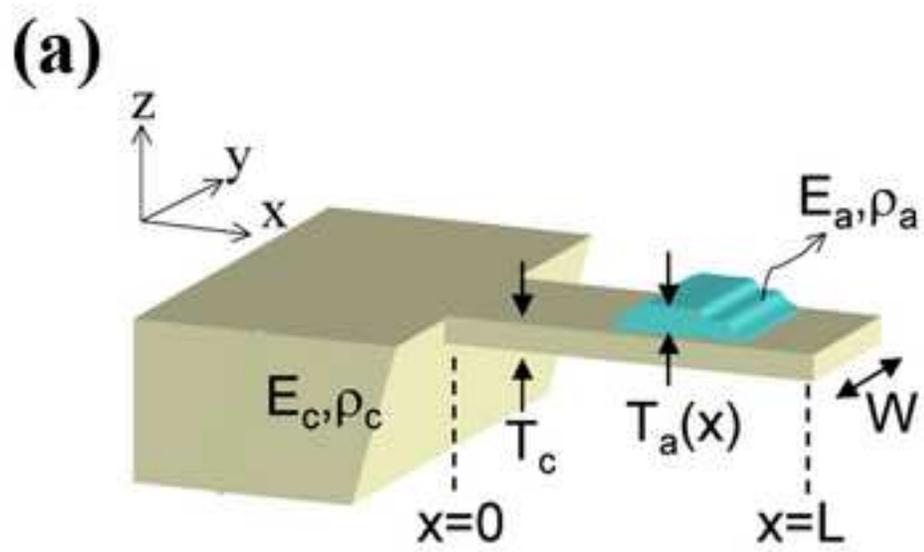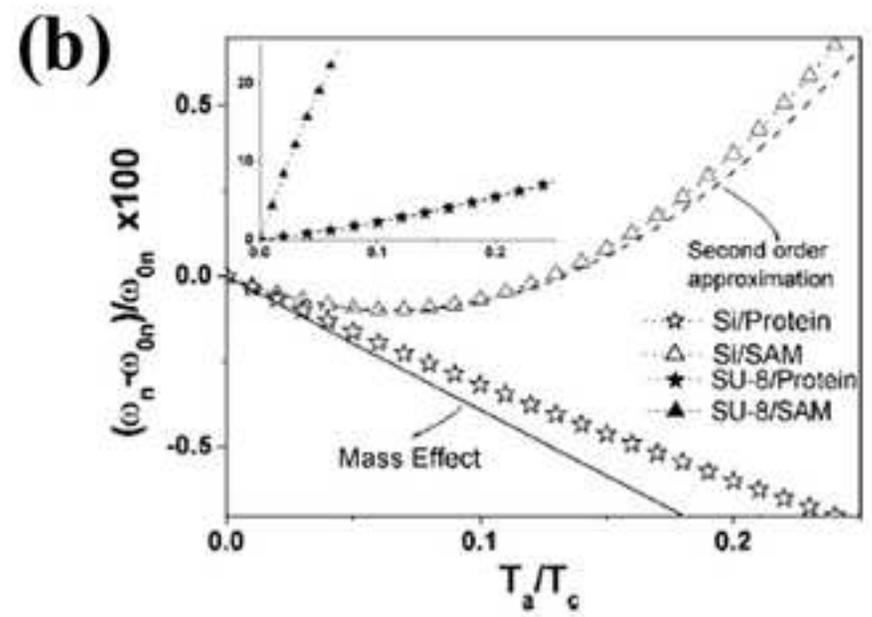

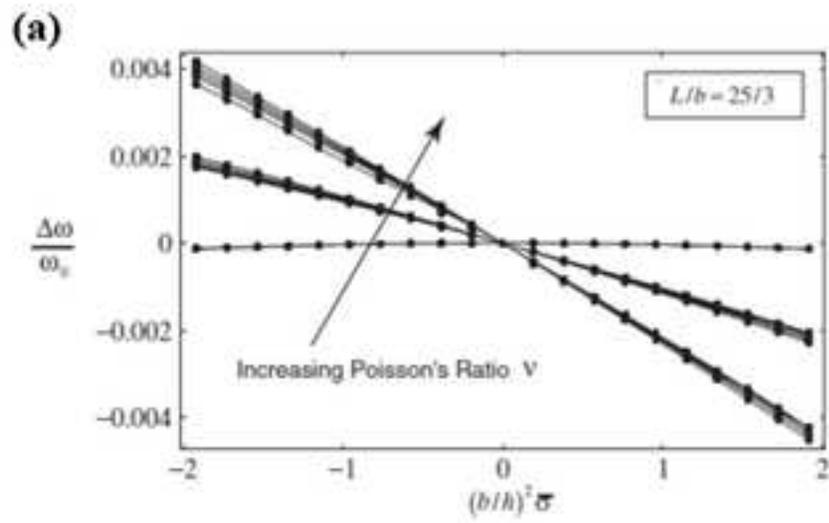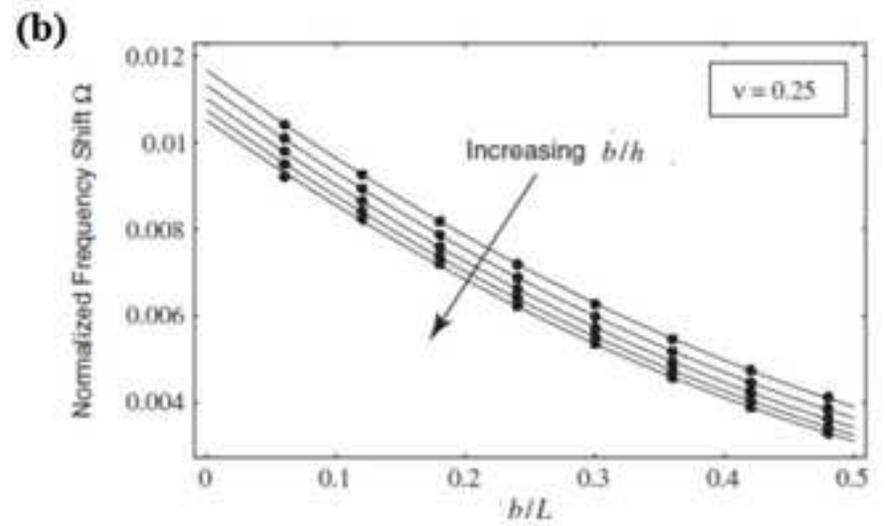

**Figure 20**
Click here to download high resolution image

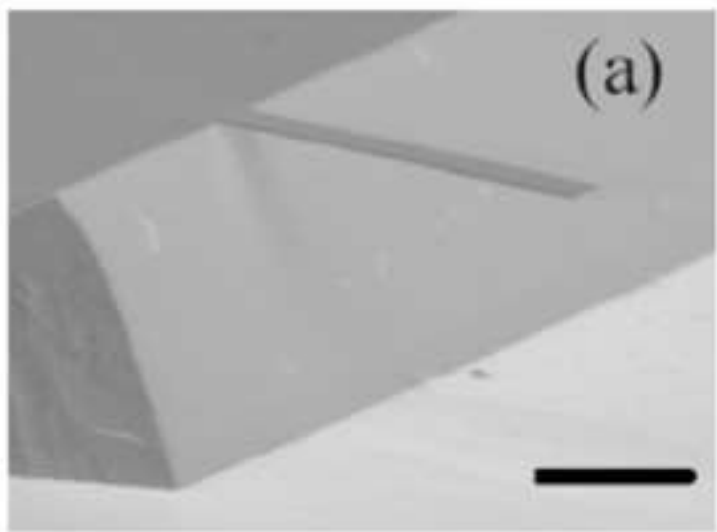
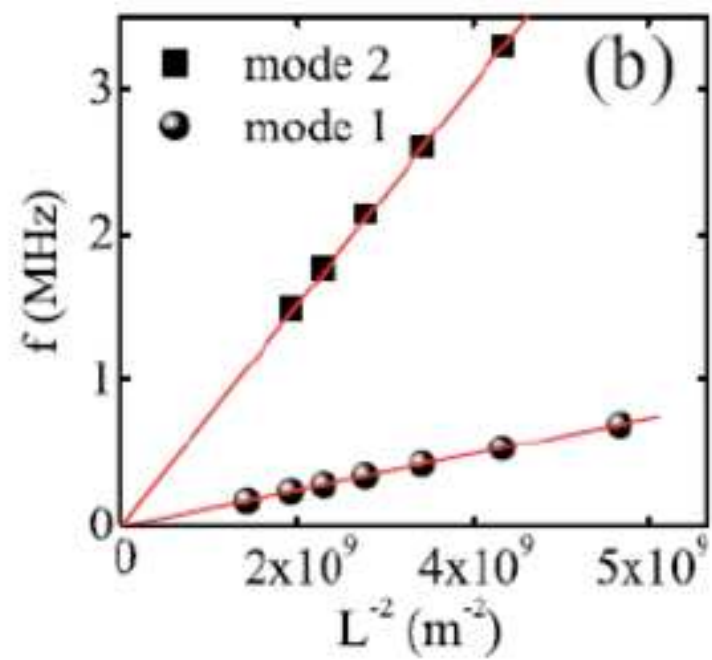
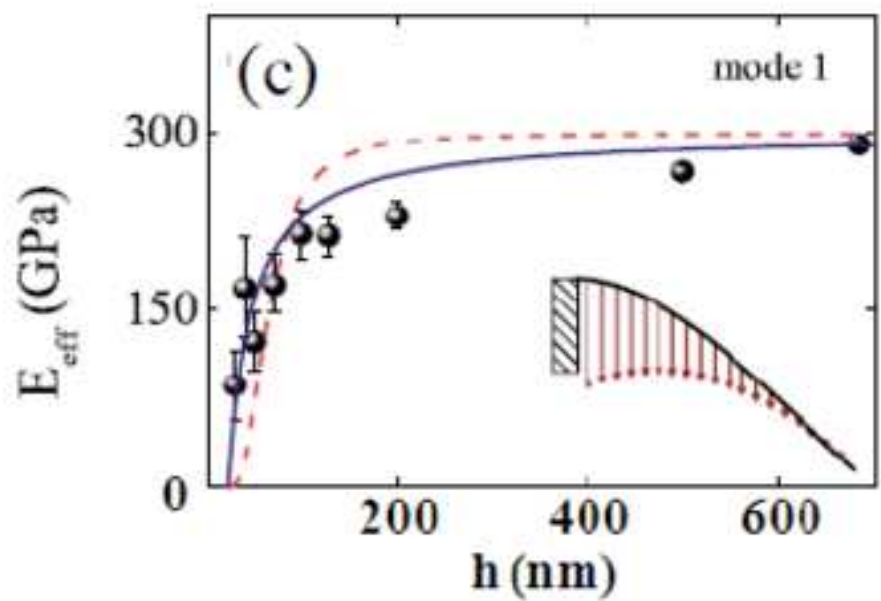
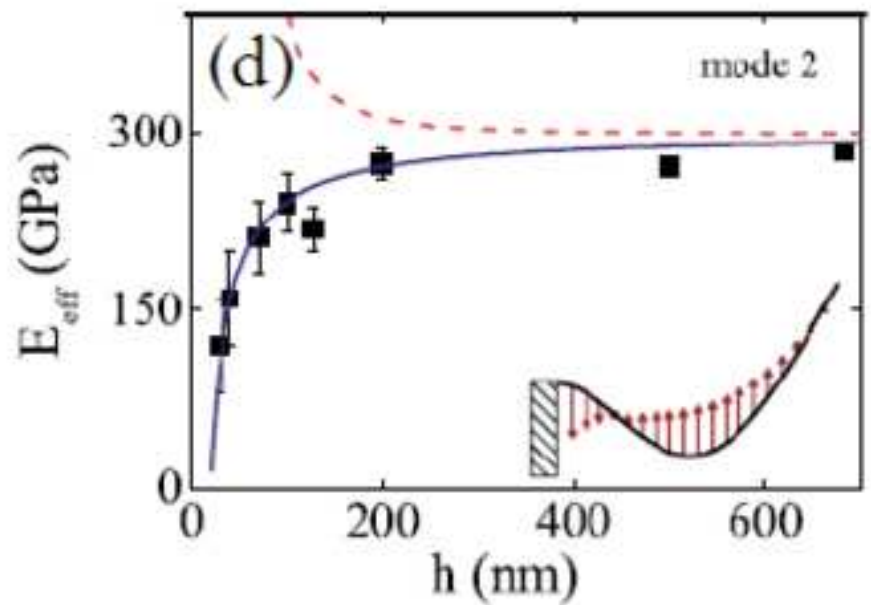

**Figure 21**
[Click here to download high resolution image](#)

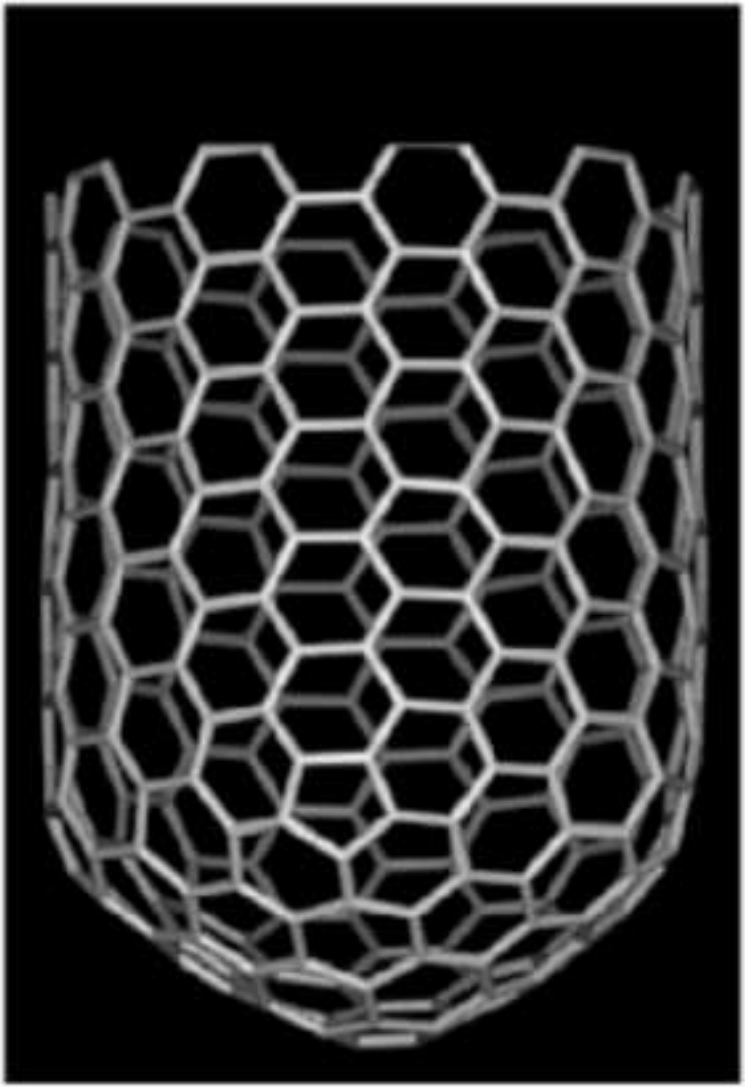
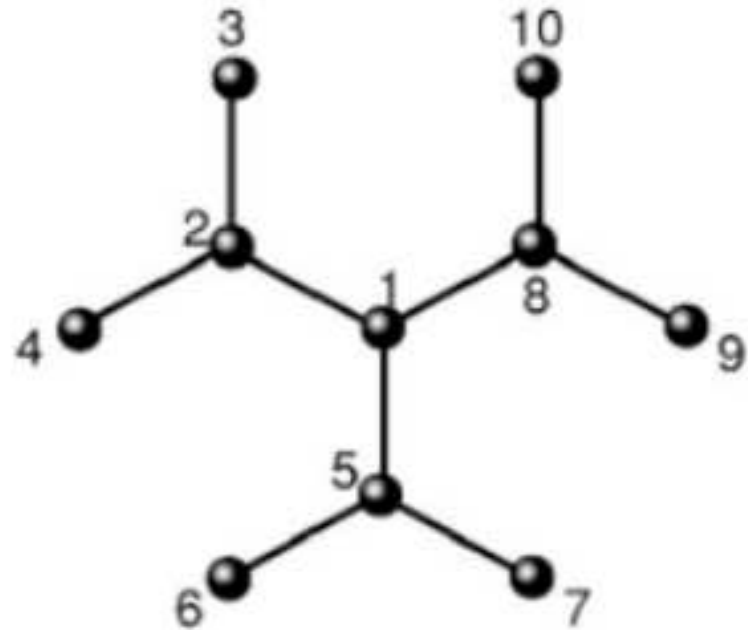

**Figure 22**
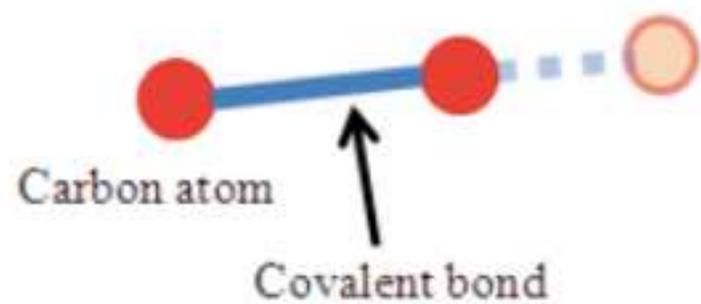

**(a) Bond Stretch**

Carbon atom / Covalent bond

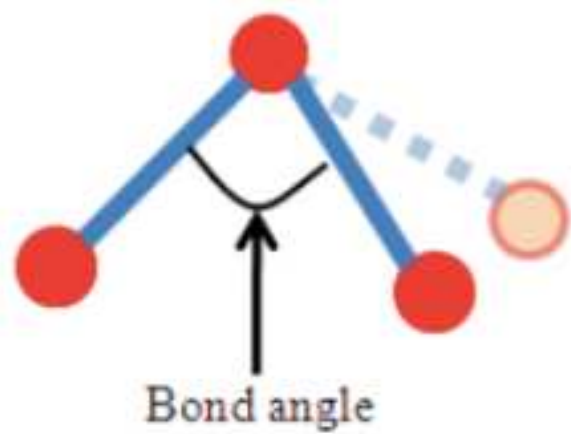

**(b) Bending of Bond Angle**

Bond angle

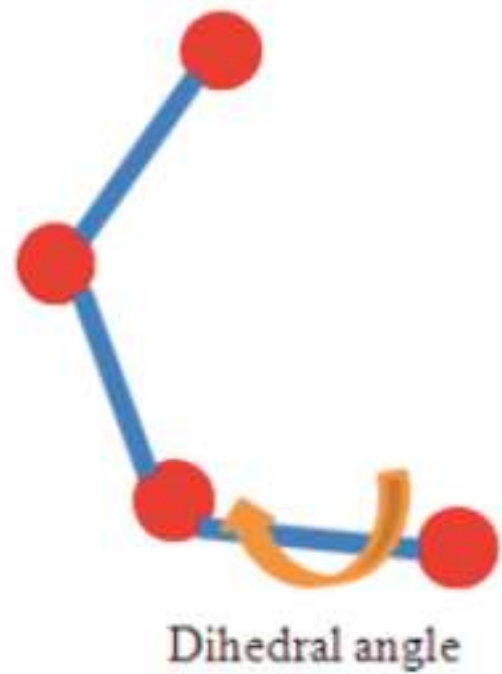

**(c) Torsion of Dihedral Angle**

Dihedral angle

**Figure 23**
Click here to download high resolution image

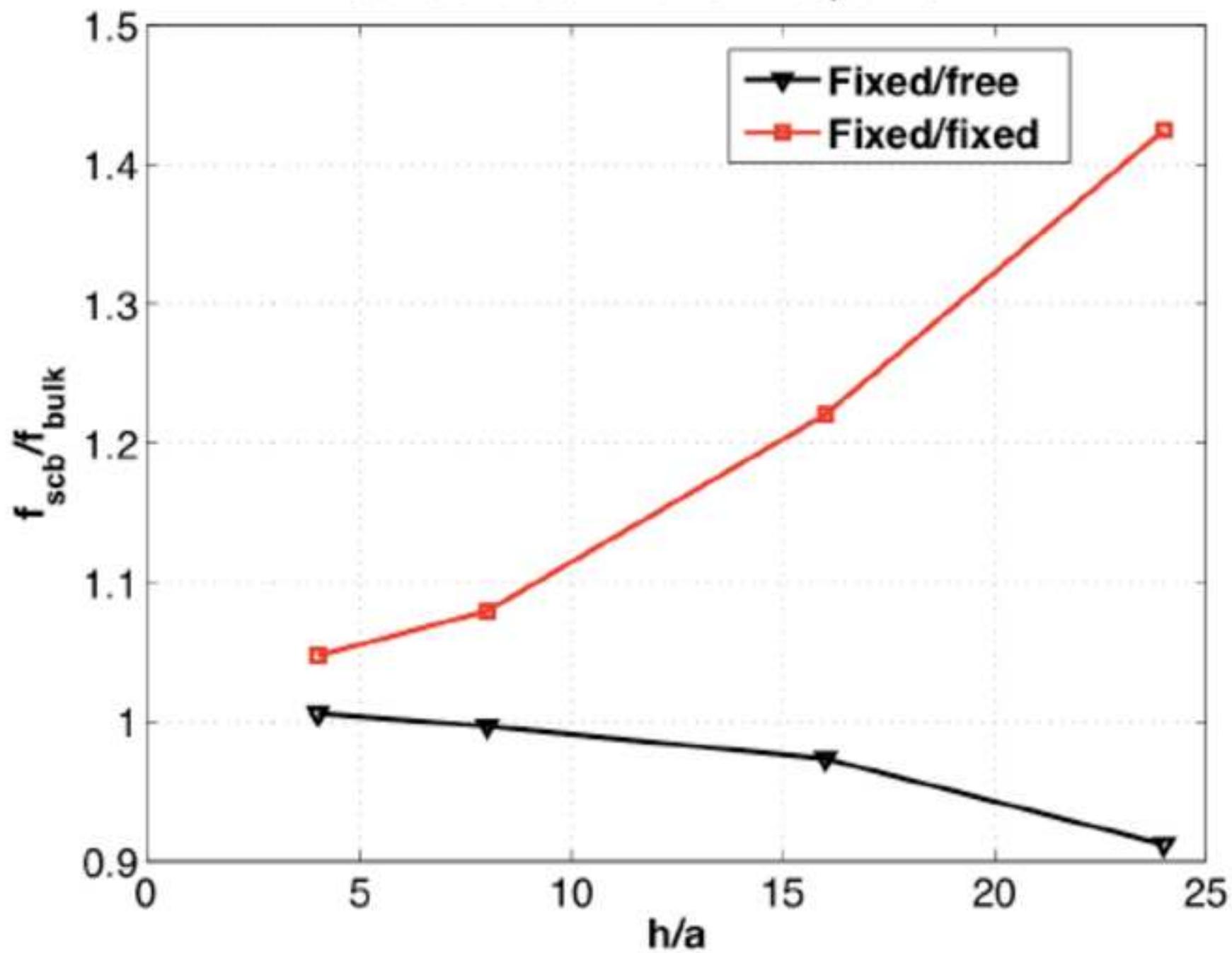

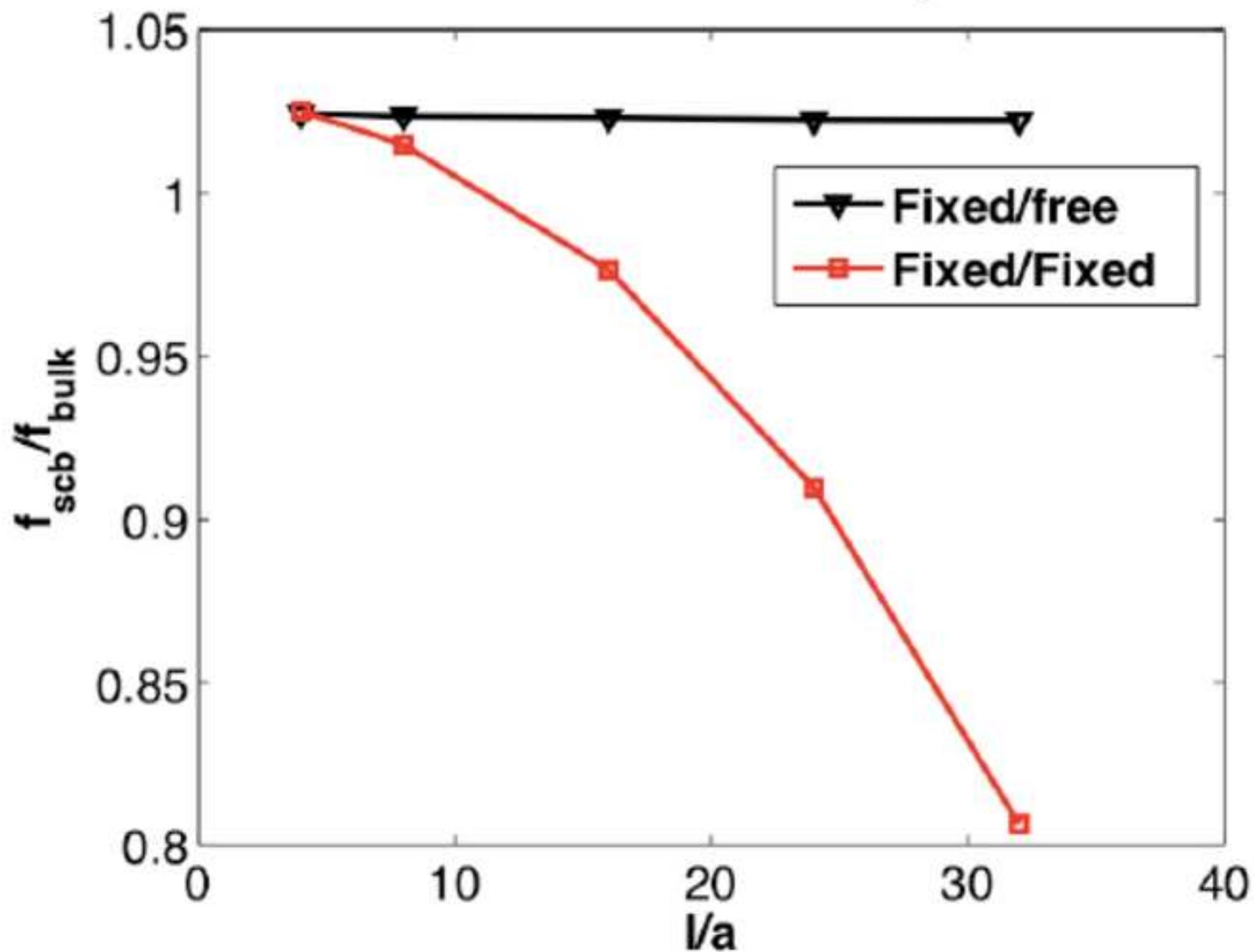

**Figure 25**
[Click here to download high resolution image](#)

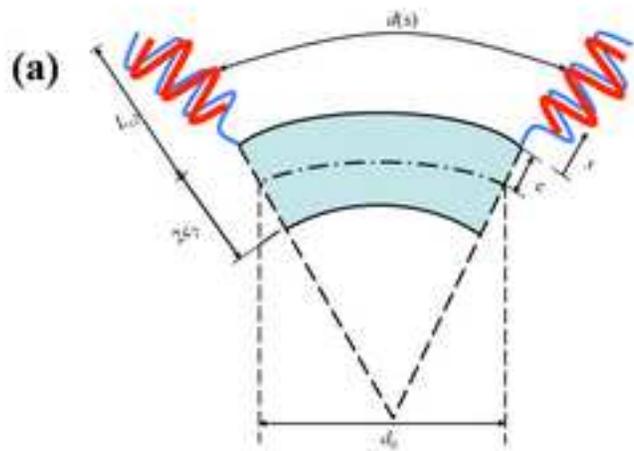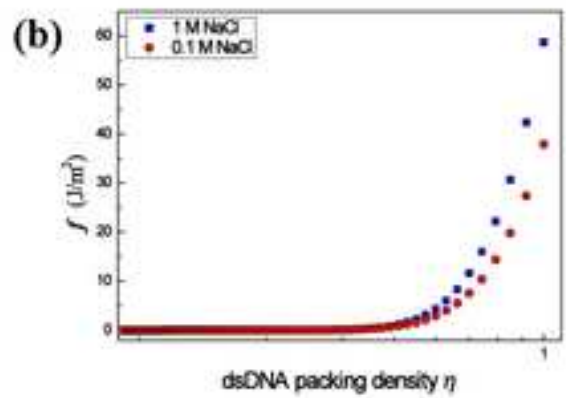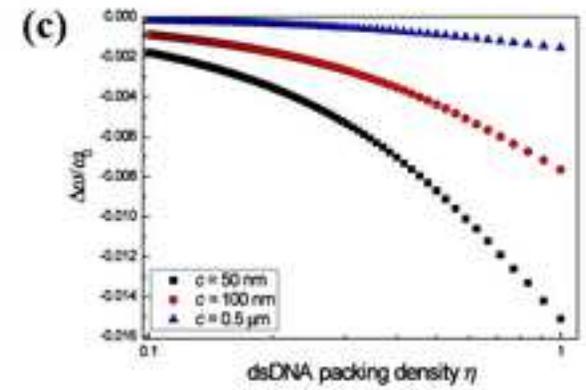

**Figure 26**
[Click here to download high resolution image](#)

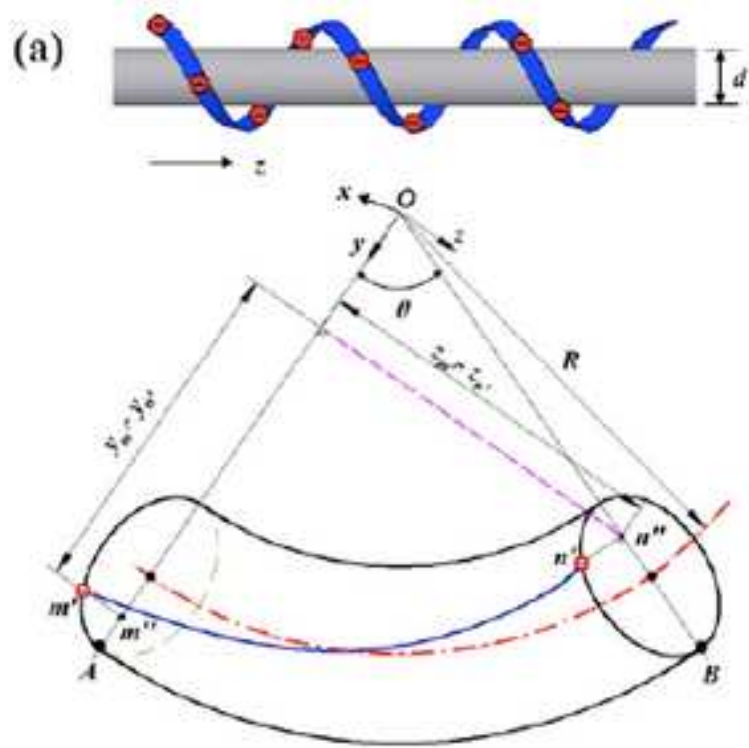
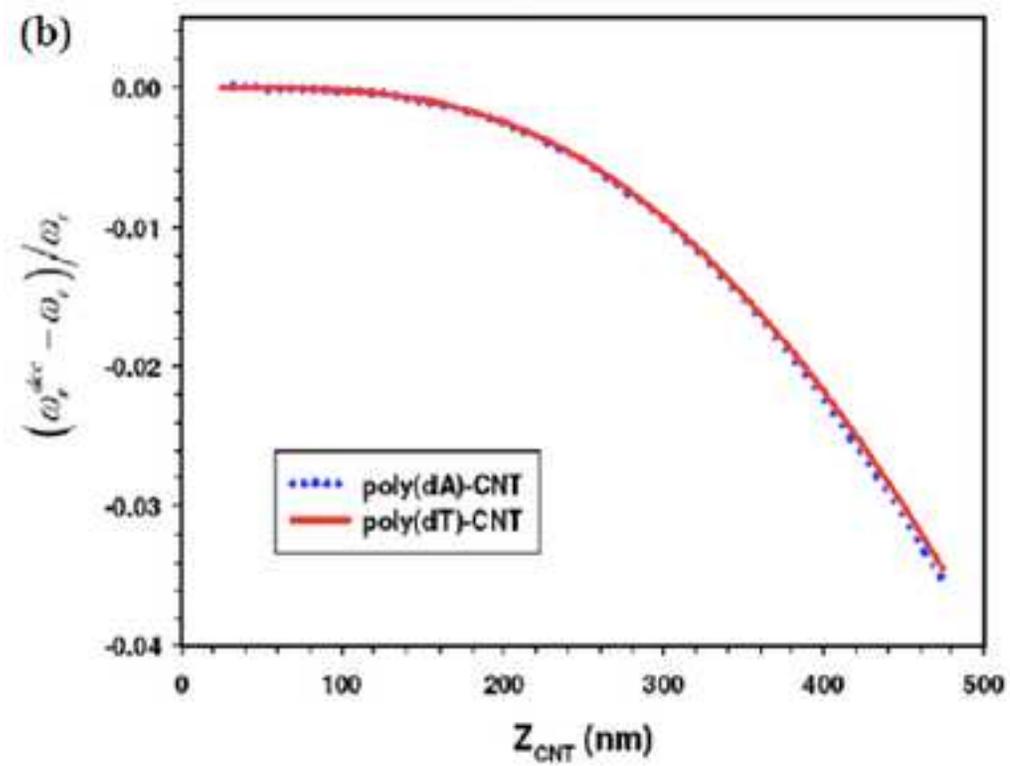